\documentclass[nolinenumbers,twocolumn]{aastex631}

\usepackage{subfigure}
\usepackage{rotating}
\usepackage{longtable}
\usepackage{pythonhighlight}
\usepackage{multirow}

\shorttitle{Variable stars in M31 stellar clusters}
\shortauthors{Smith et al.}
\graphicspath{{./}{figures/}}

\begin{document}

\title{Variable Stars in M31 Stellar Clusters from the Panchromatic Hubble Andromeda Treasury}

\correspondingauthor{Richard Smith}
\email{richard.smith@noirlab.edu}

\author[0000-0002-3346-2370]{Richard Smith}
\affiliation{NSF's National Optical-Infrared Astronomy Research Laboratory, Gemini North, 670 N Aohoku Place, Hilo, HI 96720, USA}
\author[0009-0000-5120-1193]{Avi Patel}
\affiliation{Division of Engineering and Applied Science, California Institute of Technology, Pasadena, CA 91125, USA}
\author[0000-0001-6360-992X]{Monika~D.~Soraisam}
\affiliation{NSF's National Optical-Infrared Astronomy Research Laboratory, Gemini North, 670 N Aohoku Place, Hilo, HI 96720, USA}
\author[0000-0001-8867-4234]{Puragra~Guhathakurta}
\affiliation{Department of Astronomy and Astrophysics, University of California Santa Cruz, 1156 High Street, Santa Cruz, CA 95064, USA}
\author[0009-0008-2161-0509]{Pranav Tadepalli}
\affiliation{Georgia Institute of Technology, Atlanta, GA 30332, USA}
\author{Sally Zhu}
\affiliation{Stanford University, Stanford, CA 94305, USA}
\author[0009-0007-8904-4139]{Joseph Liu}
\affiliation{University of Southern California, Los Angeles, CA 90007, USA}
\author[0000-0002-6301-3269]{L\'eo Girardi}
\affiliation{Osservatorio Astronomico di Padova -- INAF, Padova 35136, Italy}
\author[0000-0001-6421-0953]{L.\ Clifton Johnson}
\affiliation{Center for Interdisciplinary Exploration and Research in Astrophysics (CIERA) and Department of Physics and Astronomy, Northwestern University, 1800 Sherman Avenue, Evanston, IL 60201, USA}
\author{Sagnick Mukherjee}
\affiliation{Department of Astronomy and Astrophysics, University of California Santa Cruz, 1156 High Street, Santa Cruz, CA 95064, USA}
\author[0000-0002-7134-8296]{Knut A.\ G.\ Olsen}
\affiliation{NSF's National Optical-Infrared Astronomy Research Laboratory, 950 North Cherry Avenue, Tucson, AZ 85719, USA}
\author[0000-0002-7502-0597]{Benjamin~F.\ Williams}
\affiliation{Astronomy Department, University of Washington, Seattle, WA 98195}

\begin{abstract}
Variable stars in stellar clusters can offer key constraints on stellar evolution and pulsation models, utilising estimates of host cluster properties to constrain stellar physical parameters. We present a catalogue of 86 luminous (${\rm F814W}<19$) variable stars in M31 clusters identified by mining the archival Panchromatic Hubble Andromeda Treasury (PHAT) survey using a combination of statistical analysis of sparse PHAT light curves and difference imaging. We determine the evolutionary phases and initial masses of these variable stars by matching them with theoretical isochrones generated using host cluster properties from the literature. We calculate the probability of PHAT photometry being blended due to the highly crowded nature of cluster environments for each cluster-variable star, using these probabilities to inform our level of confidence in the derived properties of each star. Our 86 cluster-variable stars have initial masses between $0.8\mbox{--}67 M_{\odot}$. Their evolutionary phases span the main sequence, more evolved hydrogen- and helium-burning phases, and the post-asymptotic giant branch. We identify numerous candidate variable star types: RV Tauri variables, red supergiants and slowly pulsating B-type supergiants, along with Wolf Rayet stars, $\alpha$ Cygni and Mira variables, a classical Cepheid and a possible super-asymptotic giant. We characterise 12 cluster-variable stars at higher confidence based on their difference image quality and lower blending probability. Ours is the first systematic study of variable stars in extragalactic stellar clusters leveraging the superior resolution of the \textit{Hubble Space Telescope\/} and demonstrating the unique power of stellar clusters in constraining the fundamental properties of variable stars.

\end{abstract}


\section{Introduction}\label{sec:intro}

Stellar clusters provide valuable insights into stellar evolution due to their constituent stars thought to have largely formed from the same material at the same time, resulting in stars with homogeneous metallicities and ages. This leaves stellar mass as the only significant variable between cluster members, allowing models of stellar evolution at measured cluster age and metallicities to be tested. 

By examining clusters of different ages, we can understand the evolution of star formation in a galaxy at different epochs \citep{Johnson2017}.
However, interstellar dust dominates the galactic plane of the Milky Way, limiting the scope of stellar population studies in our galaxy \citep{marshall2006}, complicated further by differing dust densities, and thereby extinction laws, depending on sight lines \citep{Zasowski2009, Gonzalez2014, Wang2019, Saha2019, Zhang2023}. 
Extragalactic clusters have provided a solution to expanding the study of stellar populations in various environments, such as starburst galaxies (e.g. \citealt{Lim2013} and references therein). 
However, it is a challenge to resolve these distant sources into stars, including using instruments such as the \emph{Hubble Space Telescope} ({\sl HST\/}). This in turn inhibits our ability to perform photometry on sources within clusters and thereby constrain various ensemble parameters such as cluster age and metallicity.
M31 offers an ideal laboratory for studying stellar evolution, containing individual stars resolvable by \emph{HST} thanks to its close-proximity, while also hosting stars of all ages \citep{Ferguson2005, Bernard2012, Williams2017} and a large number of clusters \citep{Johnson2015}.

Variable stars in clusters are fundamental probes for studying both stellar pulsation and evolution \citep{Zhuo2021,Palakkatharappil2023}. 
A majority of the work on cluster-variable stars has focused on high amplitude variables such as 
Cepheids, which have been used to calibrate the Cepheid period-luminosity relation through main sequence fitting procedures \citep{Feast1999, Burke2002, Chow2004}. 
Period-age and period-age-color relations for Cepheids \citep{Bono2005, Anderson2016, Medina2021} have also been calibrated via  isochrone fitting of Cepheid-hosting clusters.
Recently, asteroseismology in low-amplitude variables in stellar clusters has gained increased attention due to its ability to inform models of stellar interiors and stellar evolution \citep{Templeton1997,Hekker2011}. 
In particular, such studies allow measurements of stellar masses, radii and ages \citep[e.g., red giant branch (RGB) stars in open clusters;][]{Hekker2011, Miglio2012, Pinsonneault2018}, providing further constraints for stellar models. 

Extragalactic time-domain studies of variable stars have also been pivotal in our understanding of stellar physics. \cite{Conroy2018} studied variability in M51 using \emph{HST} data to chart variability in the luminous regions of the color-magnitude diagram (CMD). \cite{Soraisam2020} examined the variability of massive stars in M31, finding increasing prevalence of variability in later spectral types, with redder stars displaying larger fluctuations than bluer stars (see their Figure~8). A key hindrance with these studies is a lack of physical parameters (e.g. mass) for the stars. Constructing a statistical sample of variables in extragalactic stellar clusters will help us better understand stellar variability by leveraging the estimates of host cluster properties providing constraints on stellar physical parameters. 

Historically, due to its large angular size, ground-based studies of variable stars in M31 are often limited in scope, focusing on field star populations in targeted OB associations or portions of the halo \citep{Massey1986, Mould1986, Pritchet1988, McConnachie2009}, though larger surveys and monitoring projects have also been conducted recently \citep[e.g.,][]{Lee2012PA1,Soraisam2020}. The resulting photometry was typically limited to only the brightest field stars in such studies due to the low angular resolution available from ground-based observations. Additionally, due to their abundance and proximity, globular clusters in the M31 halo have been extensively catalogued and studied  using ground-based telescopes \citep{Galleti2004, Peacock2010, Kang2012}. However, difficulties have arisen in resolving variable stars within clusters due to the low angular resolution of ground-based telescopes, largely as a result of atmospheric effects.

The Panchromatic Hubble Andromeda Treasury (PHAT) survey is a multi-cycle \emph{HST} program covering approximately one-third of M31's disk and most of its bulge in six different wide-band filters, from the ultraviolet to near infrared bands, at high resolution \citep{dalcanton2012, Williams2014}. More than 2000 stellar clusters have been identified in the PHAT survey data through the citizen-science Andromeda Project\footnote{http://www.andromedaproject.org} \citep{Johnson2015}. These clusters are a rich arena for studying stars in rapid phases of their evolution, such as Cepheid variables. 
Few studies have been carried out thus far to probe stellar variability in extragalactic stellar clusters. For example, by cross-matching the Pan-STARRS1 Cepheid catalog of \citet{Kodric2013} with the PHAT cluster catalog, 
\cite{Senchyna2015} identified potential Cepheids in M31 stellar clusters. However, a systematic study of variable stars in the PHAT clusters has not been done.

\begin{figure*}[!th]

  \centering
  \includegraphics[width=0.45\textwidth]{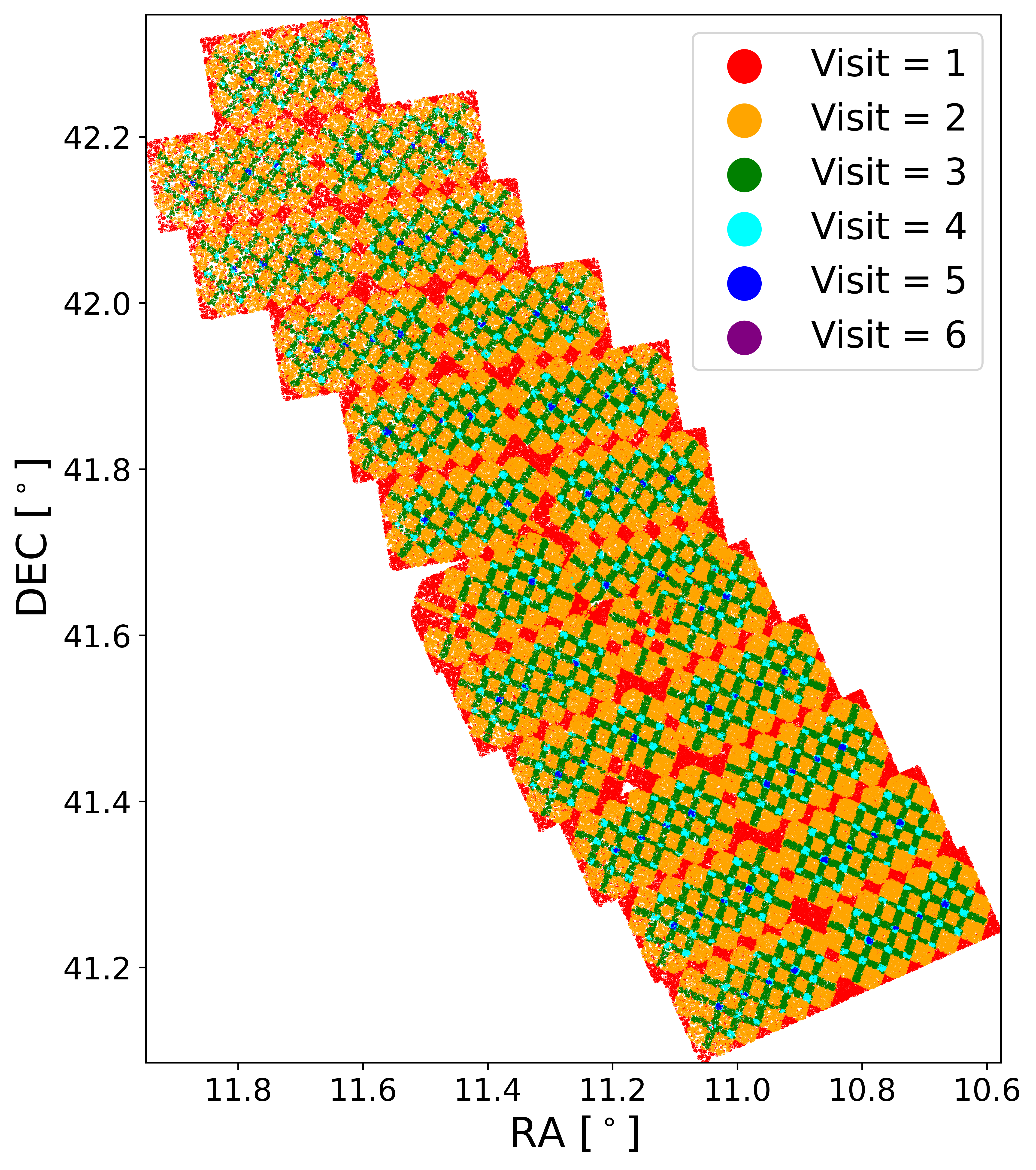}
  \includegraphics[width=0.45\textwidth]{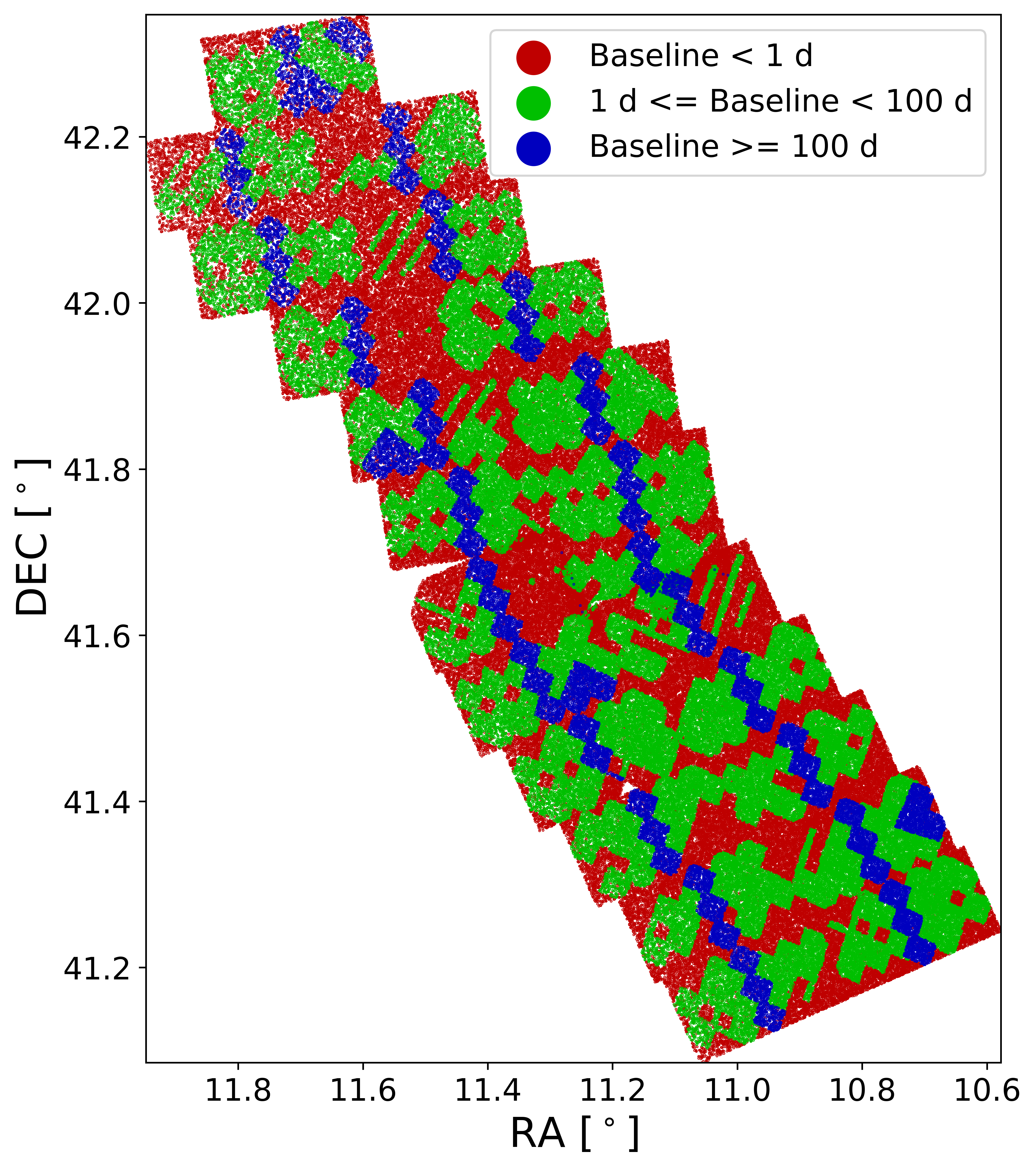}
  \caption{Left: Map of the number of visits by ACS/WFC in the F814W filter to a representative sample of $\sim200,000$ 
  stars identified across the PHAT survey footprint. 
  Right: Map of the temporal baseline of all observations in the F814W filter for the same sample of stars. We obtain similar maps when using observations from the F475W filter.}
\label{fig:PHAT}
\end{figure*}

In this work, we perform a systematic search for variable stars in M31 clusters exclusively using the PHAT survey data. We constrain both the evolutionary phase and initial mass of our resulting cluster-variable stars via isochrone fitting. We structure the paper as follows. First, we describe the PHAT survey design and highlight its time-domain component in Section~\ref{sec:data}. 
We describe the methods used to identify the cluster-variable stars in Section~\ref{sec:method}, including the difference imaging pipeline. We present the results from the application of the difference imaging pipeline and subsequent isochrone analysis of these variable stars in Section~\ref{sec:res}, along with an assessment of the effects of blending on our results. We describe the future direction of our study using the PHAT survey data in Section \ref{sec:future}, before we summarise our findings in Section~\ref{sec:con}.

We have made the tables containing the properties of the variable stars, along with their light curves, accessible via a public directory on the NSF's NOIRLab \emph{Astro Data Lab} \citep{Fitzpatrick2014DataLab} file storage system. 
A brief guide detailing how to access these data is provided in the Appendix.

\section{Data}\label{sec:data}

\subsection{PHAT as a time-domain data set}\label{phat_tda}

The \emph{HST} imaging for the PHAT survey has been conducted using the Advanced Camera for Surveys (ACS) Wide Field Channel (WFC), and the Wide Field Camera~3 (WFC3) ultraviolet/optical (UVIS) and near infrared (IR) channels, altogether providing coverage in six filters -- F275W and F336W with WFC3/UVIS, F475W and F814W with ACS/WFC, and F110W and F160W with WFC3/IR \citep{dalcanton2012}. The full survey area of $0.5~{\rm deg}^2$ is organized into 23 sub-areas called `bricks', each spanning $\sim 6\arcmin\times 12\arcmin$, which is formed by $3\times6$ contiguous WFC3/IR pointings (or `fields'). Each brick is observed in two $3\times3$ halves taken 6 months apart by ACS/WFC. Due to the relatively large \mbox{ACS/WFC} field-of-view, its pointings contain overlapping regions, leading to multiple measurements of sources in those regions and thereby facilitating variability studies.   

The photometric depth and spatial resolution (e.g., $\sim0.1\arcsec$ for the F814W filter; \citealt{Krist2003ACSPSFVariation}) make PHAT ideal for studying resolved stellar populations in clusters. Of the six available filters, 
the optical F475W and F814W bands of ACS/WFC achieve the greatest optical depth \citep[27.9 and 27.1 mag, respectively;][]{dalcanton2012}. This, along with their repetition in the survey thanks to the survey tiling strategy mentioned above, make these filters the main source of data in our study. For each object detected in the PHAT survey in a given band, its \emph{per-exposure} as well as \emph{combined} (across exposures) photometric measurements have been obtained by \citet{Williams2014} and \citet{Williams2018} using DOLPHOT \citep{DOLPHOT2000}. The combined photometric measurements are generated by summing the corresponding pixel values of the exposures after scaling for the respective exposure times and removing the most deviant values (see \citealt{DOLPHOT2000} for details).  The catalogs of both types of measurements are made available via the NOIRLab Astro Data Lab science platform \citep{Fitzpatrick2014DataLab} -- we use these catalogs for our study. 

Note that throughout this paper, we use the combined photometry for target selection and isochrone analysis, while the per-exposure photometry is used to construct light curves. 

The left panel of Figure~\ref{fig:PHAT} displays a map of the number of visits made by ACS/WFC (F814W filter) to a representative sample of $\sim$200,000 stellar sources identified in the PHAT footprint, showing that most sources have $1\mbox{--}3$ visits each. Note that the ACS/WFC exposure map shown in Figure~5 of \citet{dalcanton2012} roughly outlines a brick from this map. Each visit comprises a sequence of exposures. 
The temporal baseline -- the time between first and last observations across all visits in F814W for a given source -- over the PHAT footprint is also illustrated in the right panel of Figure~\ref{fig:PHAT}. We observe that a majority of sources have a baseline that is $<1$~d or $1\mbox{--}100$~d.

\subsection{PHAT star clusters}
\label{SubSec:StarClusters}

\begin{figure}
  \centering
  \includegraphics[width=80mm]{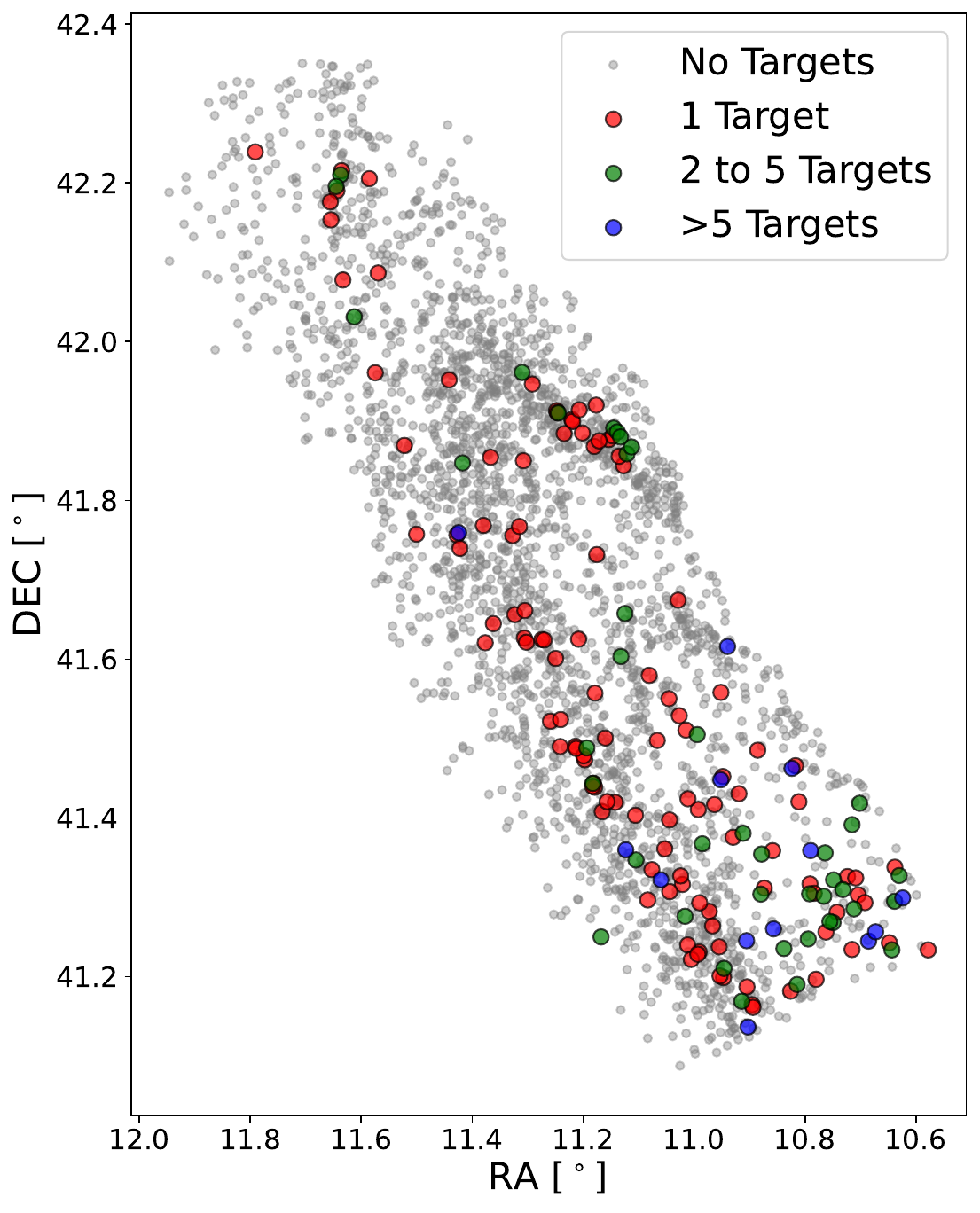}
  
  \caption{Distribution of the 2753 stellar clusters identified by \cite{Johnson2015} in the PHAT survey. Gray dots denote clusters excluded from our sample because they do not have any luminous (${\rm F814W}<19$) stars. The remaining red, green and blue circles indicate clusters with different numbers of luminous stars in our sample.}
\label{fig:targMap}
\end{figure}

We use the list of 2753 clusters catalogued by \cite{Johnson2015} in our study and query for stars from the PHAT survey projected in each cluster. The clusters are distributed across the entire PHAT footprint, as shown in Figure \ref{fig:targMap}. We identify candidate stars in each cluster using a search radius of $2\times r_{\rm eff}$, where $r_{\rm eff}$ is the half-light radius of the cluster as given in Table~C1 of \cite{Johnson2015}. We find a total of 294,981 stars within the identified PHAT clusters. 
Given the stellar crowding in these clusters, we only consider luminous stars, defined here as F814W magnitude $<19$~mag, for our variability analysis, since at higher stellar densities the limiting magnitude is brighter for ACS data \citep{Johnson2012}. 
After applying this magnitude threshold, we obtain 376 luminous stars within 166 host clusters (see Table~\ref{table:ClusterSample} for a summary of the number of stars identified at each step of our analysis).

\begin{figure*}
  \centering
  \includegraphics[width=0.52\textwidth]{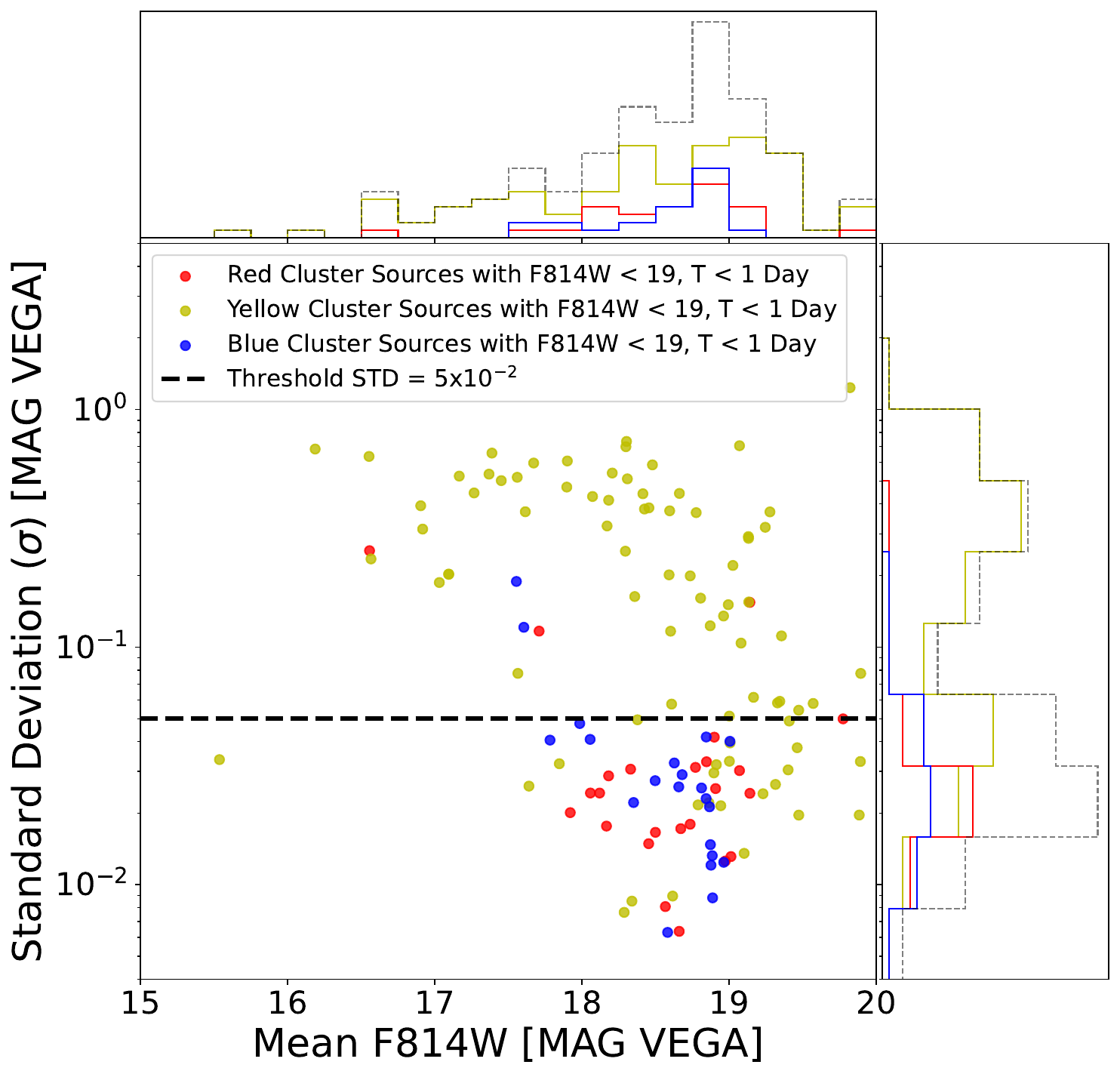}
  \includegraphics[width=0.35\textwidth]{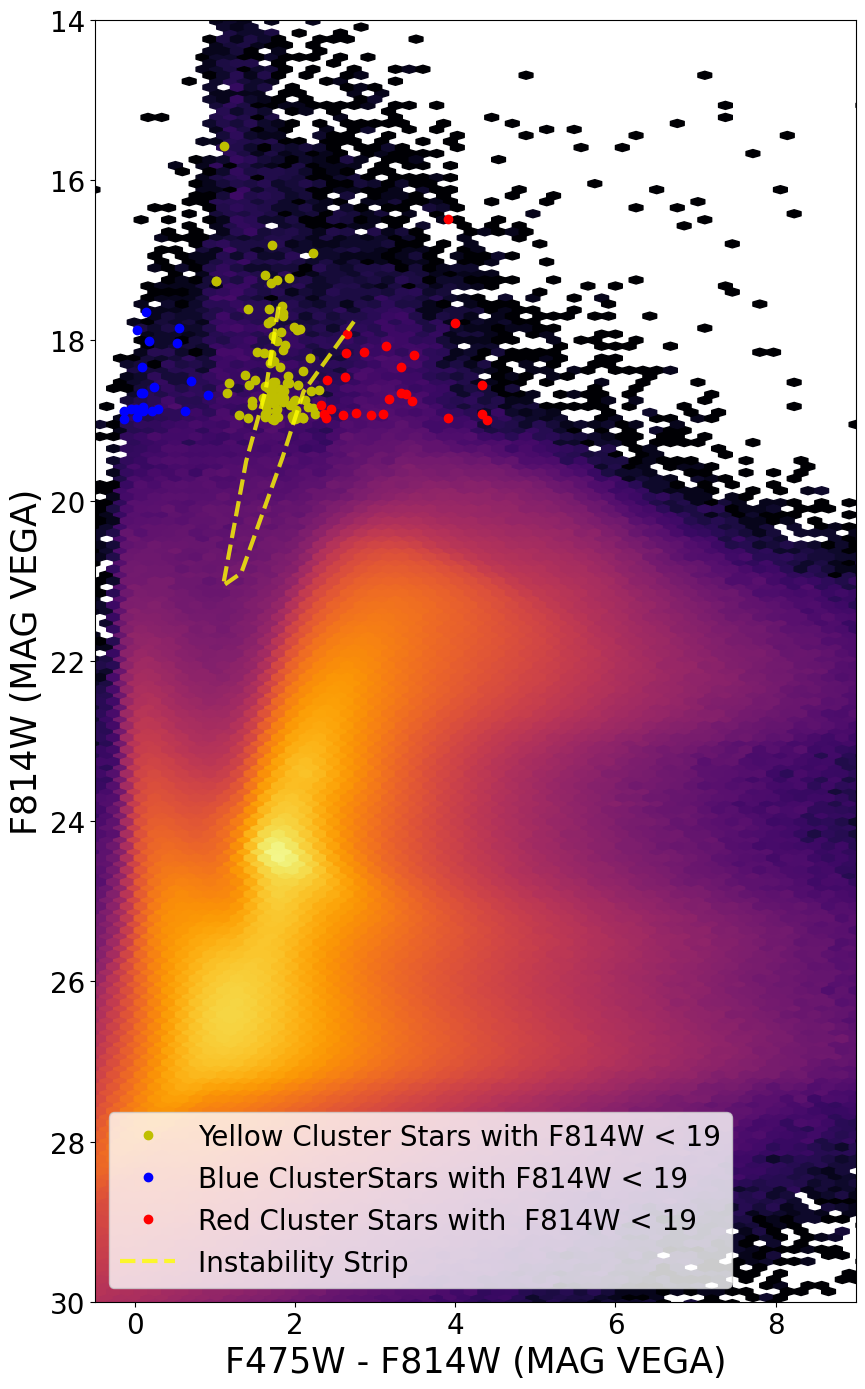}
  \caption{Left: Mean vs.\ standard deviation of the light curves of luminous (combined F814W magnitude $<19$) cluster stars with temporal baselines $<1$~d. As opposed to the combined magnitudes used to filter the luminous stars, the mean magnitudes on the x-axis are simple (i.e., unweighted) means of the light curves after applying a conservative threshold of sharpness-squared value less than 0.2 (see Section~\ref{phat_tda}). 
  Right: Color-magnitude diagram for the same luminous cluster stars. They are separated into blue, yellow and red cluster stars based on their F475W-F814W color positions on the CMD (see text for details). The background heat map is generated using the \citet{Williams2014} catalog of M31 stars. The yellow dashed line indicates the theoretical Cepheid instability strip from \cite{Fiorentino2002} for $Z=0.02$, 
  using a distance modulus of 24.47 \citep{McConnachie2005} and applying a foreground extinction of $\mathrm{A_{V} = 0.19}$ \citep[using the foreground reddening of $\mathrm{E(B-V) = 0.06}$ along the line of sight to M31;][]{Cordiner2011}{}.  
  We use stars with standard deviations $\lesssim0.05$ mag to calibrate the systematic error, chosen based on the distribution of standard deviations for presumed non-variable redder stars in the sample. 
  }
\label{fig:syserr}
\end{figure*}

\section{Method}\label{sec:method}

We present our methods for identifying variable stars, beginning with an analysis of the F814W light curves of the 376 luminous cluster stars selected above to produce a list of stars with evidence of variability. 
We then use the \emph{per-visit} images in F814W to confirm the variability of these stars via difference imaging. We define the \emph{per-visit} image of an object in a given filter as the image corresponding to a distinct brick and field combination obtained by combining all drizzled exposures at each observing epoch (visit). We download the per-visit ACS/WFC images used in this study from the Mikulski Archive for Space 
Telescopes (MAST) \citep{PHAT_MAST_2012}.

\subsection{Light Curve: Calibration of Systematic Errors}
\label{subsec:lightcurvecalibration}

The M31 stellar clusters are very crowded fields with median $r_{\rm eff}\sim0.44\arcsec$ \citep{Johnson2012}. The \emph{HST} photometry of stars within these clusters might therefore be contaminated due to crowding effects and, subsequently, the reported measurement errors may be underestimated. 
We account for this by applying a magnitude-dependent correction to the measurement errors of the per-exposure photometry values retrieved from the Astro Data Lab catalog \citep{Olsen2018AAS,Williams2018}. We refer to this correction as systematic error calibration.
We obtain this quantity by considering a sample of PHAT survey stars within M31 stellar clusters which have temporal baselines shorter than 1~d and are brighter than 19 mag in the F814W filter. 
We manually inspect the per-visit F475W and F814W filter images for each of these sources, discarding sources with images visibly affected by bad pixels to prevent introducing errors into our calibration. 
We reach a sample size of 129 stars after applying these cuts. 
We then calculate the mean magnitude and standard deviation ($\sigma$) of the F814W light curve for each of these sources after removing any measurements affected by cosmic ray strikes, bad pixels, etc.\ by thresholding on the \emph{sharpness} ($S$) stellar profile statistic available in the catalog\footnote{Note that this sharpness parameter is distinct from that described later in this work.}. The sharpness value is close to zero for a star-like profile; we use a threshold of $S^{2}<0.2$ (see also \citealt{Williams2014}).

\begin{figure*}[!t]
    \centering
    \includegraphics[width=.35\textwidth]{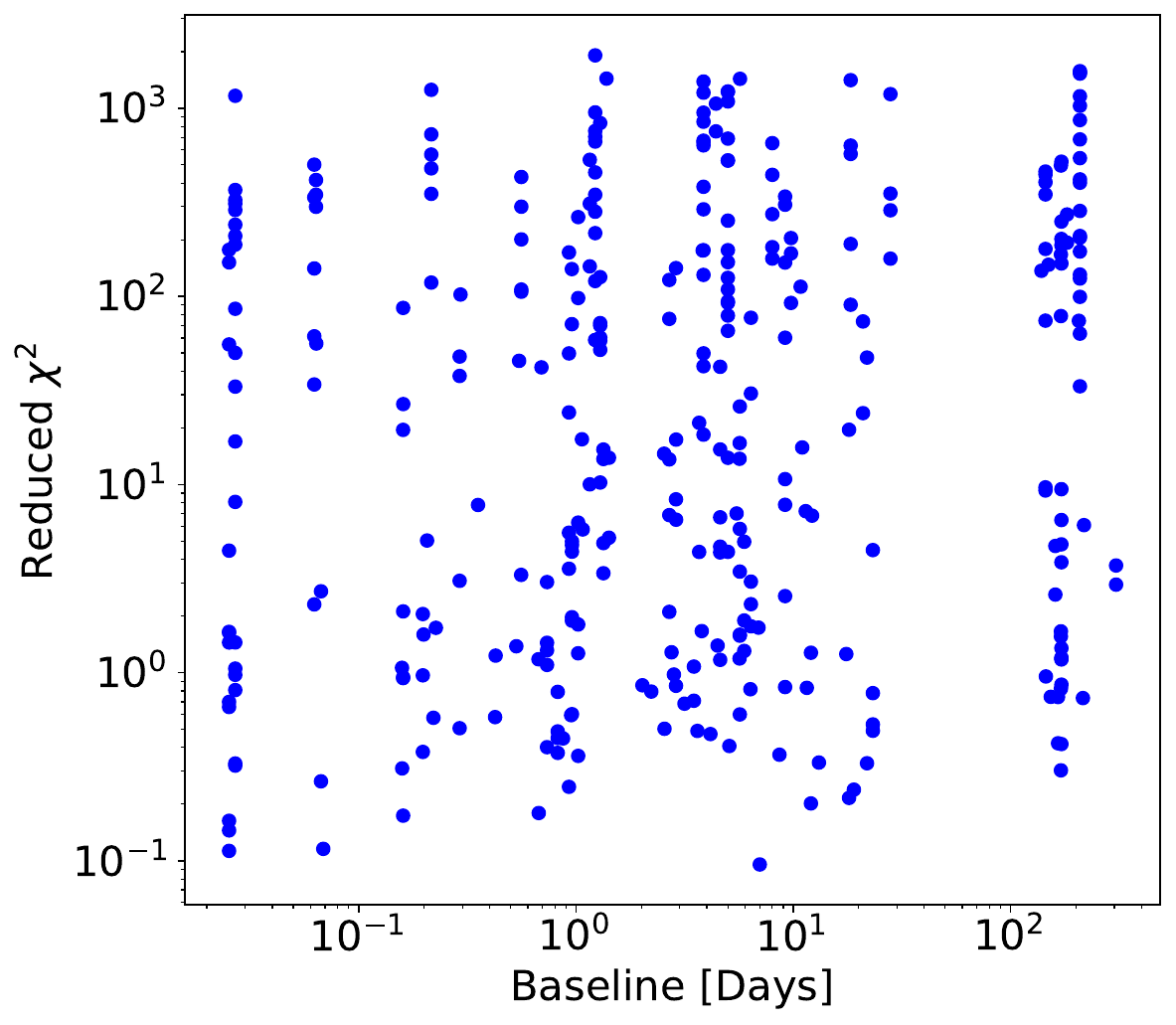}\hspace{4em} 
    \includegraphics[width=.35\textwidth]{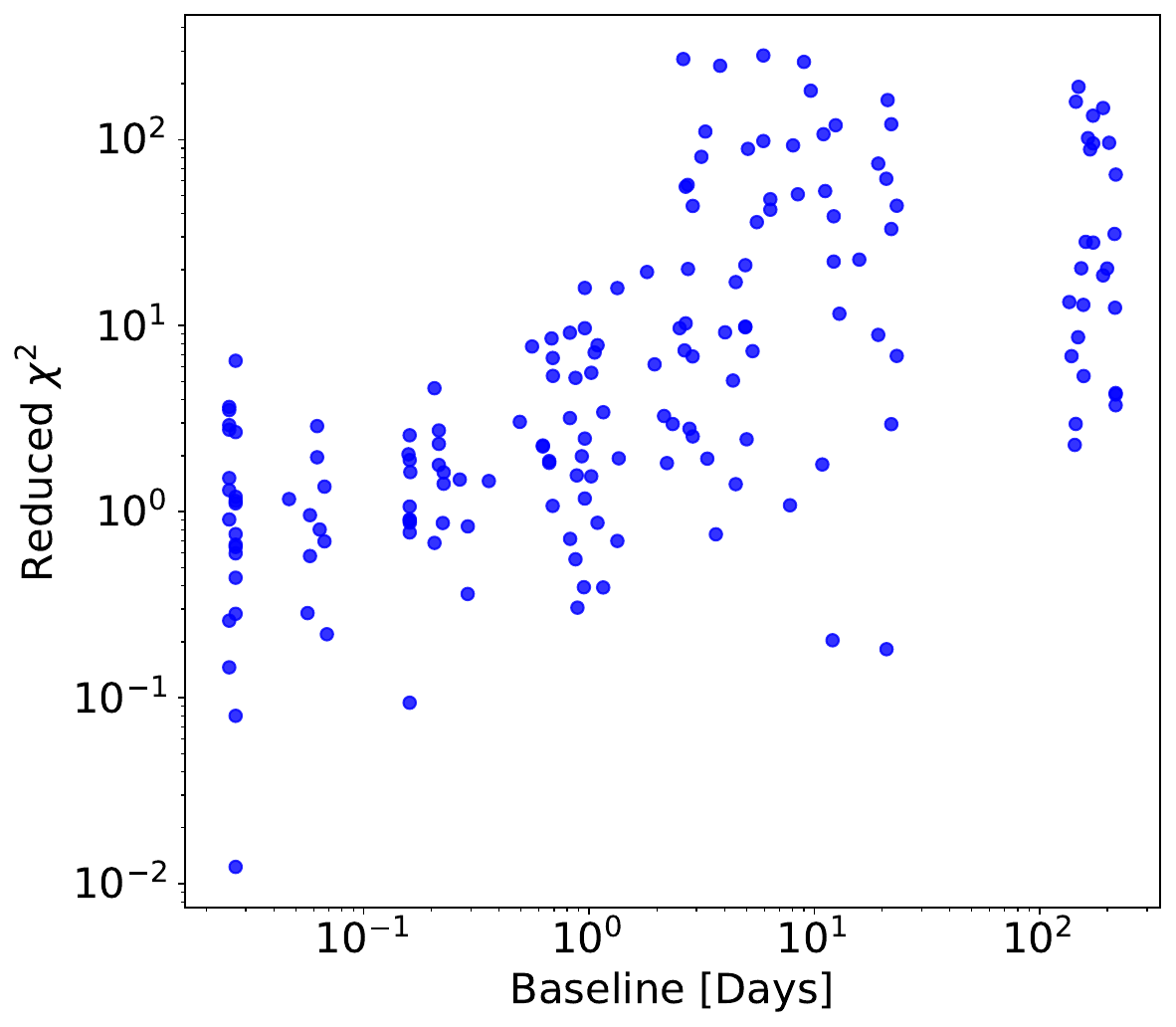} 

    \caption{
    Left: $\chi^2_{\rm red}$ vs. temporal baseline for the 358 luminous (${\rm F814W}<19$) cluster stars with more than one exposure identified in the PHAT data (see Section \ref{SubSec:StarClusters}). 
    Right: $\chi^2_{\rm red}$ vs. temporal baseline for classified field Cepheids from \cite{Wagner-Kaiser2015}. 
   }
    \label{fig:redchisquarebaseline} 
\end{figure*}

The left panel of Figure~\ref{fig:syserr} shows the standard deviation vs.\ mean magnitude for these 129 cluster stars. 
The right panel of the figure shows the F475W--F814W color vs.\ F814W magnitude for these stars, along with a heat map produced using the \cite{Williams2014} catalog of more than 117 million M31 stars identified throughout the PHAT survey. 
The marker colors for each cluster star plotted in Figure \ref{fig:syserr} are based on the distinctive plumes of stars
visible at $\rm F814W\lesssim19$~mag, denoting redder stars at/above the tip of the red giant branch ($\rm F475W-F814W\geq2.3$), intermediate-color stars ($1\leq\rm F475W-F814W<2.3$), and bluer stars ($\rm F475W- F814W<1$).
We find a clear bimodal distribution of light curve standard deviations of all the cluster stars in our calibration sample, as shown in the left panel of Figure~\ref{fig:syserr}. 
This sample of stars therefore comprises two broad groups with $\sigma$ above or below $\sim0.05$~mag: the group with higher values predominantly composed of stars with intermediate colors ($\rm1\leq F475W-F814W<2.3$) compared with the other group, which contains the majority of the redder ($\rm F475W-F814W\geq2.3$) and bluer ($\rm F475W-F814W< 1$) stars in the sample. 
Red variable stars are known to have characteristic timescales of a few hundred days \citep[e.g.,][]{Soraisam2020}, so it is significant that the majority (90\%) of the redder stars in our calibration sample with short ($<1$~d) time baseline light curves have $\sigma<0.05$~mag, placing them in the lower of the two bimodal groups. We therefore make the assumption that the group of stars with $\sigma<0.05$~mag are non-variable based on the available PHAT survey data, their measured variations instead being dominated by noise.

We use all the stars in this lower group to calibrate the systematic error as follows. For each star in this group, we use the standard deviation of its light curve as an estimate of the error on its F814W magnitude measurement ($\sigma_m$). We place the stars into four magnitude bins according to their mean light curve  magnitudes and compute the average $\sigma_m$ within each bin ($\bar{\sigma}_m$). 
We then determine the optimal systematic error to add to the reported error of all data points in the magnitude bin by identifying the correction value that minimizes the square difference between $\bar{\sigma}_m$ and $\sigma$ for data points in that bin. 
The systematic errors of the 376 luminous cluster stars are then calibrated by grouping the stars into the same magnitude bins identified for the calibration sample and applying the systematic error values to the stars in each bin.

\subsection{Light Curve: Reduced \texorpdfstring{$\chi^2$}{X2} Analysis}
\label{subsec:lightcurveanalysis}

After calibrating the systematic error for our sample of 376 stars, 
we analyze their F814W light curves for photometric variability. We identify and exclude any star with only a single exposure in its F814W light curve and also exclude measurements of the stars with sharpness parameter $S^{2}\geq0.2$ as described above, which leaves 358 stars. 
We then investigate the light curves for these remaining 358 stars by comparing each F814W observation to its mean light curve value using reduced chi-squared, $\chi^2_{\rm red}$, as a metric. The left panel of Figure \ref{fig:redchisquarebaseline} shows the $\chi^2_{\rm red}$ vs.\ the temporal baseline of the F814W light curve for these stars. As can be seen, the stars in our sample occupy the full $\chi^2_{\rm red}\mbox{--}{\rm baseline}$ space and many have $\chi^2_{\rm red}>1$ over the full range of baselines, indicating potential variability in their F814W light curves. 
In contrast, when the same $\chi^2_{\rm red}$ vs.\ baseline analysis is performed for classified field Cepheids from \cite{Wagner-Kaiser2015} using their PHAT F814W light curves, shown in the right panel of Figure \ref{fig:redchisquarebaseline}, there is a clear increase to $\chi^2_{\rm red}>2.5$ at baselines $>1$~d. 
This is concurrent with the known variability periods of classical Cepheids of $\sim1$--100~d, indicating the PHAT light curves are capturing the variability of these stars. 
We therefore apply a threshold of $\chi^2_{\rm red}>2.5$ to our sample of 358 stars to select those with light curves showing evidence of variability in PHAT observations. This results in 239 \emph{candidate} cluster-variable stars.

\subsection{Vetting of Per-Visit Images}

\begin{table}
\begin{tabular}{|c|c|}
 \hline
 Criterion & Number of stars \\
 \hline\hline 
  Full 2753 Cluster Sample & 294981 \\
\hline 
Star F814W $<$ 19~mag & 376 \\ 
\hline
Number of exposures $>1$ & \multirow{2}{*}{358} \\ 
(with $S^{2}<0.2$) & \\
 \hline
 $\chi^2_{\rm red}$ $>$ 2.5 & 239 \\
 \hline
 Number of visits $>$ 1 & 212 \\
 \hline 
\end{tabular}
\caption{\label{table:ClusterSample}Breakdown of the number of candidate cluster-variable stars identified at each stage of the vetting process described in Section \ref{sec:method}, prior to application of the DI Pipeline.}
\label{tab1:number_breakdown}
\end{table}

We visually examine the per-visit images of each of the 239 candidate cluster-variable stars identified above to investigate the behaviour and source of the detected variability at each \emph{HST} visit. We aim to identify variability caused by contaminants, such as cosmic ray strikes on the detector, as these contaminants will result in inaccurate variability measurements for the affected candidate cluster-variable stars. 
This process revealed that a large fraction of the per-visit images are affected by cosmic ray strikes. However, the manual examination of the images around the candidate cluster-variable stars proved to be a much greater challenge due to the lack of contrast in the crowded cluster environments. 
We therefore turned to using a more robust method---difference imaging (DI)---to efficiently validate our sample of candidate cluster-variable stars.

In order to carry out the DI analysis of a source, at least one pair of per-visit images is needed. We assess the  available data of the 239 candidate variable stars to find those with more than one \emph{HST} visit. We use the per-exposure photometry catalog from Astro Data Lab \citep{Olsen2018AAS,Williams2018} to identify unique observing epochs for each candidate variable star. Following this step, the number of candidate cluster-variable stars which could be processed using the DI Pipeline is reduced to 212. 
Table \ref{table:Candidates} details the properties of each of the final 212 candidate cluster-variable stars prior to the application of the DI pipeline. 

\begin{deluxetable*}{cccccc}[!t]
    \tablecaption{\label{table:Candidates}Candidate variable stars identified in M31 clusters through the vetting process described in Section \ref{sec:method}. We use the same cluster ID as in \citet{Johnson2015}. The full table is available in the supplementary material.}
       
    \tablehead{
        \colhead{Cluster ID} & \colhead{Cluster RA[Deg]} & \colhead{Cluster DEC[Deg]} & \colhead{PHAT ID} & \colhead{ PHAT RA[Deg]} & \colhead{PHAT DEC[Deg]}
    }
    \startdata
        27 & 11.016225 & 41.276218 & PHAT11.0161713+41.276141 & 11.016171 & 41.276141 \\
91 & 10.990885 & 41.231335 & PHAT10.9906814+41.231282 & 10.990681 & 41.231282 \\
93 & 11.011452 & 41.240176 & PHAT11.0114911+41.240166 & 11.011491 & 41.240166 \\
230 & 11.127033 & 41.843671 & PHAT11.1262019+41.843634 & 11.126202 & 41.843634 \\
285 & 10.945997 & 41.210538 & PHAT10.9464478+41.209956 & 10.946448 & 41.209956 \\
390 & 10.911579 & 41.380726 & PHAT10.9115373+41.380738 & 10.911537 & 41.380738 \\
403 & 11.104662 & 41.347108 & PHAT11.1047292+41.346899 & 11.104729 & 41.346899 \\
403 & 11.104662 & 41.347108 & PHAT11.1046035+41.347157 & 11.104603 & 41.347157 \\
403 & 11.104662 & 41.347108 & PHAT11.1050349+41.347152 & 11.105035 & 41.347152 \\
445 & 10.951771 & 41.448099 & PHAT10.9517399+41.448030 & 10.951740 & 41.448030 \\
445 & 10.951771 & 41.448099 & PHAT10.9516719+41.448098 & 10.951672 & 41.448098 \\
        $\cdots$ & $\cdots$ & $\cdots$ & $\cdots$ & $\cdots$ & $\cdots$ \\
    \enddata
\end{deluxetable*}

\subsection{Difference Imaging Pipeline}

We verify the potential variability detected in the light curves of our 212 candidate cluster-variable stars by developing and applying a DI pipeline to their available F814W per-visit images. 
Performing DI analysis only between images with the largest difference in light curve magnitudes may limit the potential number of confirmed variables if at least one of the frames is affected by contaminants such as cosmic ray strikes, bad detector columns, etc. 
We therefore choose to use all the images covered by the F814W light curve of a given candidate variable star, differencing the first available visit-image (which we refer to as the \emph{template frame}) and all subsequent visit-images (\emph{science frames}). 

As a given M31 stellar cluster may host multiple candidate cluster-variable stars, we generate $250~{\rm pix}\times250$~pixel ($12\farcs5\times12\farcs5$) cutouts from the visit-images centered on the \cite{Johnson2015} co-ordinates of each host cluster to be used as template and science frames. 
This cutout size also ensures that there are at least 20 bright stars around the cluster in both frames. 
This is critical as we use these non-cluster stars in image alignment and point-spread function (PSF) matching, as described in the following sections. 
Furthermore, using these cutouts minimizes the effects of ACS/WFC detector geometric distortion \citep{Ryon2023ACSWFCInstrumentHandbook} and the spatial variation of the ACS/WFC PSF \citep{Krist2003ACSPSFVariation} while also including sufficient image background for accurate background flux measurements.

\hfill \\
\subsubsection{Bad pixel mask, image alignment \& background matching} \label{subsubsec:alignmentBackgroundmatching}

We create a bad pixel mask for each pair of template and science frames for a given candidate variable star host cluster. 
Starting with an empty array with the same dimensions as the science and template cutout frames as the basis for our mask array, we identify pixels with non-physical values (i.e., non-positive values as the pixel values are in unit of electron/sec) 
in either the template or science frame and add them to the mask array. 
We then mask the surrounding +/- 5 pixels around each of these pixels in order to exclude potential lower-level effects unaccounted for in image processing. 
We apply this bad pixel mask to both science and template frames in the pair, repeating the above process for each of the template-science frame pairs for the host cluster. 

Once masked, we transform the frames for a given host cluster onto the same coordinate plane by reprojecting each science frame onto the template frame using the \texttt{reproject} library from \texttt{astropy} and their respective WCS header information. 
We then ensure subpixel alignment between the template and science frames using \texttt{astroalign} \citep{Astroalign}, which estimates the affine transformation between the two frames by matching asterisms in the frames.

Background matching of each pair of aligned-science and template frames is performed by first subtracting the template frame from the former to produce an \emph{intermediate difference image}.  
We then use \texttt{Background2D} from the \texttt{photutils} package \citep{photutils1.8.0} to generate a 2D background map from the intermediate  difference image.  
The \texttt{boxsize} and \texttt{filtersize} parameters of \texttt{Background2D} are adjusted manually for each pair of frames on a case-by-case basis through multiple iterations to optimize the background estimation by checking for any abnormal behaviour in the 2D map (under- or over-smoothing of the image background and cluster residuals). 
We find a \texttt{boxsize} of 10 and \texttt{filtersize} of 5 are generally applicable to avoid creating large `dipoles'---immediately adjacent hemispheres of significant over- and under-subtraction caused by over-smoothing of local maxima---in the intermediate difference images.  
The resulting 2D background map is then added to the template frame.

\subsubsection{PSF-matching}
\label{subsubsec:PSFmatching}

To assess the quality of the observed PSF in each pair of science and template frames, we identify a matching set of bright stars away from the bright cluster core using \texttt{DAOStarFinder} in \texttt{photutils} \citep{photutils1.8.0}. To this end, we filter bright stars in each frame with magnitudes $>5\sigma$ (standard deviation) above the median magnitude of all stars in the frame. We then apply a cut based on radial distance from the image centre to only consider sources in the outer 20th percentile of the frame. We cross-match the identified stars from the template and aligned science frame to keep those detected in both frames.

The observed PSF of each matched star is then modelled as a two dimensional Gaussian using \texttt{Gaussian2D} from \texttt{astropy}. The full width at half maximum (FWHM) of the model PSF minor axis is used as a zeroth-order estimate of the resolution since it provides a more consistent estimate in case of PSF elongation along one axis due to blending of stars. 
We apply sigma-clipping to the distribution of model-PSF FWHMs to exclude sources with poor model fits. 
We then compare the mean FWHM value for all bright matching stars between the science and template frames. The frame with the smaller value (i.e., higher resolution) is convolved with a kernel generated using the adaptive Bramich routine \citep{AdaptBramich} of the   \texttt{optimal image subtraction} (OIS) library \citep{martinberoiz2020} to match the PSF between the pair of science and template frames. 
To prevent the bright, crowded core region of the cluster from adversely affecting the OIS kernel construction, we incorporate an optional mask covering the cluster core into the OIS software. Applied only during construction of the convolution kernel, the size of this additional square-shaped cluster mask is manually varied between 1 to 15 pixels along each axis depending on the radial extent of the cluster in each frame. 
Once the convolution kernel is applied, the convolved frame is subtracted from the less-resolved frame to produce the difference image for the two per-visit frames of the given host cluster. 

The general DI procedure described above can be summarized as:
\begin{equation}
    D = I - (R \otimes K)
    \label{DIeqn}
\end{equation}
where $I$ is the less-resolved frame, $R$ is the higher resolution frame, $K$ is the convolution kernel determined from the OIS Bramich routine, and $D$ is the difference image. We apply these steps to all visit-image pairs of the 212 candidate variable star host clusters to obtain clean DI frames with minimal artifacts.

\subsubsection{DI Source Detection}
\label{SubSubSec:DISourceDetection}

To confirm the variability of each of our 212 candidate variable stars, we identify significant residuals (at the location of the star) in the difference images produced for each star. 
We first obtain an estimate of the local background in a circular annulus around the host cluster. 
We set the inner and outer radii of the annulus based on the $0\farcs11$ FWHM of the ACS instrument \citep[e.g.,][]{Santiago2010}, using an inner radius of 10$\times$FWHM and outer radius of 20$\times$FWHM 
to include the local background of the cluster without including candidate cluster-variable stars. 
We then use the mean of the pixel values within the annulus area as the estimate of the local background. 

We next use the \texttt{DAOStarFinder} routine of \texttt{photutils} to detect sources at $5\sigma$ above the local background in the difference images. Any detections within a $0\farcs11$ radius of a candidate variable star position in the difference image are designated as potential counterparts to that candidate variable star. 
If no detections are found in this first pass, the image is inverted and the \texttt{DAOStarFinder} search repeated to allow for negative residuals to be detected. 
Candidate cluster-variable stars with no corresponding sources within a $0\farcs11$ radius of their position in all of their host cluster difference images are excluded from further analysis. 
From our sample of 212 candidate cluster-variable stars, we identify counterpart difference-image sources for 89 stars using this \texttt{DAOStarFinder} source search, confirming their variable nature.

\begin{figure}
    \centering
    \includegraphics[width=8cm]{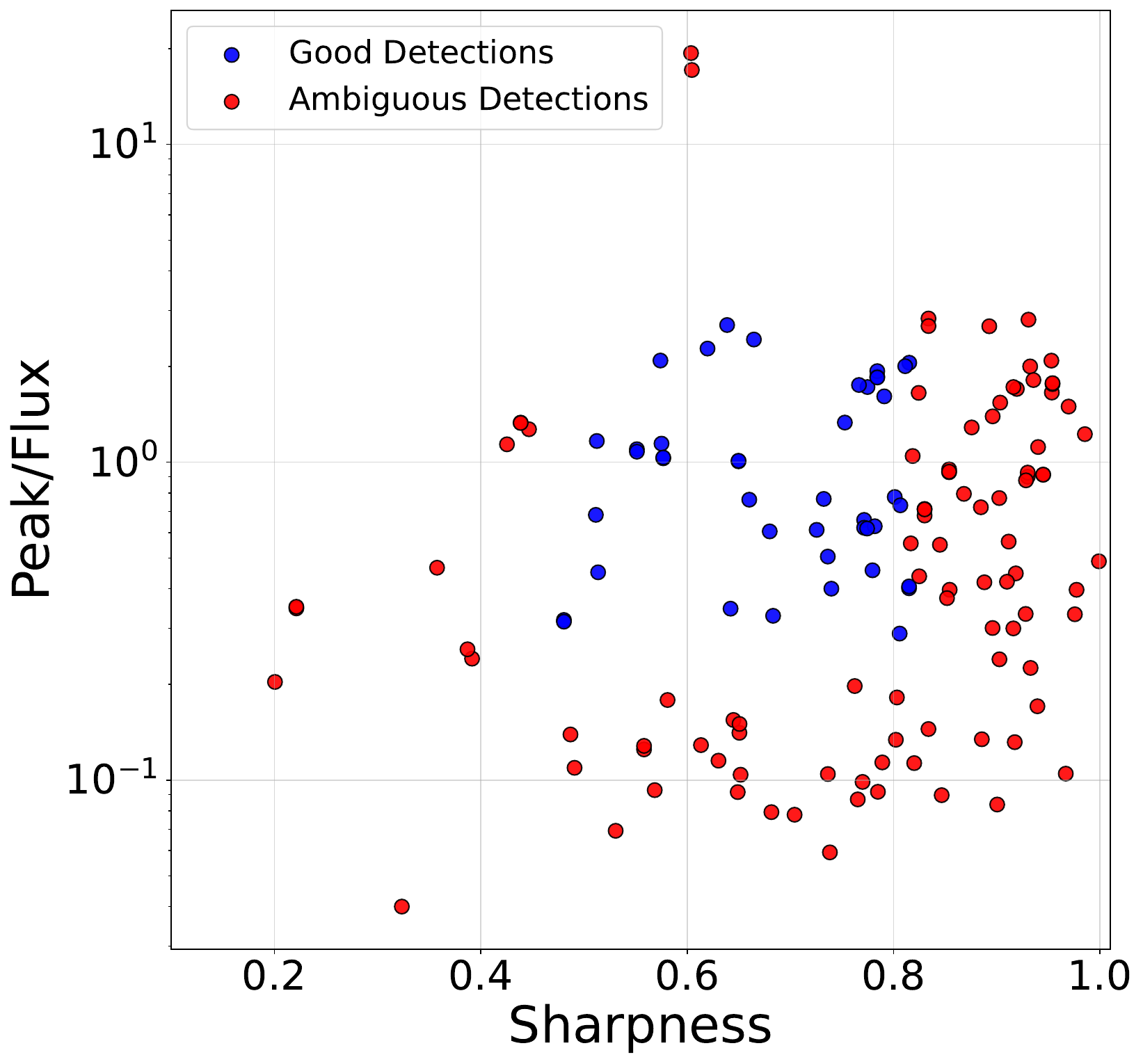} 
    \caption{Sharpness vs peak/flux parameter values from \texttt{DAOStarFinder} for all sources detected within a $0\farcs11$ radius of a candidate cluster-variable star position across all host cluster difference images, divided into \emph{Good} and \emph{Ambiguous} categories as described in Section \ref{SubSubSec:DISourceDetection}.}
    \label{fig:peakfluxvssharp}
\end{figure}

We now use the \texttt{DAOStarFinder}\footnote{\url{https://photutils.readthedocs.io/en/stable/api/\\photutils.detection.DAOStarFinder.html}} \texttt{sharpness}, \texttt{peak} and \texttt{flux} output parameters for the source detection(s) associated with each confirmed variable star as metrics to categorize the source detection(s) into one of two groups, based on the source exhibiting stellar-like behaviour (`\emph{Good}') and those which do not  (`\emph{Ambiguous}'), e.g. due to potential contamination by difference imaging artifacts. 
We determine \emph{Good} source detections to be those with $0.26 < peak/flux < 10.84$ (between the 30th and 99th percentile) and $0.48 < sharpness < 0.82$ (between the 9th and 59th percentile) when considering the \texttt{DAOStarFinder} parameters of all the detections associated with cluster-variable stars, as shown in Figure~\ref{fig:peakfluxvssharp}. Source detections with one or both metrics outside these ranges are classified as \emph{Ambiguous}. 
Any candidate cluster-variable star with at least one \emph{Good} detected source is then designated as having \emph{Good} (stellar-like) difference image source detection behaviour: otherwise, the star is deemed to have \emph{Ambiguous} source detection behaviour.

\begin{figure*}         
   \fig{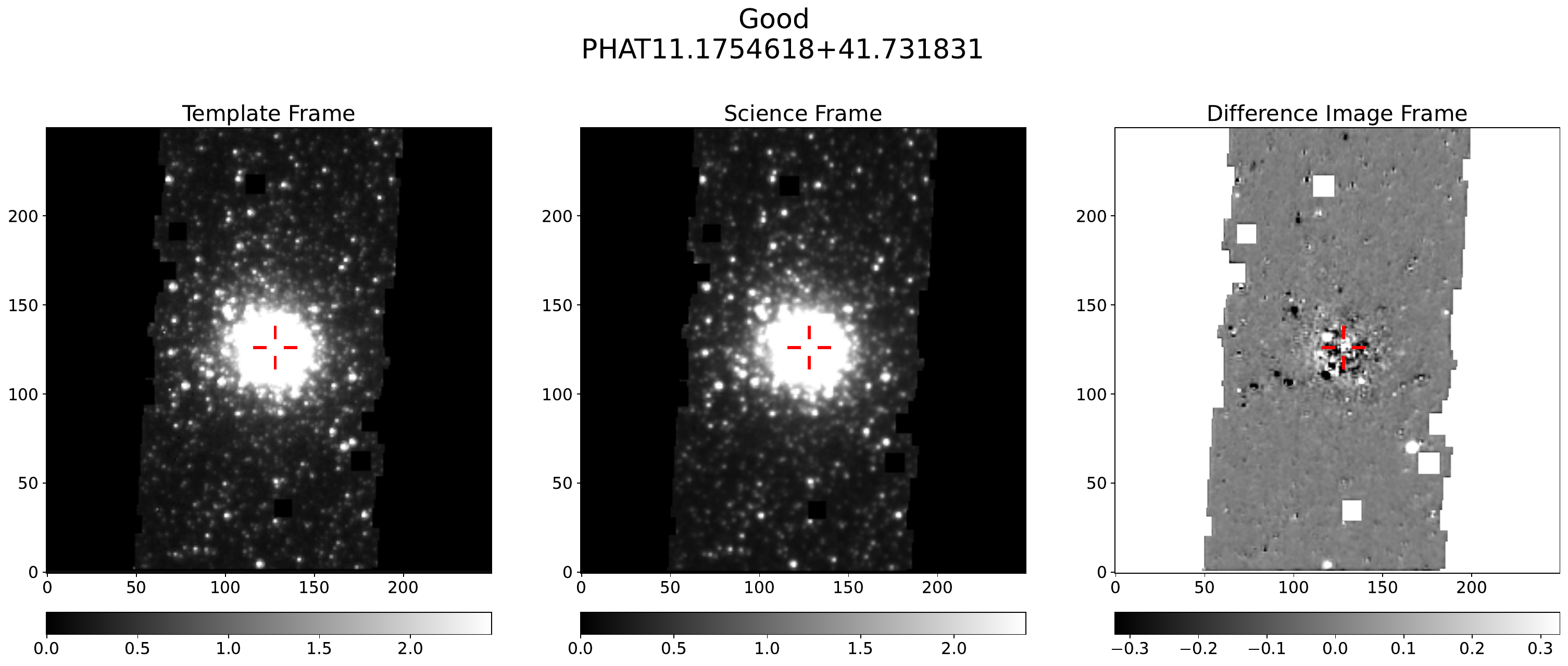}{1\textwidth}{(a)}
    \vspace*{0cm}
    \fig{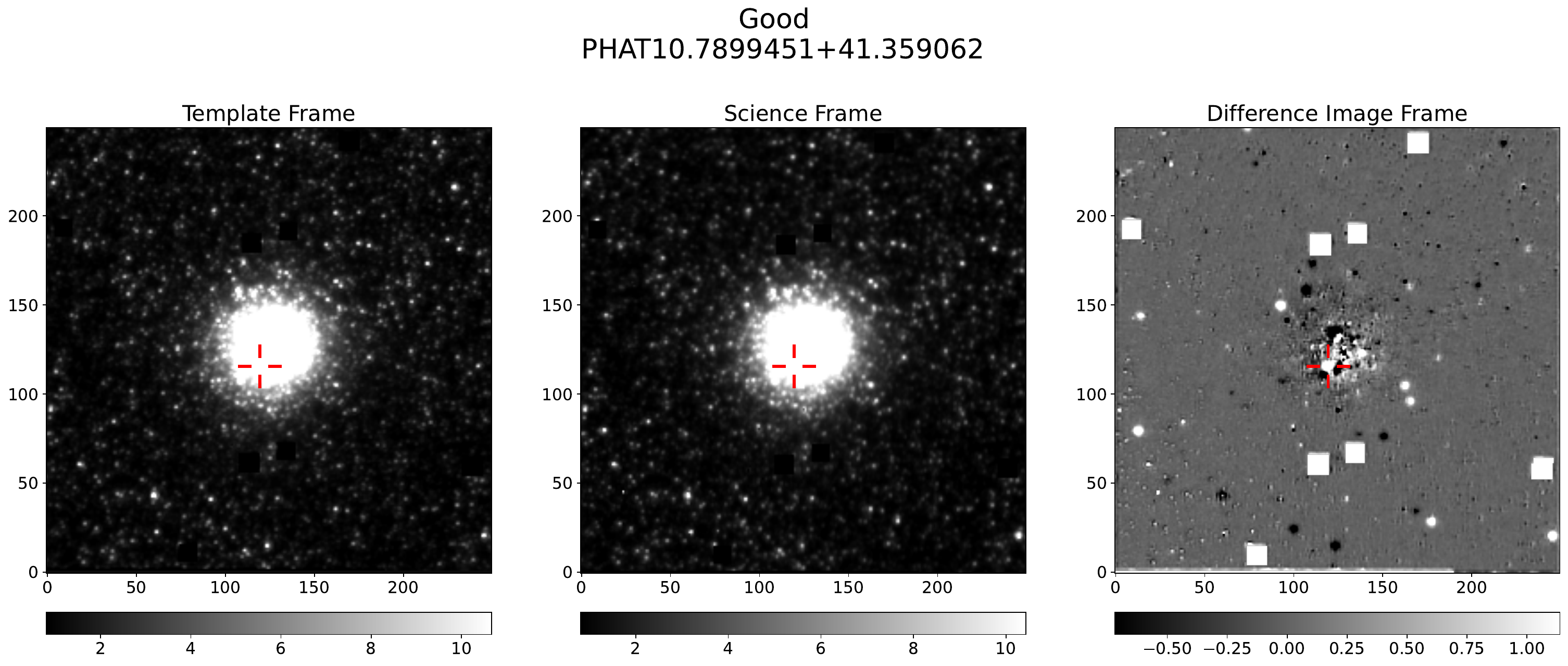}{1\textwidth}{(b)}
    \caption{Examples of \emph{Good} cluster-variable star source detections as described in Section \ref{SubSubSec:DISourceDetection}, with the difference image being the science minus template frame. 
    The red cross-hair in each frame indicates the location of the detection.}
    \label{fig:DI_Good}
\end{figure*}

\begin{figure*}
    \fig{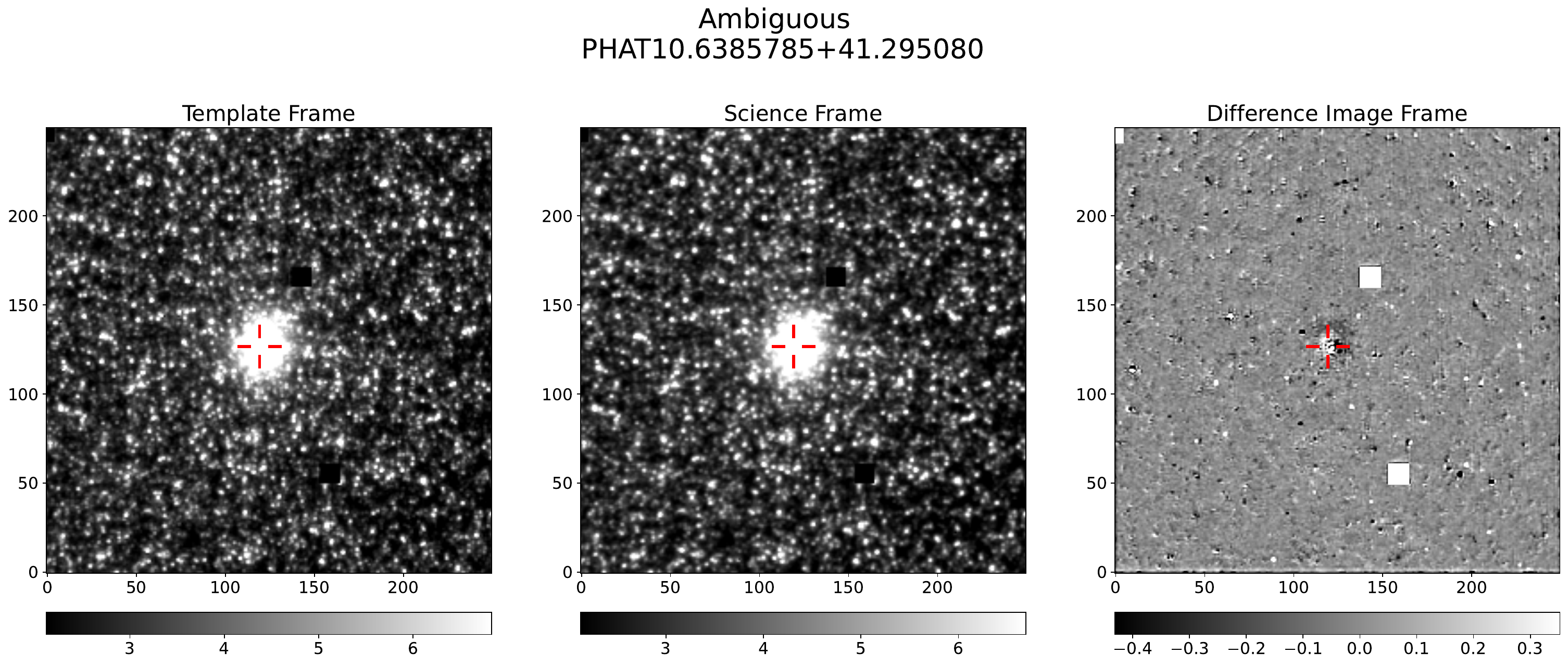}{1\textwidth}{(a)}
    \fig{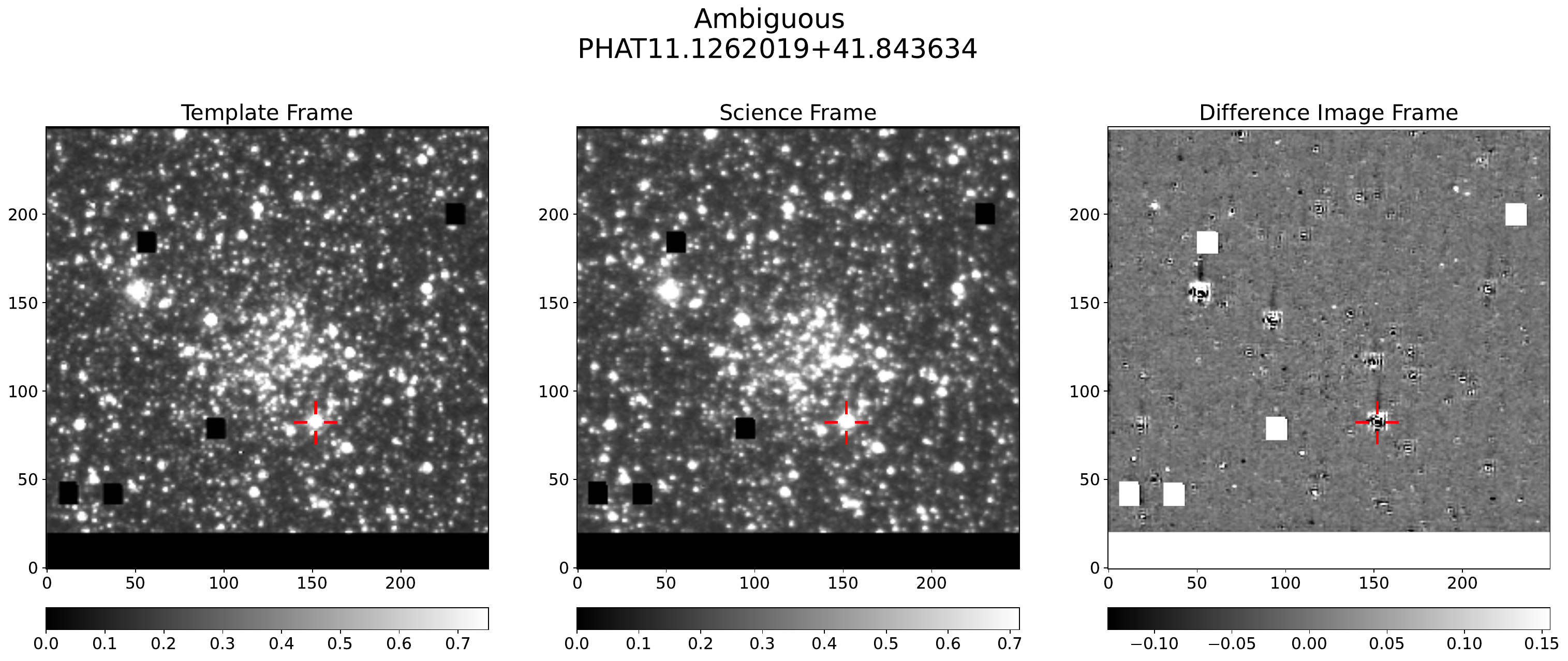}{1\textwidth}{(b)}
    \caption{Examples of \emph{Ambiguous} cluster-variable star source detections as described in Section \ref{SubSubSec:DISourceDetection}, with the difference image being the science minus template frame. 
    The red cross-hair indicates the location of the detection.}
    \label{fig:DI_Ambiguous}
\end{figure*}

\begin{figure}[htp!]
    \centering
    \subfigure{{\includegraphics[width=7cm]{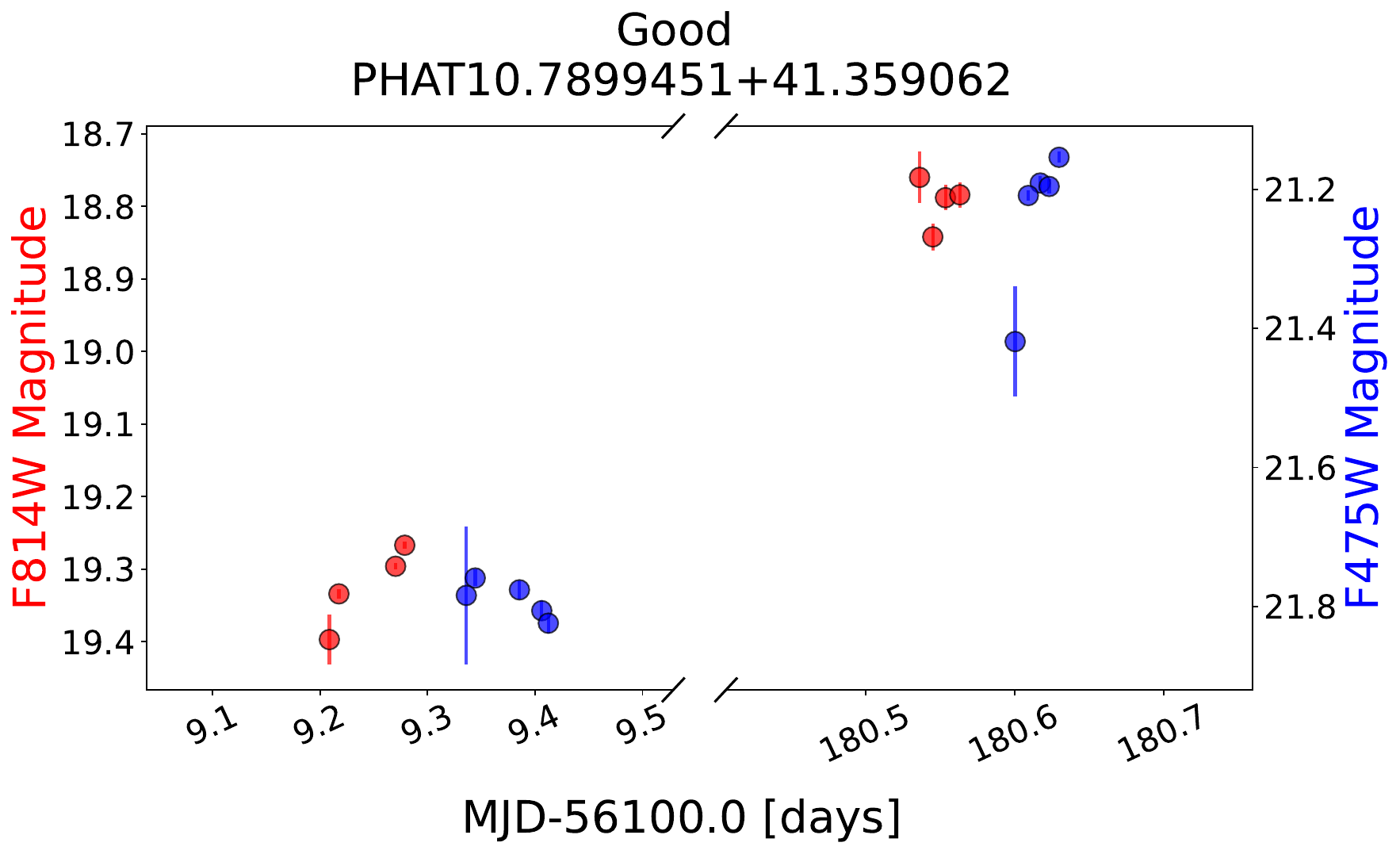} }}
    \subfigure{{\includegraphics[width=7cm]{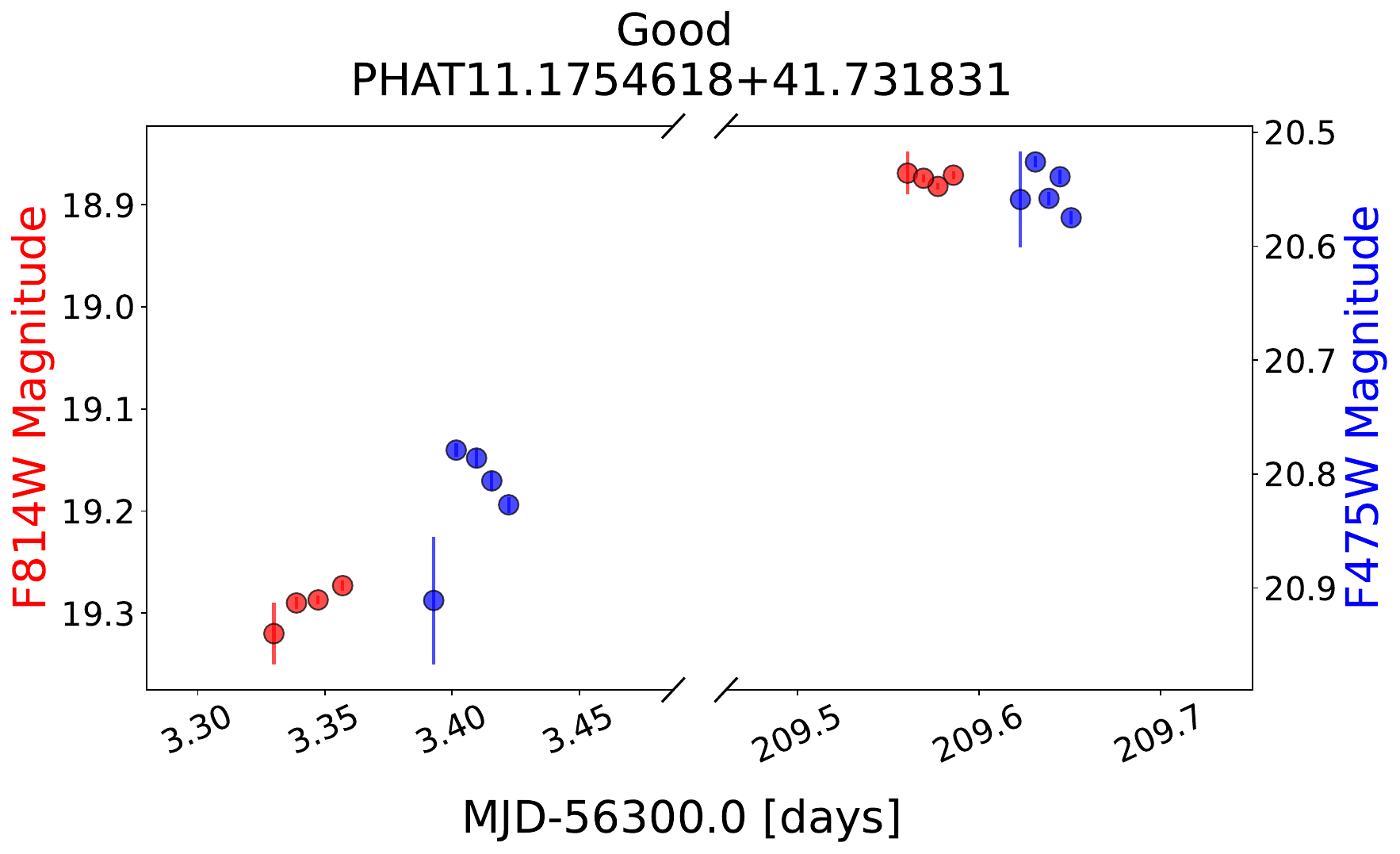} }}
    \caption{F814W and F475W light curves for the two example `Good' cluster-variable stars shown in Figure \ref{fig:DI_Good}.}
    \label{fig:good_lcs}
\end{figure}

\begin{figure}[htp!]
    \centering
    \subfigure{{\includegraphics[width=7cm]{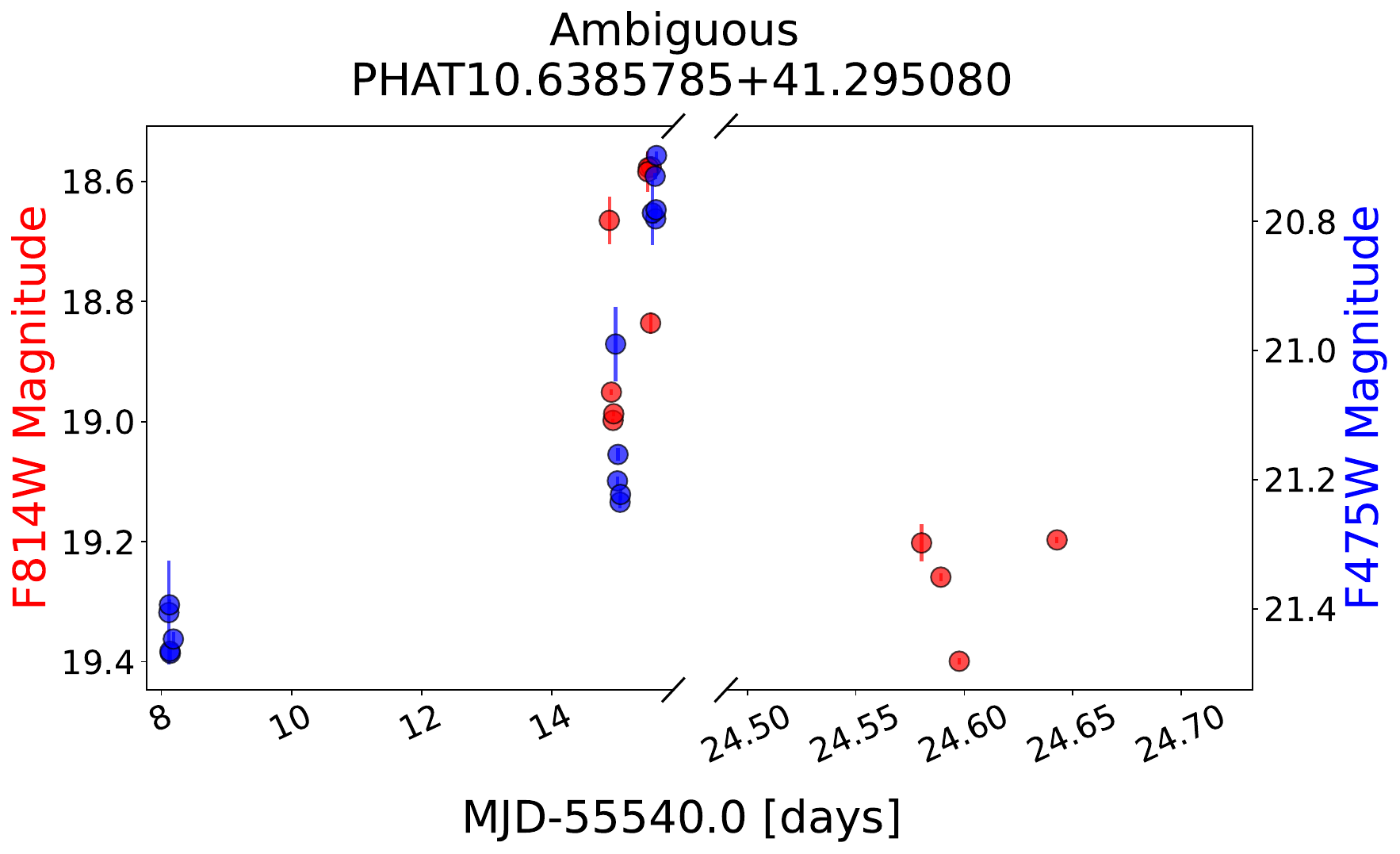} }}
    \subfigure{{\includegraphics[width=7cm]{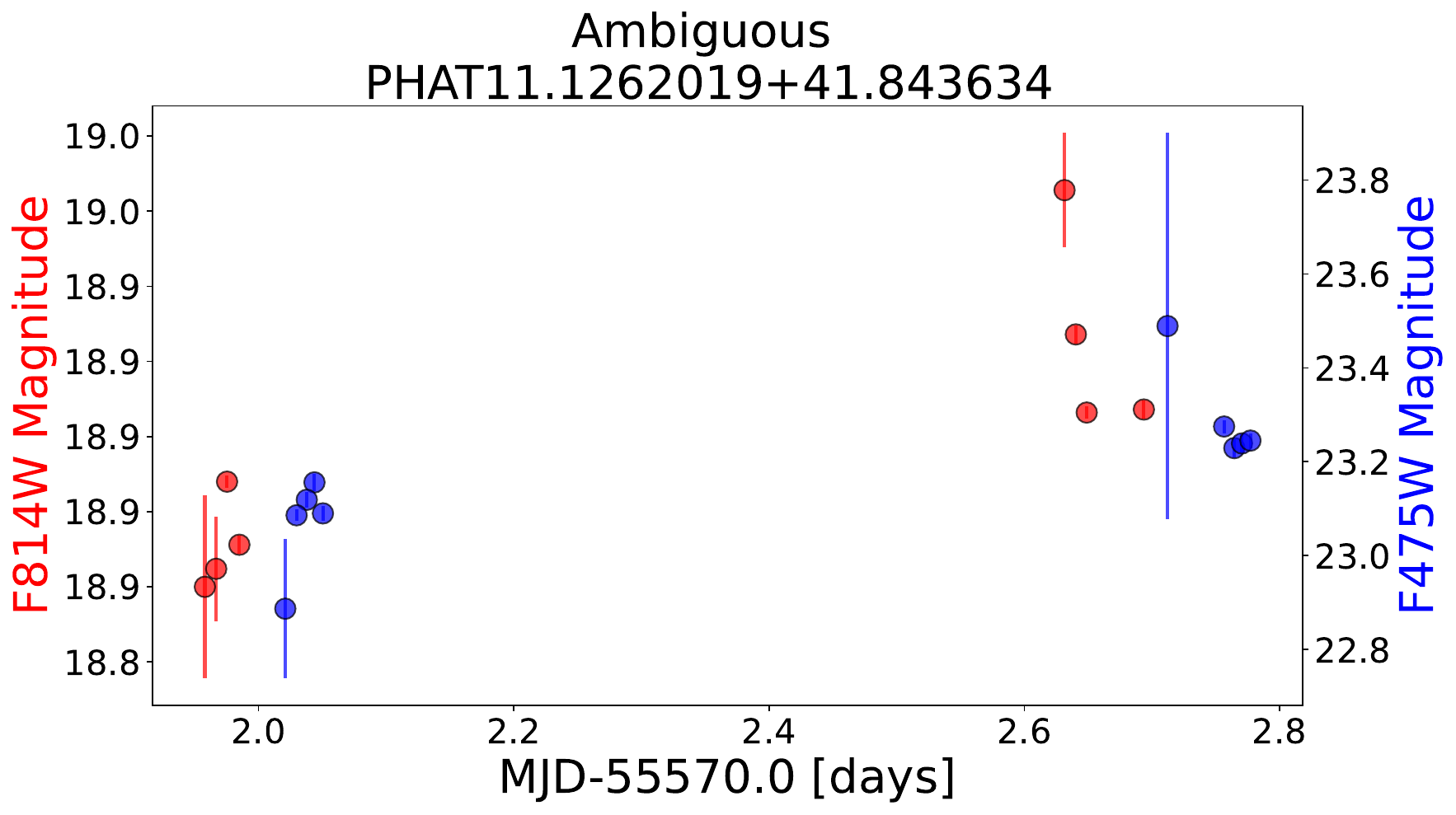} }}
    \caption{F814W and F475W light curves for the two example `Ambiguous' cluster-variable stars shown in Figure \ref{fig:DI_Ambiguous}.}
    \label{fig:ambiguous_lcs}
\end{figure}

Examples of template, science and difference image frames for \emph{Good} and \emph{Ambiguous} detection classifications are shown in Figures \ref{fig:DI_Good}, \ref{fig:DI_Ambiguous}, respectively, with light curves for the corresponding cluster-variable star candidates shown in Figures \ref{fig:good_lcs} and \ref{fig:ambiguous_lcs}.
After applying the DI Pipeline to the 212 variable star candidates and performing the subsequent source detection analysis, we are able to confirm 89 cluster-variable stars in 50 host clusters.
Of these, we find 38 stars with at least one \emph{Good} source detection and 51 stars with \emph{Ambiguous} source detections based on the thresholds outlined above.

\subsection{Chance coincidence with field stars} 

Before analysing the 89 cluster-variable stars further, it is prudent to assess the potential of contamination of our sample by luminous field stars. 
To do this, we first identify the number of bright ($\rm F814W<19$) field stars detected across the entire PHAT survey footprint, finding a total of 7,635 stars. 
By making the simplifying assumption that these field stars are uniformly distributed across the PHAT survey area, we are able to obtain a stellar density of $\approx1.22\times10^{-3}$ per sq.\ arcsec for these bright stars. 
Based on the radii of each of the M31 stellar clusters tabulated in the \citet{Johnson2015} catalog, we find the total area of the 2753 clusters to be $\approx3260$ sq.\ arcsec. 
The expected number of luminous field stars appearing in these clusters by chance is then $\approx4$. 
Using Poisson statistics, we find a negligible (close to zero) probability of randomly finding 89 or more luminous field stars in the clusters; 
the probability of finding more than 9 of such field stars in clusters by chance is less than 1\%. We therefore conclude that it is highly unlikely that the majority of the luminous cluster-variable stars we have identified are contaminant field stars.

Nevertheless, we cannot rule out that a small fraction of these stars are in fact field stars. To investigate this possibility,   
we search for potential foreground stars in our sample of 89 cluster-variable stars using available Gaia DR3 data accessible through \textit{Astro Data Lab} and the method of \citet{Barmby2023} (see Appendix~\ref{App:ForegroundStars} for details). We find 3 likely foreground stars, which leaves a final sample of 86 confirmed cluster-variable stars in 47 host clusters, of which 37 stars have at least one \textit{Good} source detection and 49 stars have \textit{Ambiguous} source detections, as described above.

\section{Results}\label{sec:res}
We have identified 86 cluster-variable stars in 47 host clusters through the application of our DI pipeline. 
We now proceed to infer the evolutionary phases and initial masses of these stars by matching each star to the most probable point on an isochrone generated using the derived properties of the host cluster. 
We also consider the blending effects on the PHAT photometry, using representative samples of likely- and less likely to be blended stars in the PHAT clusters to calculate the probability of each cluster-variable star in our sample being blended based on their available properties.
We then use the inferred fundamental parameters of the stars to classify the type of variability, qualified by the source detection behaviour and blending probability of each cluster-variable star.

\subsection{Isochrone Generation}\label{subsec:MISTIsochrones}

We use the MESA Isochrones \& Stellar Tracks (MIST) \citep{Dotter2016MISTPaper0,Choi2016MISTPaper1} Web Interpolator\footnote{\url{https://waps.cfa.harvard.edu/MIST/interp_isos.html}} to generate an isochrone for each variable star-hosting cluster. 
To do this, we use estimates of age, extinction ($A_V$) and metallicity for the PHAT clusters provided by \cite{deMeulenaer2017}, which they obtain by comparing the integrated broadband photometry of a given cluster with a 4-dimensional grid of models parameterized by cluster age, mass, extinction, and metallicity
\citep[][and references therein.]{deMeulenaer2015,deMeulenaer2017}{}{}.
In their analysis of the derived metallicities, \cite{deMeulenaer2017} show that there is a high prevalence of solar or super-solar metallicity values in younger ($< 1$ Gyr) stellar clusters, while older ($\geq 1$ Gyr) clusters exhibit a broader metallicity distribution. This motivates keeping the metallicity of younger ($<1$ Gyr) clusters fixed to solar value ($Z = 0.02$) while allowing the metallicity of older clusters to vary. These authors also show that there is reasonable agreement between their results and those of \citet{Caldwell2011}, which are based on spectroscopic observations, for the age, mass, and metallicity of the clusters common to both studies, barring extinction (see \citealt{deMeulenaer2017}).

We adopt the \cite{deMeulenaer2017} properties derived with non-solar metallicity if its resulting age estimate is  $>1$~Gyr and otherwise select the properties derived using solar metallicity.  
The selected age and metallicity values for each host cluster are used to generate an extinction-free MIST isochrone for that cluster. Stellar rotation is known to play an important role in stellar evolution \citep[e.g.,][]{MeynetMaeder2000,Hirschi2004}. 
We assume that rotation takes place but with a moderate effect on the evolutionary tracks, by adopting a velocity ratio of $v_{\rm initial}/v_{\rm critical}=0.4$ \citep[][and references therein]{Choi2016MISTPaper1}{}.

We transform the $A_V$ estimate of the given cluster derived by \cite{deMeulenaer2017} to the appropriate wavelength-dependent extinction for the PHAT bands, $A_{\lambda}$, using the conversion factors from \cite{Wang2022} (1.17 for F475W, 0.65 for F814W) and apply it to the observed magnitudes of each variable star in the cluster to correct for extinction. 
We adopt a distance modulus of 24.47 for M31 \citep{McConnachie2005} to covert apparent magnitudes of the variable stars into absolute magnitudes.

\subsection{Isochrone Matching}
\label{subsec:PosteriorCalc}

We use the generated isochrone of each host cluster to obtain the most likely properties for each of our variable stars using a modified version of the Bayesian method presented by \cite{daSilva2006}. We calculate a posterior probability for each datum along the host cluster isochrone given the observed color and magnitude properties of the star, following:
\small
\begin{equation}
    P(M^{'}_{init})\ \propto \ \phi(M^{'}_{init})\ \exp{\left[ -\frac{(x - x^{'})^{2}}{\sigma_{x}^{2}} \\-\frac{(y - y^{'})^{2}}{\sigma_{y}^{2}} \right]}
    \label{eq:OccuProb}
\end{equation}
\normalsize
\noindent where we use a Salpeter \citep{Salpeter1955} initial mass function (IMF) $\phi(M^{'}_{init})$ as the prior for the initial stellar mass of the isochrone datum, $M^{'}_{init}$; $x$ and $y$ are the observed $\rm F475W-F814W$ color and F814W absolute magnitude of the variable star, respectively; $x^{'}$ and $y^{'}$ are the corresponding theoretical $\rm F475W-F814W$ color and F814W absolute magnitude for the isochrone datum, respectively.

\begin{figure}[htp!]
    \centering
    \subfigure{{\includegraphics[width=7.5cm]{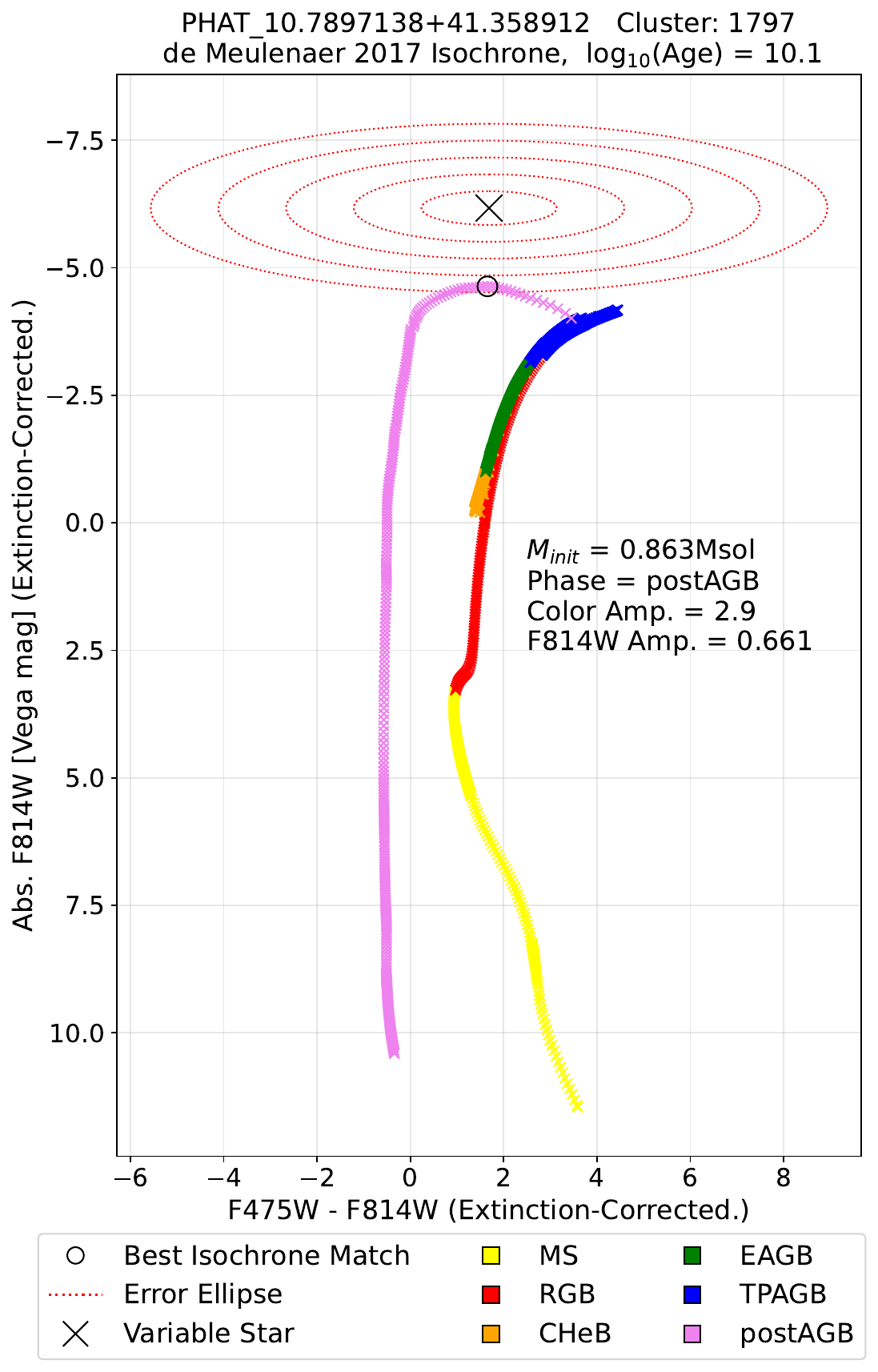} }}%
    \caption{An example F475W-F814W vs.\ F814W absolute magnitude CMD showing how the error ellipse of the variable star is scaled to intersect with the host cluster isochrone. 
    }
    \label{fig:isofit}
\end{figure}

We estimate the variance of the observed properties, $\sigma_{x}^{2}$ and $\sigma_{y}^{2}$ in Equation \ref{eq:OccuProb}, 
by first considering that the positions of our cluster-variable stars on the CMD are unlikely to represent their equilibrium states based on the sparsely-sampled PHAT measurements. To facilitate the matching of our variable stars to their host cluster isochrones, we construct \emph{error ellipses} using estimates of their observed amplitudes in F814W magnitude and $\rm F475W-F814W$ color space. For each variable star, we compute the largest difference between measurements comprising its F814W light curve ($\Delta m_{\rm F814W}$) and adopt it as the radius along the vertical axis of its error ellipse. 
Note that $\Delta m_{\rm F814W}$ provides only a lower bound to the light curve amplitude of each star in our sample since the sparse PHAT observations may not cover the full extent of their light curves. 
We further estimate the color amplitude ($\Delta c_{{\rm F475W},{\rm F814W}}$) of each variable star in our sample by identifying the pair of F475W and F814W measurements from its corresponding light curves that produce the largest color value. This $\Delta c_{{\rm F475W},{\rm F814W}}$ value is used as the radius along the horizontal axis of its error ellipse. 
We then place the error ellipse on a CMD, centered on the variable star, along with its host cluster isochrone and scale the size of the ellipse until it intersects or surpasses one or more points on the isochrone. We use these scaled error values as estimates of the variance on each observed parameter in Equation~\ref{eq:OccuProb}. 
Figure~\ref{fig:isofit} shows an example $\rm F475W-F814W$ vs.\ F814W absolute magnitude CMD for a variable star in our sample, demonstrating the scaling of its error ellipse until it intersects with its host cluster isochrone.

\begin{figure}
    \centering
    \subfigure{{\includegraphics[width=7.5cm]{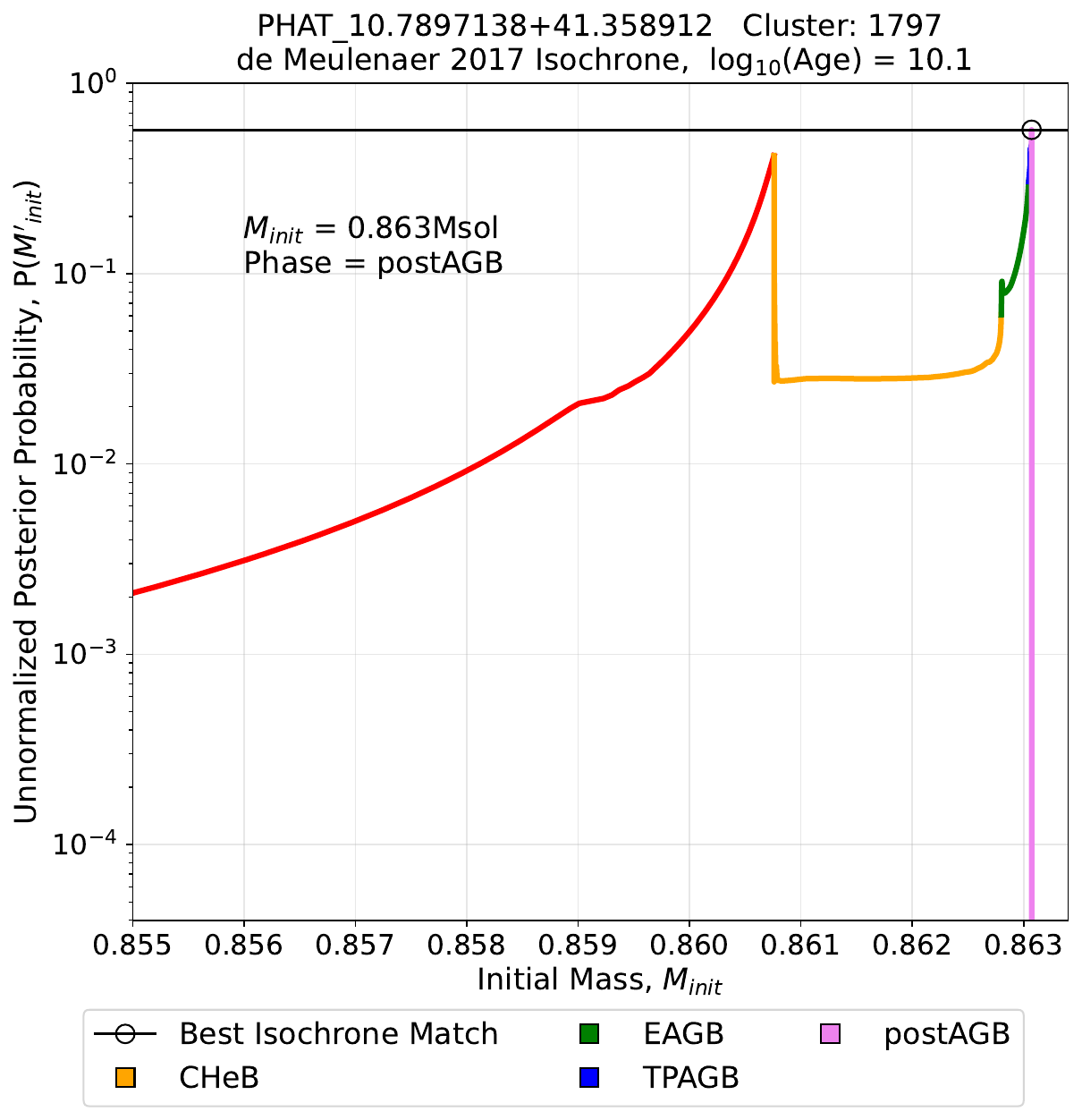} }}%
    \caption{
    Posterior probability distribution of initial mass for the variable star shown in Figure~\ref{fig:isofit}, demonstrating the narrow initial mass range over which the post-main sequence evolution occurs for the older ($\gtrsim 1$~Gyr) isochrones. In such cases, the posterior probability is driven by the likelihood (see Section~\ref{subsec:PosteriorCalc}).}
    \label{fig:posterior}
\end{figure}

After computing the posterior probability distribution for a given cluster-variable star using Equation \ref{eq:OccuProb}, we identify the maximum a posteriori isochrone datum and adopt its initial stellar mass and evolutionary phase to be those of the variable star.  
Figure \ref{fig:posterior} shows the unnormalized posterior distribution for the same cluster-variable star in Figure~\ref{fig:isofit}.
The distribution shown in Figure~\ref{fig:posterior} is typical of the distributions produced for the older ($\gtrsim 1$ Gyr), highly evolved variable stars in our sample. The post-main sequence evolution for their host cluster isochrones occurs over a narrow range of initial masses ($< 0.33 M_{\odot}$). 
This leads to a near-negligible effect from the IMF prior, the posterior probability instead being driven by the magnitude- and color-dependent likelihood term in Equation~\ref{eq:OccuProb}. In such cases, the most probable datum on the isochrone ends up being the point closest to the variable star in color-magnitude space. Hence, despite the longer lifetimes of the early asymptotic giant branch (EAGB) and thermally pulsing asymptotic giant branch (TPAGB) phases, we obtain the post-asymptotic giant branch (post-AGB) phase as the most probable result for the variable star in Figure~\ref{fig:posterior} since it is closest to the datum with that phase (Figure~\ref{fig:isofit}); see, however, the caveats in Section \ref{subsec:caveats}.

\begin{figure*}
    \fig{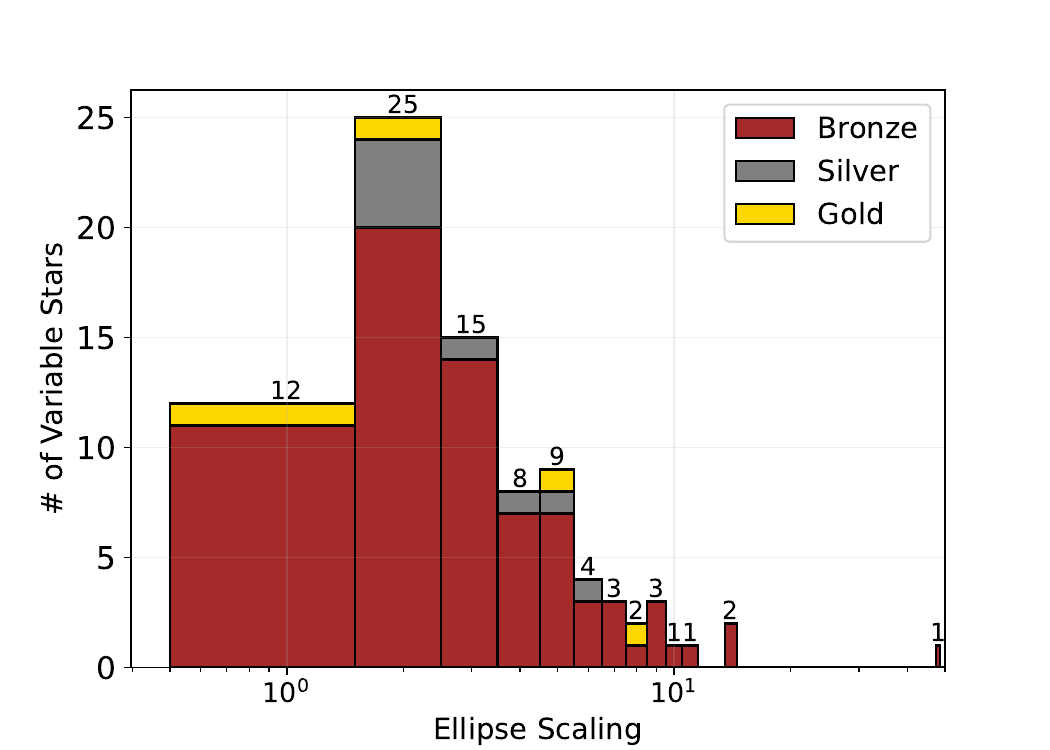}{0.45\textwidth}{(a)}
    \gridline{
    \fig{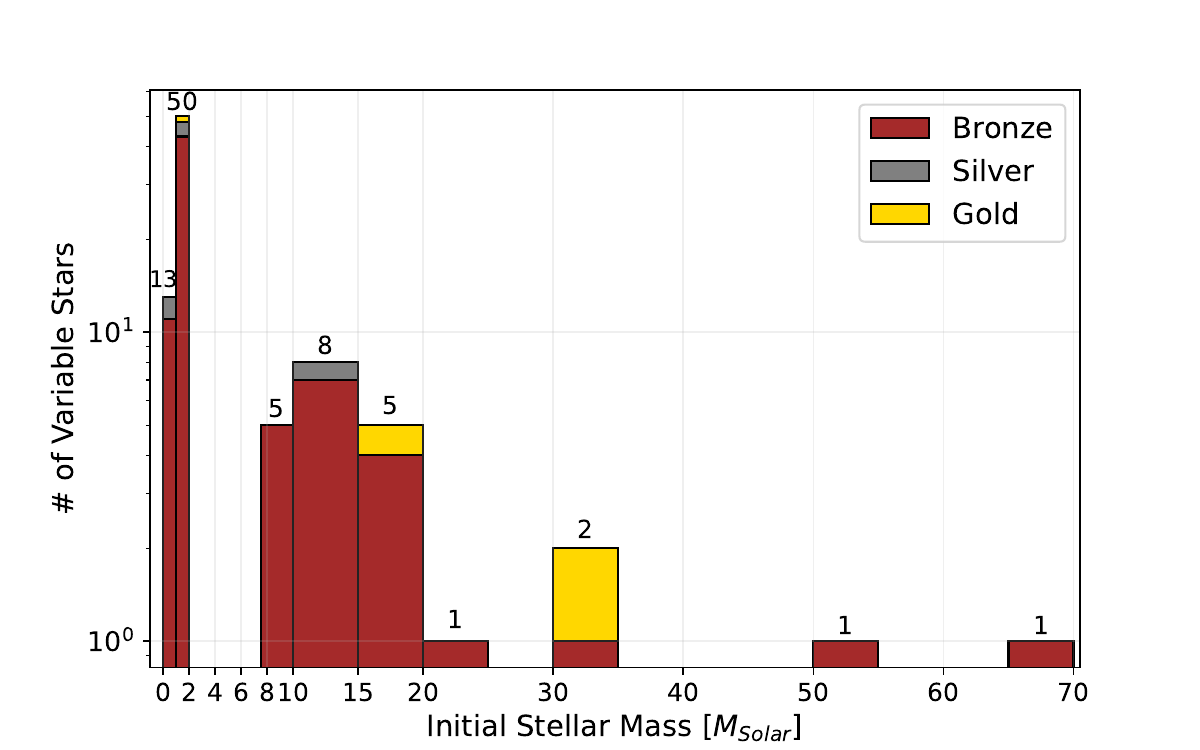}{0.51\textwidth}{(b)} 
    \fig{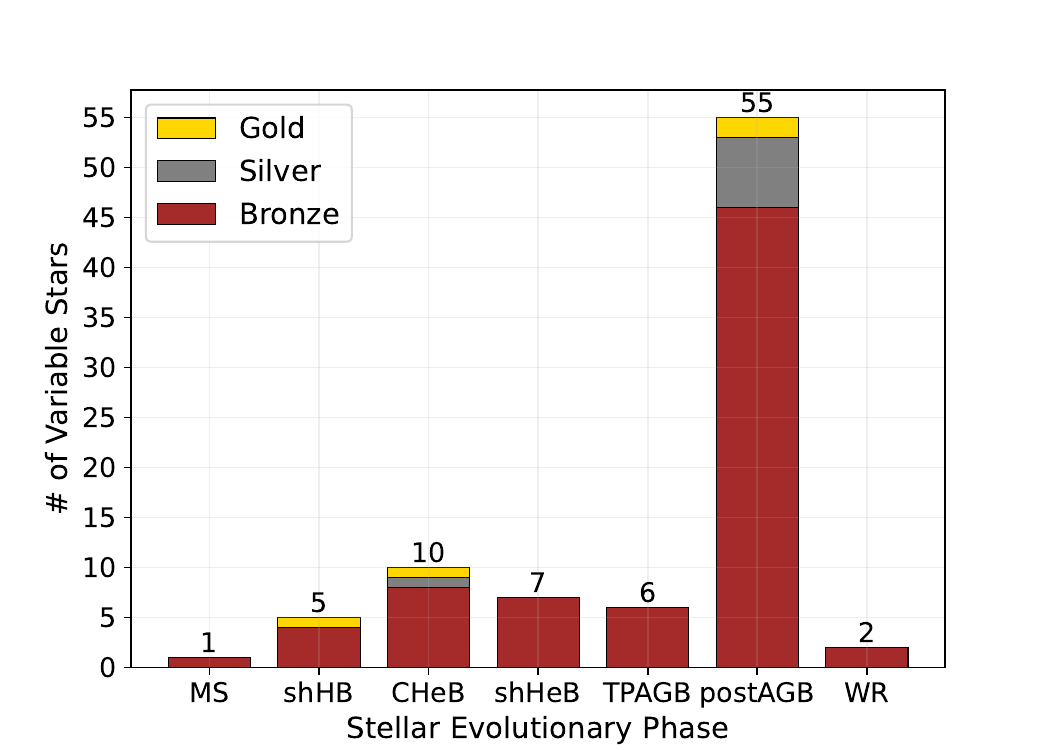}{0.45\textwidth}{(c)}}
    \caption{Distributions of stellar parameters derived from the isochrone analysis  described in Section \ref{subsec:MISTIsochrones}: (a) error ellipse scaling factor; (b) stellar initial mass, $M_{init}$; (c) stellar evolutionary phase. The \emph{Gold}, \emph{Silver} and \emph{Bronze} classification of each cluster-variable star is defined in Section \ref{subsec:BlendingAnalysis}.}
    \label{fig:iso_hists}
\end{figure*}

It is also prudent to investigate how varying host cluster properties (age, metallicity and extinction) as well as distance modulus within  their measured uncertainties affects the resulting isochrone matches.
However, formal uncertainties are not available for the host cluster properties measured by \cite{deMeulenaer2017}. 
As noted above and in their Figure~9, \citet{deMeulenaer2017} highlight the consistency between their cluster metallicity values and those of \citet{Caldwell2011} for clusters common to both studies. 
We use this consistency to identify a single cluster -- number 3801 (Johnson et al. cluster ID) -- with metallicity measurements in both studies consistent with $Z \sim 0.02$ modal value of the 47 confirmed variable star host clusters in our sample. 
We then use the difference between measured host cluster age, metallicity and extinction values obtained by Caldwell et al.\ and de Meulenaer et al.\ as the uncertainty estimates on these properties, along with \cite{McConnachie2005} distance modulus ($24.47 \pm 0.07$).
We repeat the MIST isochrone generation and matching process described above for every combination of discrete upper, middle and lower limit values of cluster 3801 age ($\mathrm{log(Age)} \in 9.8,~11.7,~13.6$), metallicity ($\mathrm{Z} \in -0.1,~0.1,~0.3$), extinction ($\mathrm{E(B-V)} \in 0.07,~0.24,~0.41$) and distance modulus (\mbox{$\mathrm{(m-M)} \in 24.40,~24.47,~24.54$}), examining the isochrone matches produced for its confirmed cluster-variable star.

The resulting isochrone matches show no change in evolutionary phase with changing host cluster properties compared to those obtained using de Meulenaer et al.\ values. 
The remaining isochrone-derived stellar properties show the greatest change as a result of varying cluster age. 
Matches to the youngest cluster isochrones ($\mathrm{log(Age)} = 9.8$) return more luminous stars (${\rm F814W}\sim-5.0$) with a broader range of stellar colors ($1.1\leq{\rm F475W}-{\rm F814W}\leq1.7$) and higher initial stellar masses \mbox{($1.1~M_{\odot}< M_{init} <1.3~M_{\odot}$)}. Matches to older cluster isochrones ($\mathrm{log(Age)} \in 11.7,~13.6$) produce fainter stars ($-4.6\leq{\rm F814W}\leq-4.4$) with bluer colors (~$0.9\leq{\rm F475W}-{\rm F814W}\leq1.4$) and smaller initial stellar masses ($0.8~M_{\odot}< M_{init} <1~M_{\odot}$).  
We find the youngest isochrones ($\mathrm{log(Age)} = 9.8$) produce matches with increasingly blue stellar colors as host cluster metallicity increases, while older cluster isochrones return increasingly red stellar colors. Increasing host cluster metallicity increases both initial mass and luminosity of the variable star for host cluster isochrones of all ages.
We find no change in isochrone-derived stellar parameters attributable to varying extinction or distance modulus.
Based on this representative host cluster example and using informal uncertainties on host cluster properties, we determine that uncertainties in host cluster age and, to a lesser extent, metallicity dominate the isochrone matching process implemented in this study; see, however, caveats in Section \ref{subsec:caveats}.

Figure~\ref{fig:iso_hists} shows the distribution of error-ellipse scale factors for the 86 confirmed variable stars, as well as the resulting distributions of initial stellar mass and evolutionary phase obtained from the isochrone analysis above. In terms of error-ellipse scaling factors, 37 of the 86 variable stars (43\%) require scaling factor $<3$, increasing to 60 (70\%) requiring scaling factor $<5$. Broken down by phase, we find the 86 cluster-variable stars in our sample comprise 1 main sequence (MS), 5 hydrogen-shell burning (shHB), 10 core helium-burning (CHeB), 7 helium-shell burning (shHeB), 6 thermally pulsing asymptotic giant branch (TPAGB), 55 post-AGB and 2 Wolf-Rayet (WR) stars. The initial masses of the stars in our sample range between $0.8\mbox{--}67~M_{\odot}$. 
We caution that these results remain subject to the effects of blending, as described in Section~\ref{subsec:BlendingAnalysis}, as well as additional caveats detailed in Section \ref{subsec:caveats}.   

Table \ref{table:finalisochroneresults} contains the results of all 86 variable stars, including observed and isochrone-derived $\rm F475W-F814W$ color and F814W absolute magnitudes, stellar evolutionary phase and initial mass, in addition to the age, extinction and metallicity properties of their host clusters obtained from \cite{deMeulenaer2017} used in isochrone generation.

\subsection{Prevalence of blending}
\label{subsec:BlendingAnalysis}

Since we derive the properties of the cluster-variable stars using their PHAT photometry in the previous section, we must consider the effects of light blending in their photometry. 
The effect of this blending is to boost the measured magnitude of a source, which can produce a near-vertical `plume' of stars on the host cluster CMD in cases of extreme blending.
Fully accounting for this blending is difficult and beyond the scope of this work. 
Instead, we assess the probability of each cluster-variable star being dominated by blending based on the measured properties of blended and unblended stars in the PHAT clusters. 
We identify two parameters to help describe the potential level of blending for each cluster star in the PHAT survey: the F814W \texttt{crowding} value produced by \texttt{DOLPHOT} \citep{DOLPHOT2000} during PHAT survey source detection \citep[described in][]{dalcanton2012, Williams2014}{}, and the distance \texttt{$D_{cluster}$} between the star and its host cluster center. 

\begin{figure*}
    
    \gridline{
    \fig{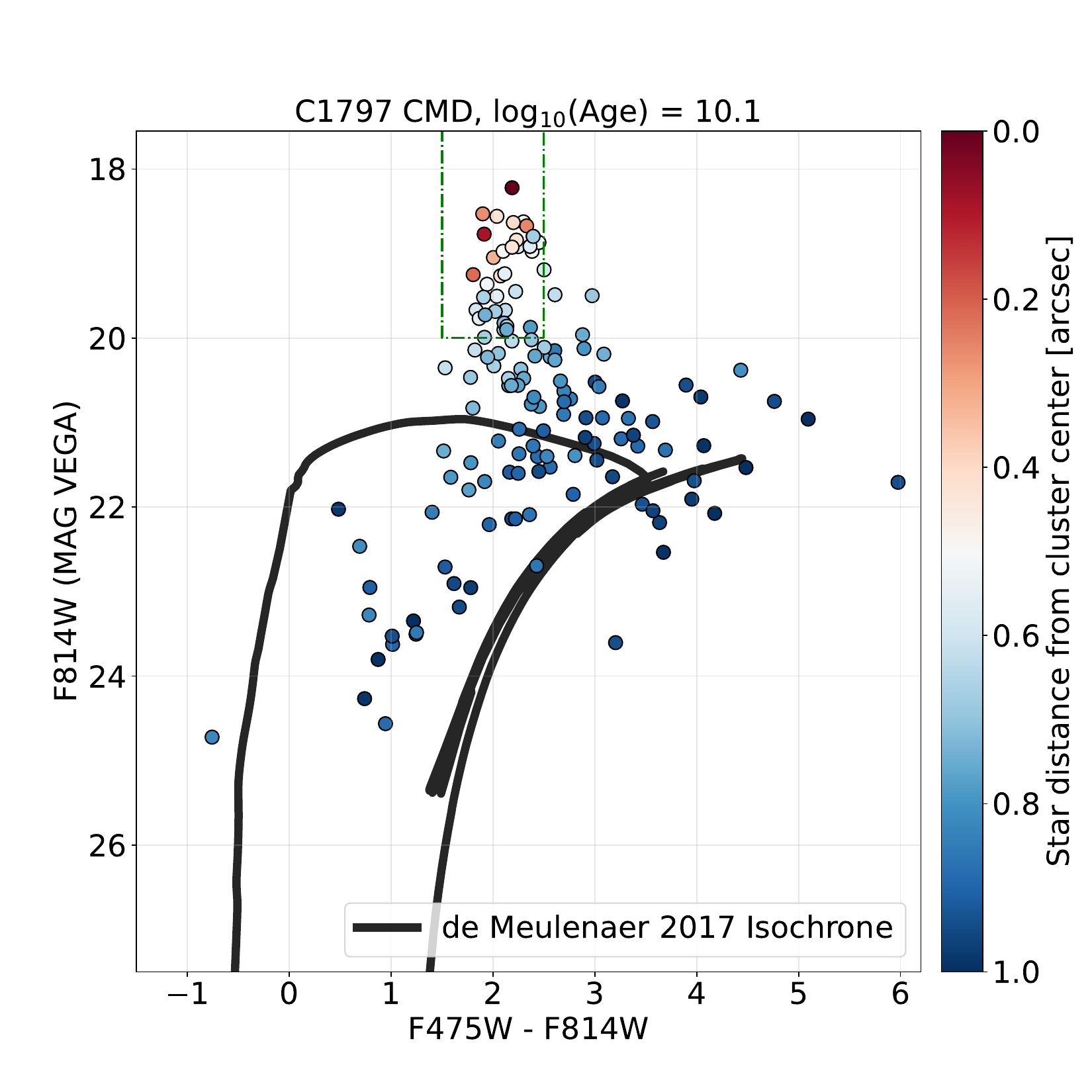}{0.45\textwidth}{(a)} 
    \fig{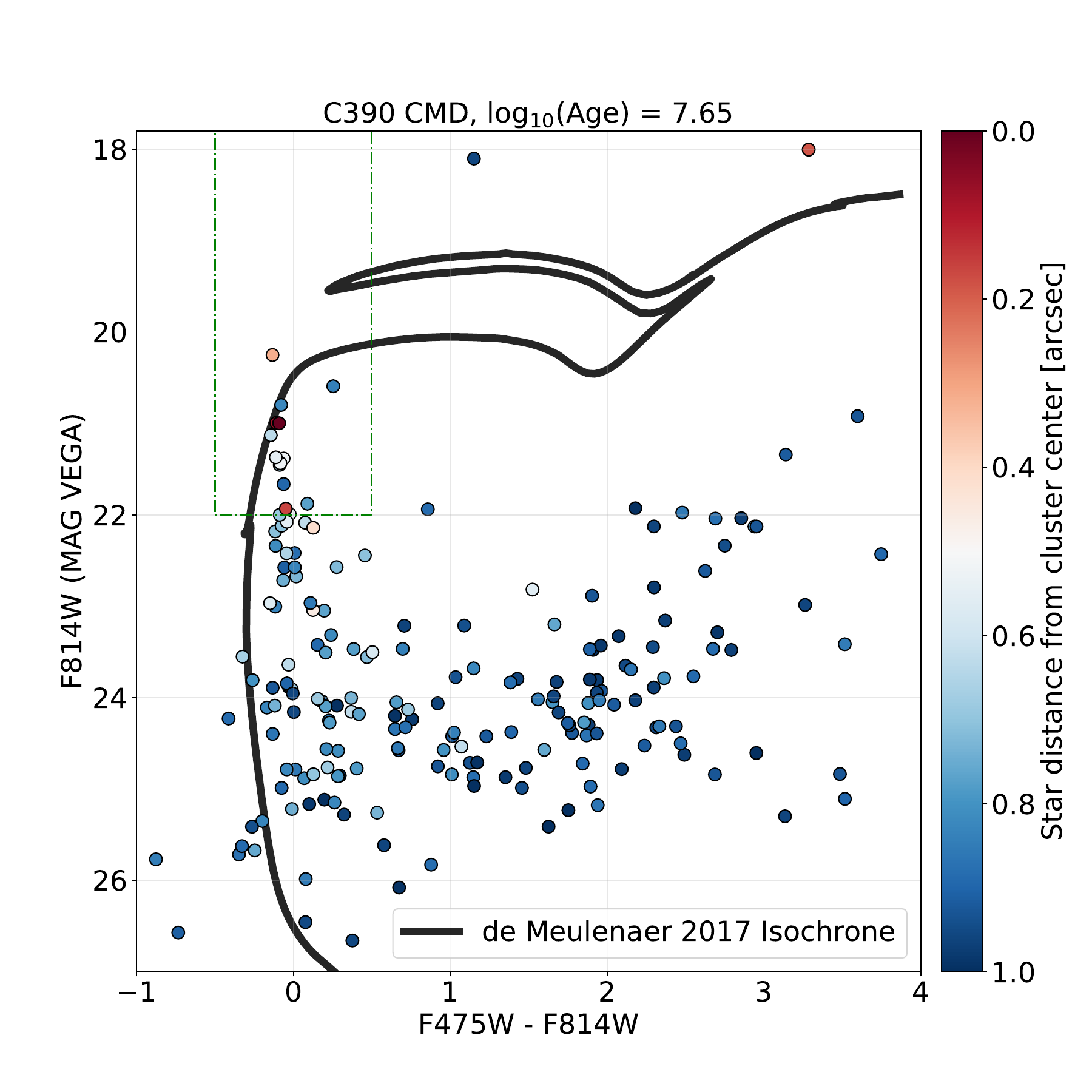}{0.45\textwidth}{(b)}
    }
    \fig{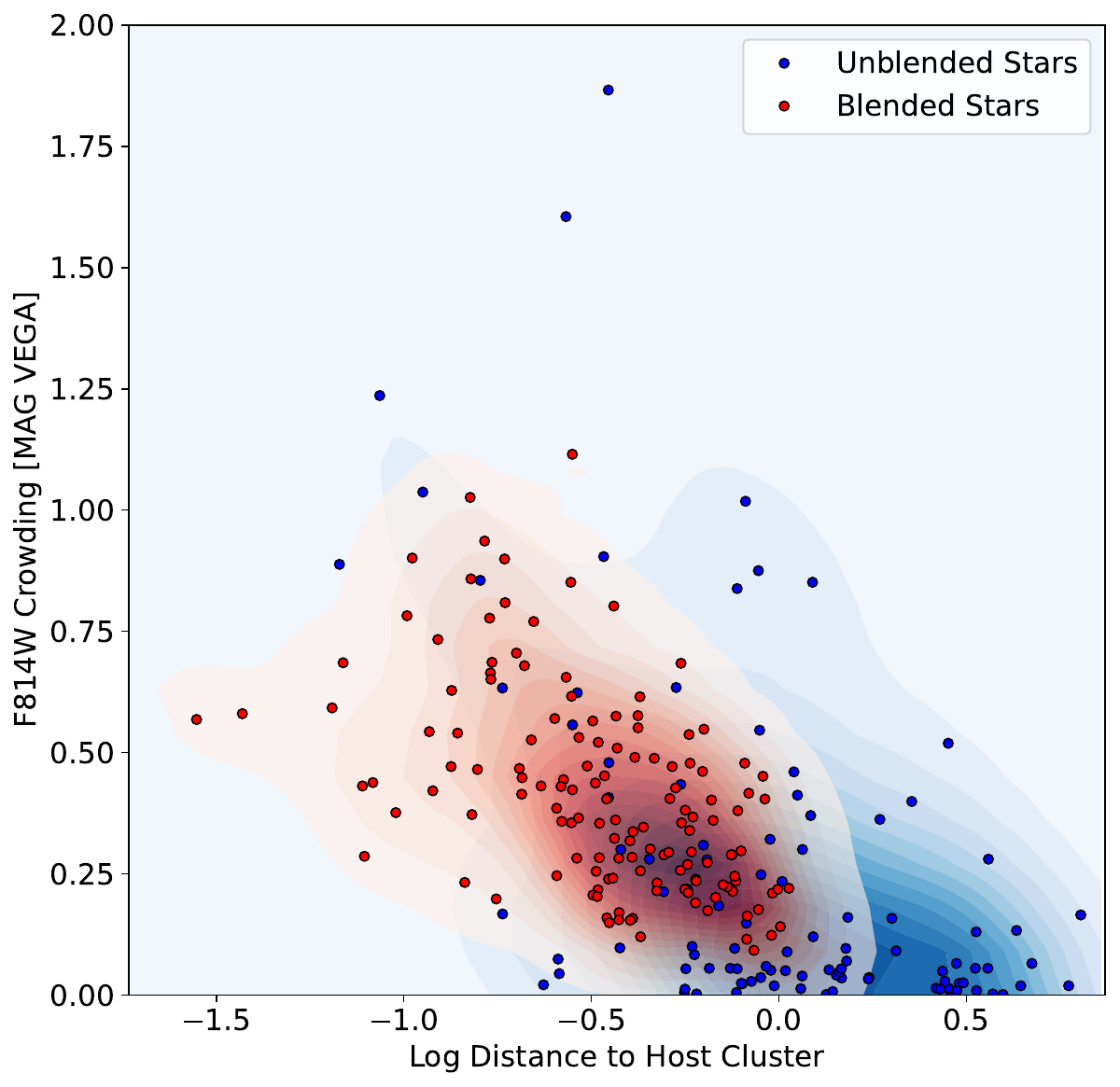}{0.45\textwidth}{(c)}
    
    \caption{CMDs of clusters 1797 (a) and 390 (b) (Johnson et al. cluster IDs), demonstrating the selection of blended (`plume') and unblended stars described in Section \ref{subsec:BlendingAnalysis} (areas enclosed by green dot-dashed lines). The \texttt{$D_{cluster}$} vs. F814W \texttt{crowding} parameter values for the blended and unblended cluster stars is shown in (c), along with their Kernel Density Estimates (KDEs) (red and blue contours, respectively). These KDEs are sampled to calculate the likelihood of each cluster-variable star being blended, following Equation \ref{eq:BlendingProb}.
    }
    \label{fig:BlendedPlume}
\end{figure*}

We then characterise the \texttt{crowding} and
\texttt{$D_{cluster}$} behaviour at varying degrees of blending by selecting representative samples of likely and less likely to be blended cluster stars from the PHAT survey. 
We select our blended star sample by first identifying globular clusters in the PHAT survey with age estimates $>10$~Gyr, their greater stellar densities making blending effects in the photometry of detected constituent stars more likely.
We produce a CMD for each globular cluster with age $>10$~Gyr along with its isochrone, generated as described in Section \ref{subsec:MISTIsochrones}, which we use to identify the `plume' of blended stars in the cluster, if present. 
We select such stars from those clusters with an unambiguous plume above its isochrone as our blended sample.  
For our unblended star sample, we consider younger (ages $<1$~Gyr), and therefore lower stellar density, PHAT survey clusters. From these young clusters, we select stars which are in good agreement with their host cluster isochrone to use as our unblended sample. 
Example cluster CMDs showing the regions used to select blended and unblended stars are shown in Figure \ref{fig:BlendedPlume}. 
Our final samples comprise 141 blended stars from 5 old globular clusters and 101 unblended stars from 7 young clusters.

We obtain the \texttt{crowding} and \texttt{$D_{cluster}$} parameters for both sets of sampled stars, which are plotted in Figure~\ref{fig:BlendedPlume}. A clear correlation between decreasing \texttt{$D_{cluster}$} and increasing \texttt{crowding} can be seen for the blended stars. We then perform Kernel Density Estimations (KDEs) for the blended and unblended groups using the \texttt{gaussian\_kde} package from the \texttt{scipy.stats} library. The resulting KDEs are also shown in Figure~\ref{fig:BlendedPlume}. 
The blended KDE (red contours) exhibits the greatest density between $0\farcs2<\ $\texttt{$D_{cluster}$}$\ < 1\arcsec$ and $0.15 <\  $\texttt{crowding}$\ < 1$ mag. The unblended KDE (blue contours) extends to larger \texttt{$D_{cluster}$} values, with highest densities occurring for \texttt{crowding}$\ \lesssim 0.15$ mag, though with clear overlap between blended and unblended KDEs. 

\begin{figure}
    \centering
    \includegraphics[width = 7.5cm]{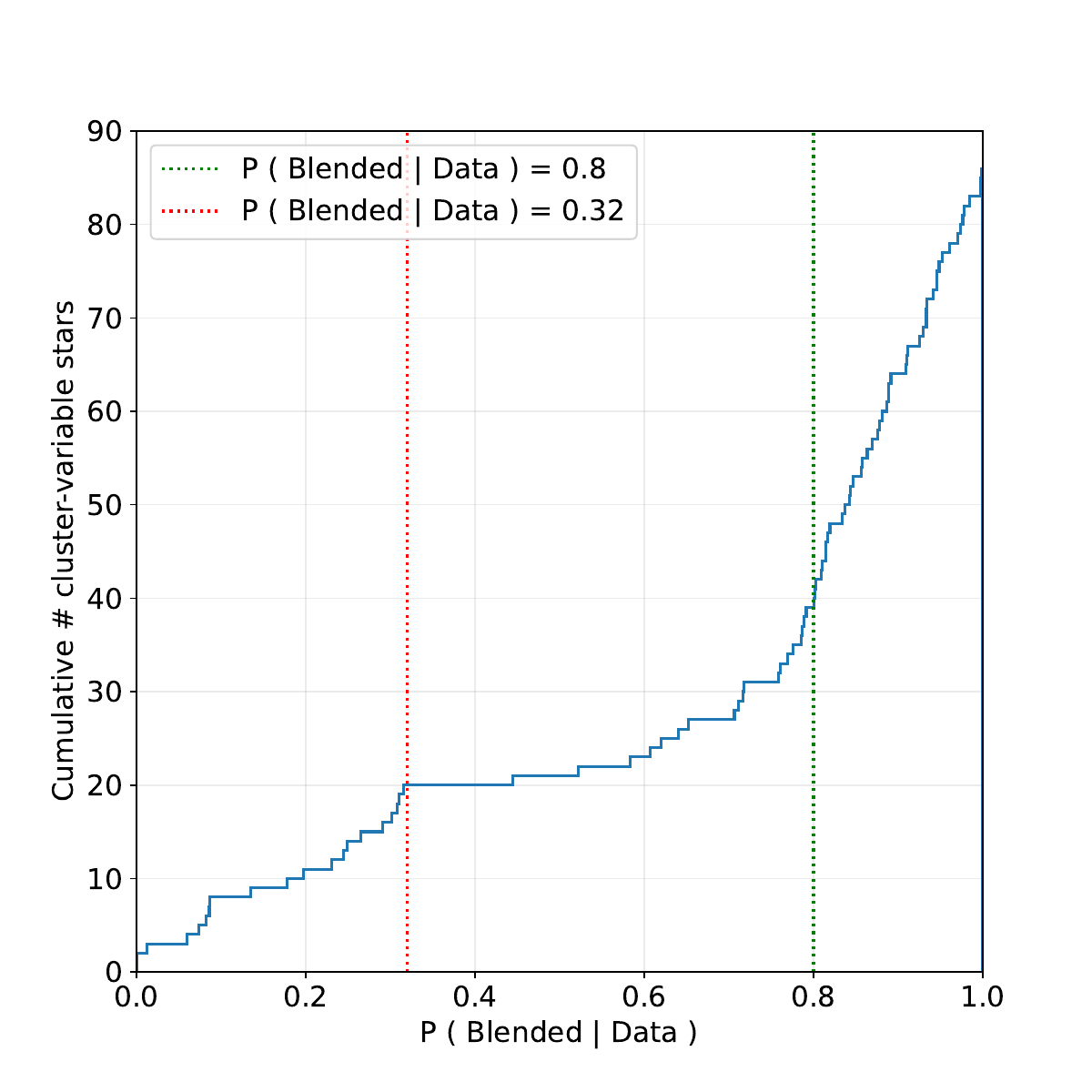}
    \caption{Cumulative distribution of blending probabilities for the 86 cluster-variable stars in our sample. The red and green vertical dotted lines denote the adopted threshold probability values between \emph{likely}/\emph{highly likely blended} and \emph{least likely}/\emph{likely blended} categories, respectively, as described in Section \ref{subsec:BlendingAnalysis}.}
    \label{fig:BlendingPosterior}
\end{figure}

We now use the blended and unblended KDEs to calculate the a posteriori blending probability for each of the 86 cluster-variable stars in our sample, following:
\begin{equation}
P(B | (x,y)) = \frac{P((x,y)| B)}{P((x,y)| B) + P((x,y)| U)}
    \label{eq:BlendingProb}
\end{equation}
\noindent where $x$ and $y$ are the \texttt{$D_{cluster}$} and \texttt{crowding} values of the cluster-variable star, respectively; $P((x,y)| B)$ is the likelihood of the star being blended, obtained by evaluating the blended KDE at $(x,y)$; $P((x,y)| U)$ is the likelihood of the star being unblended, obtained by evaluating the unblended KDE at $(x,y)$.

We examine the resulting distribution of a posteriori blending probabilities for the 86 cluster-variable stars shown in Figure \ref{fig:BlendingPosterior} to determine thresholds  by which we describe our level of confidence in the measured PHAT photometry being dominated by a single source based on measured PHAT source behaviours.
Based on this distribution, we determine:
\begin{itemize}
    \item $P(B | (x,y)) < 0.32$: star is \textit{least likely blended}
    \item $0.32 \leq P(B | (x,y)) < 0.8$: star is \textit{likely blended}
    \item $P(B | (x,y)) \geq 0.8$: star is \textit{highly likely blended}
\end{itemize}
\noindent Of the 86 cluster-variable stars, we find 20 to be \emph{least likely blended}, 19 \emph{likely blended} and 47 \emph{highly likely blended}.

We now incorporate the blending probability category for each of our cluster-variable star into a final classification, combining blending probability and our difference image source detection analysis described in Section \ref{SubSubSec:DISourceDetection}, as follows:
\begin{itemize}
    \item \textit{Gold}: `Good' detection \& \textit{least likely blended}.
    \item \textit{Silver}: `Good' detection \& \textit{likely blended}.
    \item \textit{Bronze}: 
    `Ambiguous' detection \& \textit{highly likely blended}.
\end{itemize}
\noindent Our sample of 86 cluster-variable stars therefore comprises 4 \emph{Gold}, 8 \emph{Silver} and 74 \emph{Bronze} quality results. 
These final classifications are shown in the ellipse scaling, initial stellar mass and stellar phase distributions in Figure \ref{fig:iso_hists}. 
Based on the `Good' difference image source detections and lower probability of blended photometry, we assign a higher level of confidence to the 12 \emph{Gold} and \emph{Silver} class stars in our sample. 

\begin{figure*}
    \centering
    \includegraphics[scale = 0.38]{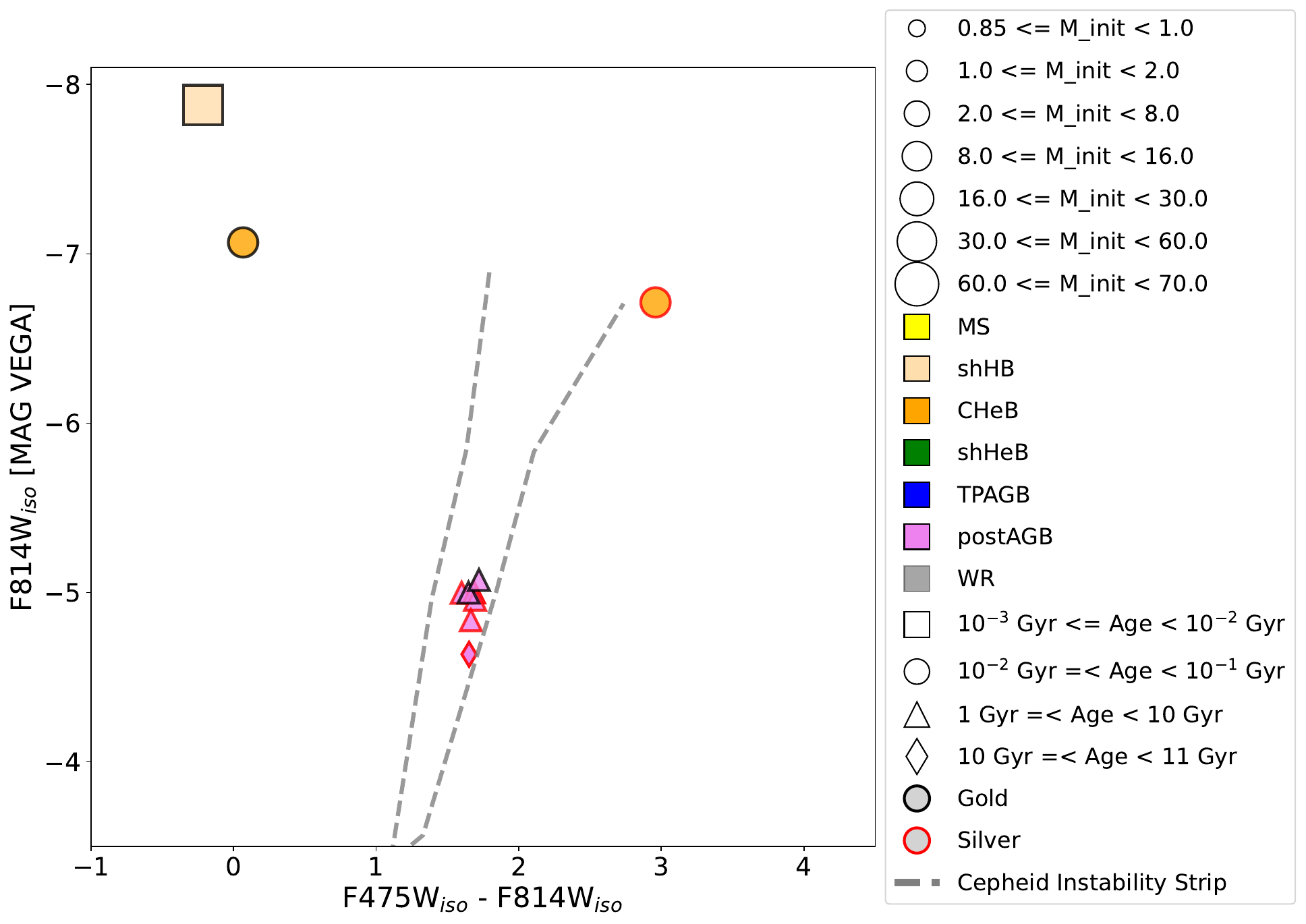}
    \caption{F475W-F814W color-absolute magnitude diagram of the 4 \emph{Gold} (black-edged points) and 8 \emph{Silver} (red-edged points) confirmed M31 cluster-variable stars, using the theoretical color and magnitude values from the most likely point on their host cluster isochrone. Stellar evolutionary phase, initial mass and host cluster age are denoted by marker color, marker size and marker shape, respectively. 
    The theoretical Cepheid instability strip is obtained from \cite{Fiorentino2002}.}
    \label{fig:MIST_CMD_GoldSilver}
\end{figure*}

\begin{figure*}
    \centering
    \includegraphics[scale = 0.38]{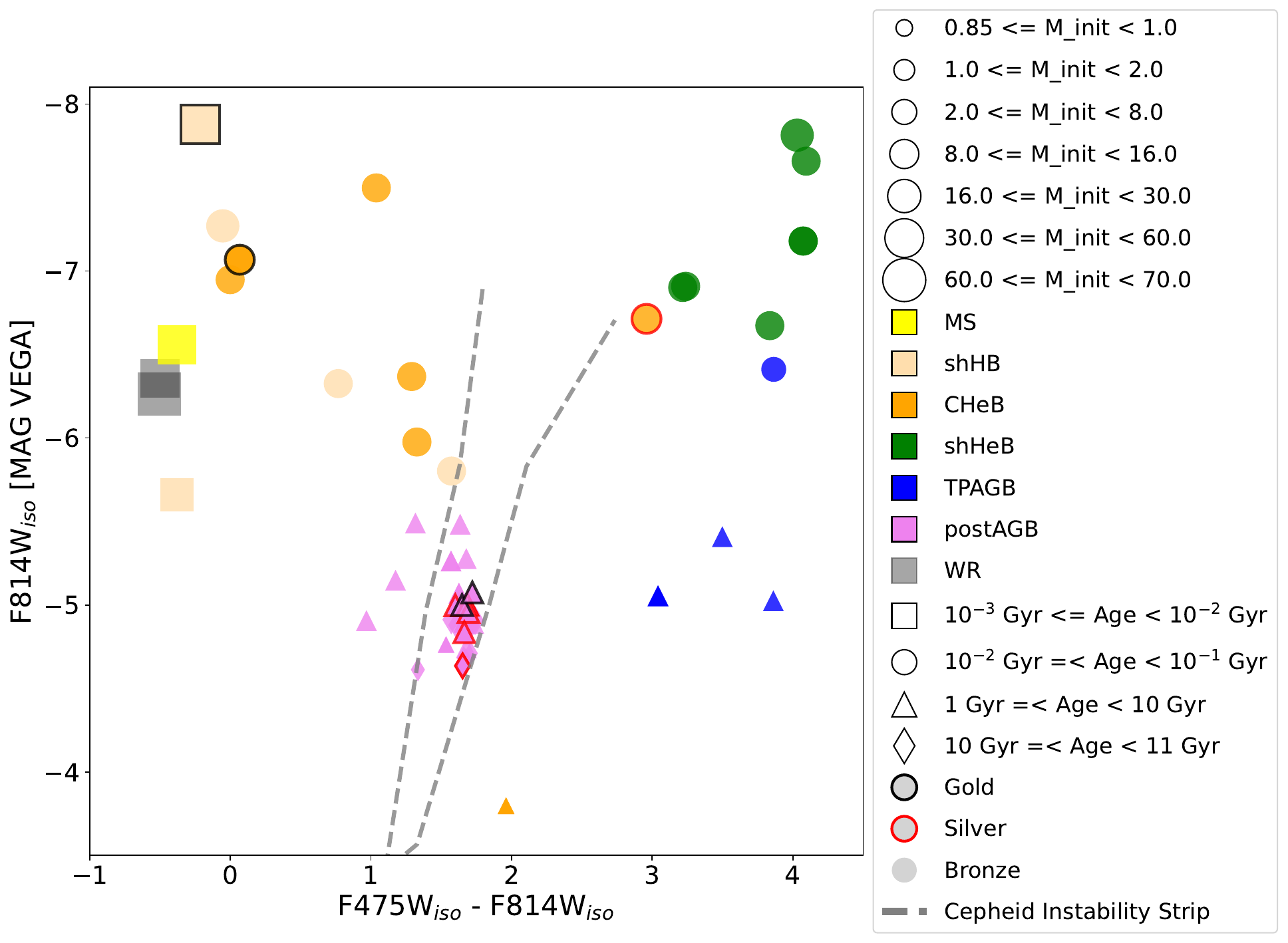}
    \caption{F475W-F814W color-absolute magnitude diagram of all 86 confirmed M31 cluster-variable stars, using the theoretical color and magnitude values from the most likely point on their host cluster isochrone. Stellar evolutionary phase, initial mass and host cluster age are denoted by marker color, marker size and marker shape, respectively. Marker edge colors denote their overall categorisation into \emph{Gold} (black-edged points), \emph{Silver} (red-edged points) and \emph{Bronze} (no edge color), as described in Section \ref{subsec:BlendingAnalysis}.  
    The theoretical Cepheid instability strip is obtained from \cite{Fiorentino2002}.}
    \label{fig:MIST_CMD_all}
\end{figure*}

As previously noted, the observed PHAT photometry of our cluster-variable stars is unlikely to represent their equilibrium states due to the sparsely-sampled PHAT measurements. 
We therefore produce CMDs of our cluster-variable stars using the theoretical color and magnitude values of the most probable datum on their host cluster isochrone, constituting their most likely equilibrium state based on the available PHAT data. 
Figure \ref{fig:MIST_CMD_GoldSilver} shows the theoretical isochrone $\rm F475W-F814W$ color vs.\ F814W absolute magnitude CMD for these 12 \emph{Gold} and \emph{Silver} cluster-variable stars using the color and magnitude values from the most probable datum on their host cluster isochrone; the figure also includes their initial masses and evolutionary phases. 
Similarly, Figure \ref{fig:MIST_CMD_all} shows the theoretical isochrone $\rm F475W-F814W$ color vs.\ F814W absolute magnitude CMD for all 86 cluster-variable stars in our sample. 
We refer to the positions of each cluster-variable star on these CMDs, in addition to their light curve properties, initial mass and evolutionary phase, when determining the type of variability shown by each star in the following section.

\subsection{Variable Star Classification}
\label{subsec:VarStarClass}

We are able to assign variable-star classifications to the 86 confirmed cluster-variable stars in our sample using the combined information of their light curve amplitudes, colors, brightness values and, for the first time, evolutionary phases and initial masses. 

\begin{figure*}[htp!]
    \gridline{
    \fig{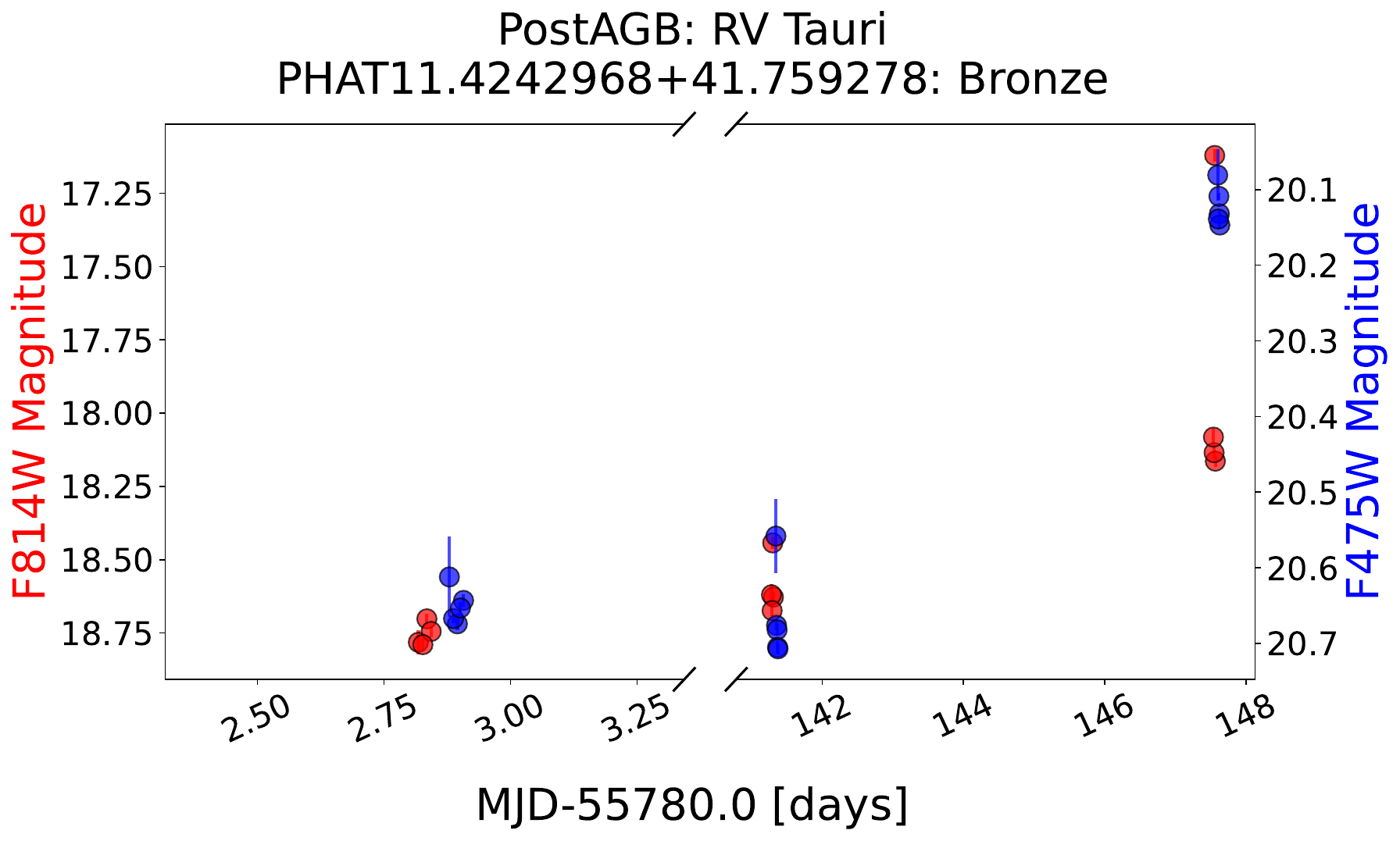}{0.45\textwidth}{(a)}
    \fig{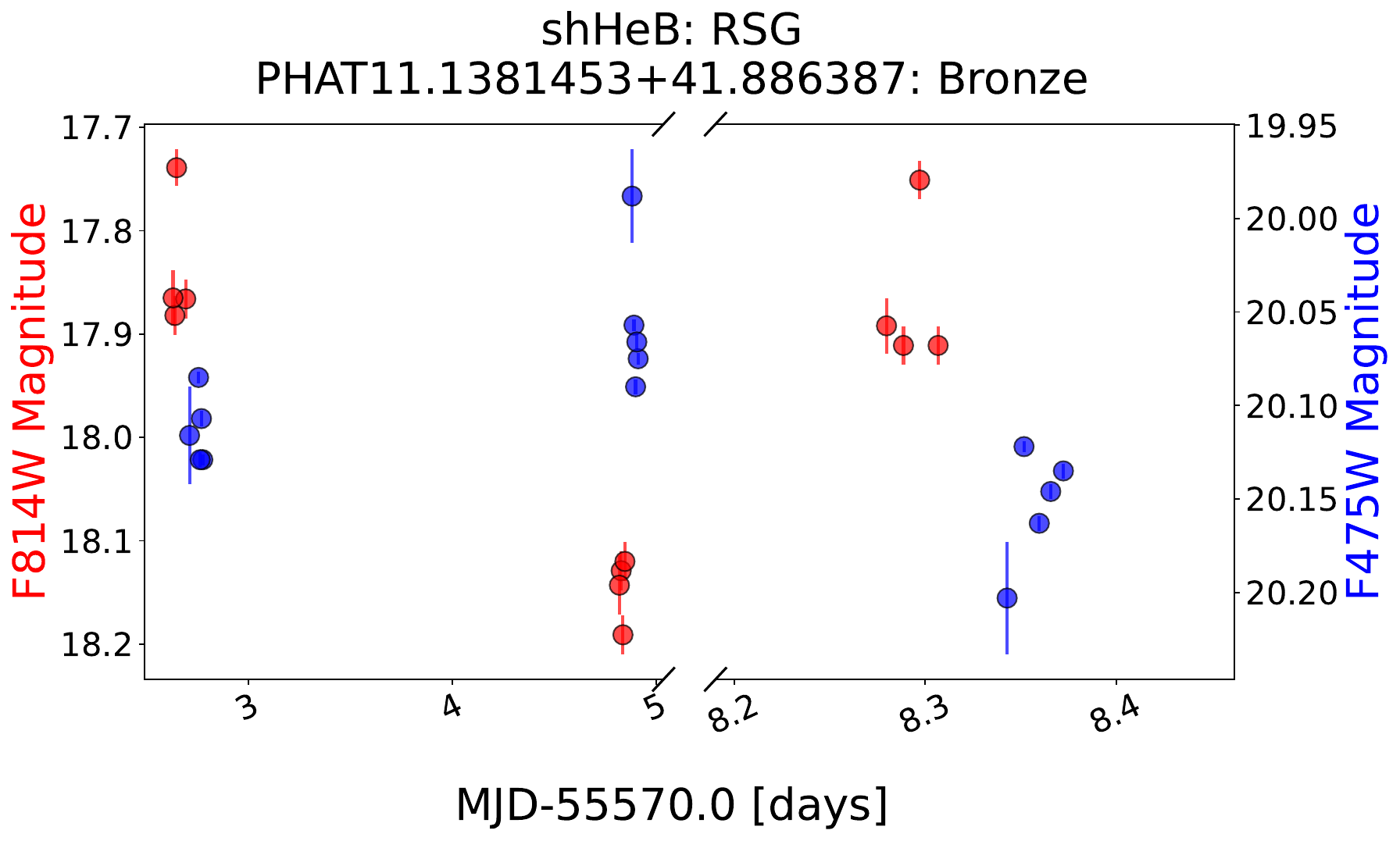}{0.45\textwidth}{(b)}}
    \gridline{
    \fig{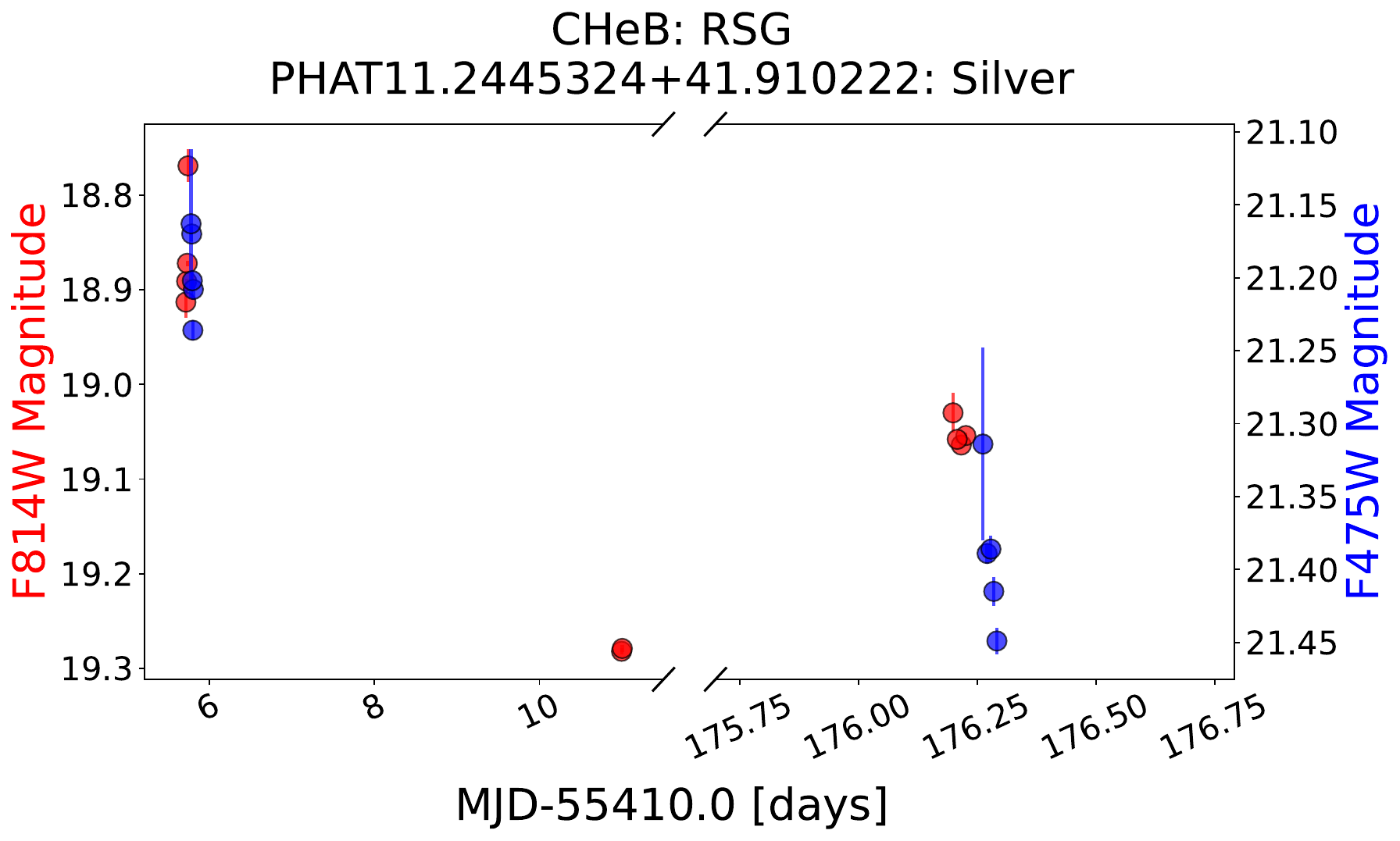}{0.45\textwidth}{(c)}
    \fig{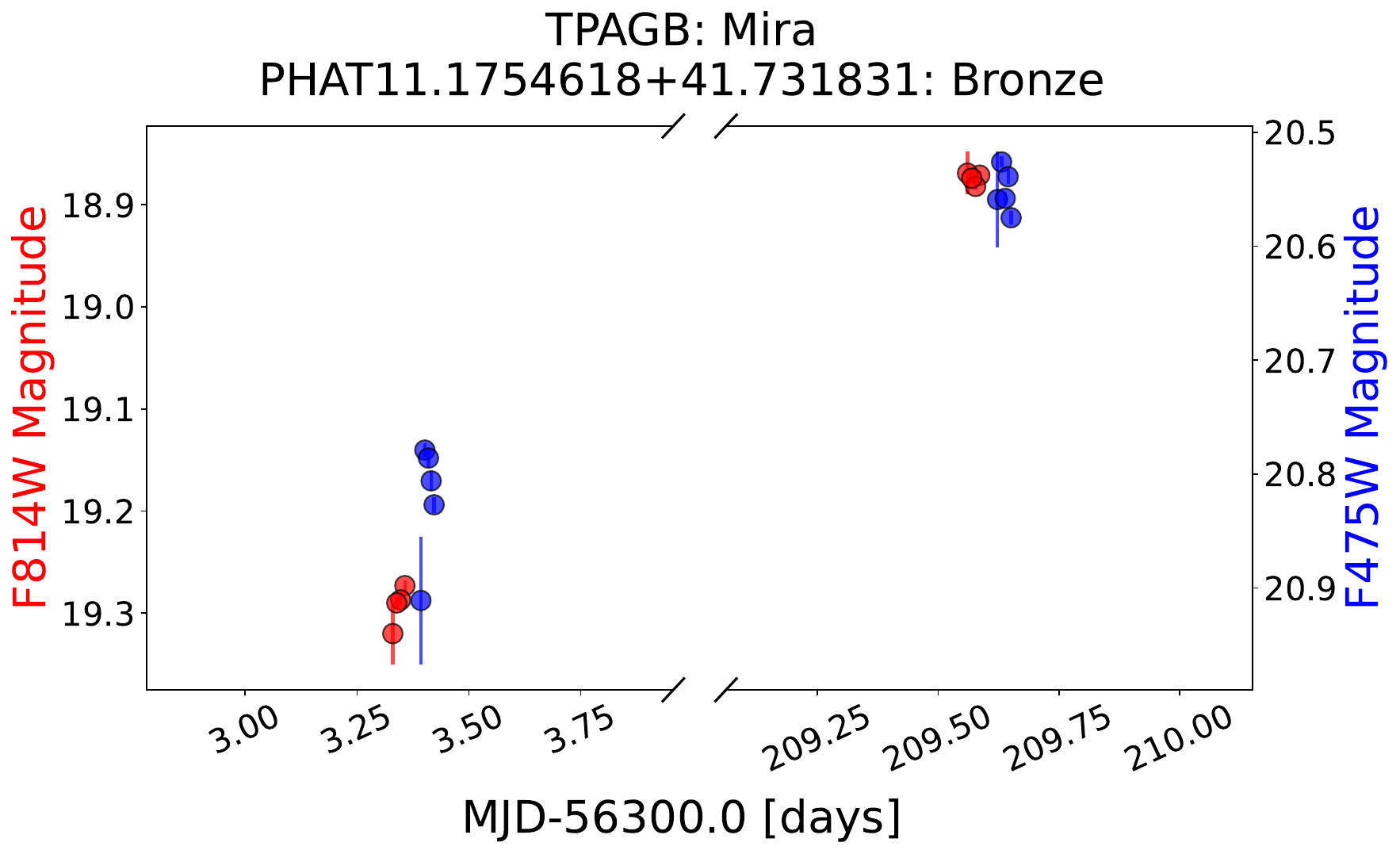}{0.45\textwidth}{(d)}}
    \gridline{
    \fig{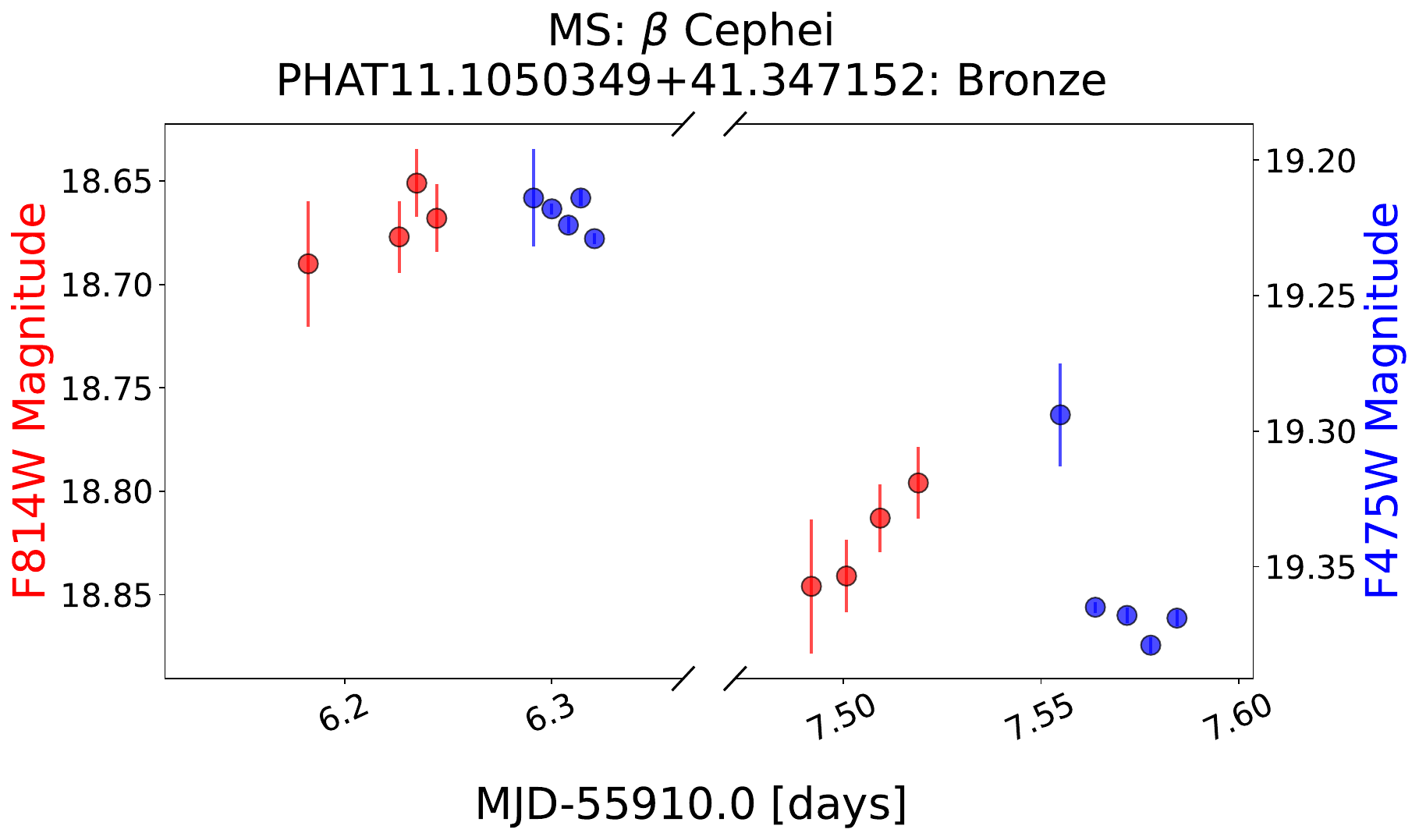}{0.45\textwidth}{(e)}
    \fig{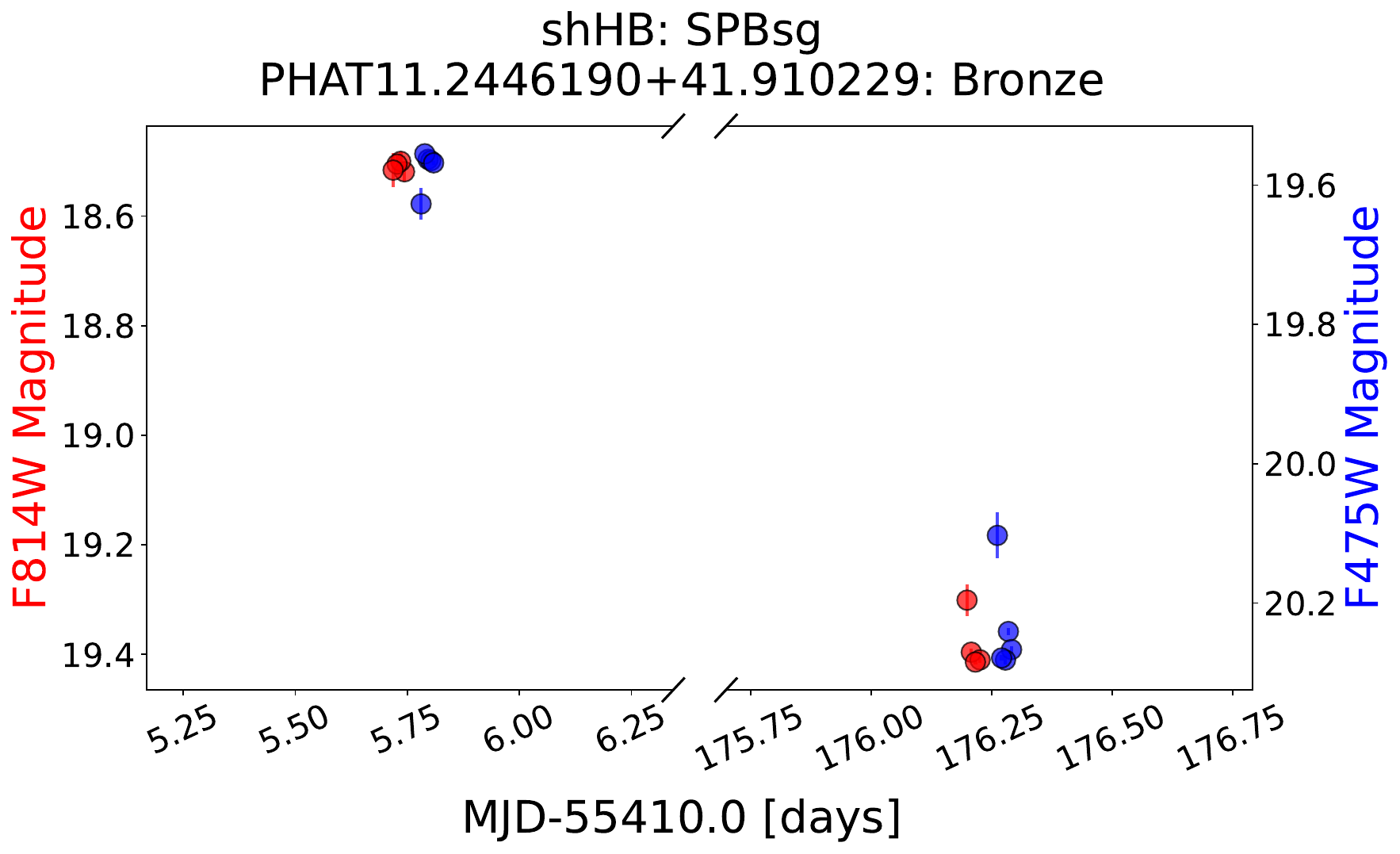}{0.45\textwidth}{(f)}}
    \caption{Light curves representing our variable stars in different evolutionary phases and variability classes including (a) RV Tauri, (b) RSG (c) classical Cepheid, (d) Mira, (e) $\beta$ Cephei (f) SPBsg.}
    \label{fig:lc_examples}
\end{figure*}

Beginning with the most numerous stars in our sample, the 55 variable stars in the post-AGB phase make up the majority of our cluster-variable star sample, with initial masses in the range $0.8\mbox{--}1.6~M_{\odot}$ and median initial mass {$\sim1.1~M_{\odot}$}. 
The most likely current stellar masses for the 55 RV Tauri stars in our sample ($0.53~M_{\odot}\leq~M_{current}<0.59~M_{\odot}$) are consistent with those typical of post-AGB stars \citep[$0.55\mbox{--}1.1~M_{\odot}$;][]{Groenewegen2017b}.
Based on their initial mass range and late evolutionary phase, we tentatively classify all 55 post-AGB stars as RV Tauri type variables.
Despite occupying approximately the same region of the CMD as classical Cepheids, as shown in Figures~\ref{fig:MIST_CMD_GoldSilver} and \ref{fig:MIST_CMD_all}, RV Tauri stars are the luminous, long-period extension of Type-II cepheids \citep{Jura1986, Alcock1998, Kiss2007, Kiss2019}. 
They have periods of $\sim20$~-150~d and a characteristic light curve exhibiting alternating deeper and shallower minima \citep{Wallerstein2002,Kiss2007,Jurkovic2021TypeIICepheids}, potentially indicating a double-mode pulsation mechanism \citep[][]{Takeuti1983}{} or chaotic behaviour \citep{Buchler1987}. 
Infrared (IR) excesses in RV Tauri observations have been attributed to circumstellar dust, the presence and nature of which can inform the singular/binary origin of the star. 
More luminous RV Tauri stars with disc-type excesses and initial stellar mass $>1~M_{\odot}$ are likely the product of binary interactions \citep{Manick2018}, while less luminous, dusty RV Tauri stars are likely post-RGB binaries with lower current masses identified by \citet{Kamath2016}.  
RV Tauri stars with no IR excess may be either a single post-AGB with initial stellar mass $\leq1.25~M_{\odot}$, with any circumstellar dust having dissipated $\sim1000$~yr after the onset of its post-AGB phase \citep{Manick2018}, or a binary post-RGB with initial stellar mass $<1~M_{\odot}$ in which the star evolves sufficiently slowly for the dust to have dissipated. 
Spectral energy distributions of the RV Tauri stars in our sample are required to identify any IR excess and diagnose their potential origins.
Alternatively, some of these post-AGB stars could be Yellow Semi-Regular variables (SRds). SRd variables are closely related to RV Tauri stars, occupying the same magnitude and color space and sharing similar initial masses, however any deep-shallow alternating light curve pattern is indistinguishable \citep{Percy2022}. 
The light curve for one of the post-AGB stars in our sample is shown in Figure \ref{fig:lc_examples}, demonstrating that the sparsely sampled PHAT survey data alone is insufficient to allow identification of RV Tauri behaviour for the stars in our sample.
That said, the distinctive RV Tauri light curve behaviour would potentially allow for easy verification for these stars using longer baseline follow-up observations designed around the $\sim20$--150~d pulsation periods of these targets.

While post-AGB stars comprise the majority of stars in our sample despite their rarity caused by the short-lived nature of this evolutionary phase, we note that 46 of our post-AGB stars are classified as \emph{Bronze}, of which 40 are \emph{highly likely} and 5 \emph{likely} affected by blending, and 7 are classified as \emph{Silver}, all of which are \emph{likely} affected by blending. 
We therefore attribute lower confidence to these 53 \emph{Bronze} and \emph{Silver} post-AGBs than the remaining 2 \emph{Gold}, \emph{least likely blended} post-AGB stars in our sample, shown with black-edged markers in Figure \ref{fig:MIST_CMD_GoldSilver}. 

The single red (${\rm F475W}-{\rm F814W}\sim3$) CHeB star with ${\rm F814W}\sim-6.7$ mag in Figures~\ref{fig:MIST_CMD_GoldSilver} and \ref{fig:MIST_CMD_all} is a \emph{Silver}-class result, with an initial stellar mass of $11.8~M_{\odot}$. 
The red color and increased luminosity of this CHeB star are consistent with the properties of red supergiant (RSG) stars. 
RSGs are the direct progenitors to Type II-P supernovae, with initial stellar masses ranging from $9~M_{\odot}$ to $\sim30\mbox{--}40~M_{\odot}$ \citep{MeynetMaeder2000,Humphreys2020} and semi-regular variability over long periods of up to a few thousand days, potentially making characterising their variability a challenge \citep{Heger1997,Soraisam2018}. 
The variability of RSGs is thought to be driven by the interaction of atmospheric convection and the \emph{$\kappa$-mechanism} in a zone of hydrogen ionization within their atmospheres. 
The $\kappa$-mechanism is a cyclical process of pulsations within a stellar atmosphere in which a layer of material increases in opacity as its temperature and density increases (such as the hydrogen ionization layer moving radially inwards in an RSG star), resulting in an increase in radiation pressure on the layer from the stellar core that pushes the layer back out.

The 7 \textit{Bronze} shHeB variable stars in our sample are also luminous (${\rm F814W}<-6.5$ mag) and show red colors (${\rm F475W}-{\rm F814W}>3$). 
The initial stellar masses of these stars range between $9\mbox{--}17~M_{\odot}$, which are again consistent with known properties of RSG variable stars outlined above \citep{Heger1997,Soraisam2018,Humphreys2020}.  
We therefore also designate these 7 evolved, red, helium-shell burning stars as RSG variables. An example light curve for one of the shHeB RSG stars in our sample is included in Figure \ref{fig:lc_examples}. 

A previous study by \cite{Johnson2012} identified about 15 candidate RSG clusters (i.e. clusters hosting RSGs) using data available from the first year of the PHAT survey. To check whether the 8 RSGs from this work correspond to those in this previous study, we cross-match the host clusters of our RSGs with those from \cite{Johnson2012} using a radius of $1\arcsec$, finding 5 matches. 
We thus confirm 5 of the Johnson et al.\ RSG candidates via variability analysis and find 3 new RSGs in PHAT survey clusters.

The \emph{Bronze}-class luminous (${\rm F814W}\sim-6.4$), red (${\rm F475W}-{\rm F814W}\sim3.9$) TPAGB star is distinct from the other TPAGB cluster-variables in our sample and appears consistent with the properties of the RSGs outlined above, as shown in Figure \ref{fig:MIST_CMD_all}. 
Its late evolutionary stage, increased luminosity and high initial mass of $\sim8 M_{\odot}$ are consistent with this star being a super-AGB (SAGB) variable candidate.
SAGBs span the low-to-high stellar mass range ($\sim6 - 12M_{\odot}$), with higher mass SAGBs being potential supernova progenitors while lower mass SAGBs evolve to white dwarf stars \citep{Doherty2017,Hiramatsu2021}. 
SAGBs occupy similar color and magnitude space as RSGs, making them nearly indistinguishable bar the potential exception of RSGs having lower variability  \citep{Doherty2017}. 
While we classify this SAGB candidate as a \emph{Bronze} result, 
our blending analysis shows it as \emph{least likely} to be affected by blending, making it an intriguing candidate SAGB star for follow-up observation and verification through measurement of its thermal pulsation period.

The remaining 5 TPAGB stars in our sample all have significantly fainter magnitudes (${\rm F814W}\gtrsim-5.5$ mag) and low initial masses in the range $1.0 \mbox{--} 1.3 M_{\odot}$.  
All 5 of these fainter TPAGBs are \emph{Bronze} class, with 4 being \emph{highly likely blended} and one being \emph{likely blended}.
The late evolutionary phase and low initial mass of these stars appear consistent with the properties of Mira variables \citep[e.g.][]{Ireland2004} -- though only one star has a F814W light curve amplitude close to that of the 2.5~mag optical magnitude threshold indicative of Mira variability at $1.9$ mag \citep{Mattei1997Miras,Willson2012,Iwanek2021}, the PHAT amplitudes of the other 4 stars ranging between $0.29 - 0.72$ mag.
The single TPAGB with Mira-like F814W amplitude is among the \emph{highly likely blended} grouping, meaning blending effects may contribute to this increased amplitude.  
However, the sparseness of light curve data available in this study only provides a lower limit on the light curve amplitudes for stars in our sample, particularly compared to the potential $\sim150$ -- 1000~d periods of Miras. 
The characteristic Mira variability of $>2.5$ mag in optical wavelengths should make verification of these TPAGB stars as Mira variables relatively straightforward using observations designed to account for the longer variability periods of Miras \citep{Iwanek2021}.

We find that the 3 CHeB stars residing within the same host cluster return identical theoretical isochrone color-magnitude properties of ${\rm F475W}-{\rm F814W}\sim 2$ and ${\rm F814W}\sim -3.8$, as shown in Figure \ref{fig:MIST_CMD_all}. 
Based on their colors, fainter magnitudes and core helium-burning evolutionary phase, these stars appear consistent with BL Herculis variable stars. 
BL Herculis variables are the least luminous members of the Type II Cepheids, moving blue-to-red along the asymptotic giant branch with a helium-burning core, unlike the higher luminosity later-stage W Virginis Type II Cepheids which undergo shell helium burning \citep[][]{Bhardwaj2020}{}.
BL Herculis pulsate with the shortest observed periods of Type II Cepheids at $\lesssim5$ days \citep{Wallerstein2002,Soszynski2008}. 
However, the $\sim1 M_{\odot}$ initial mass of the 3 CHeB stars in our sample is higher than the $\sim0.5\mbox{--}0.6 M_{\odot}$ typically assumed for BL Herculis variables \citep[e.g.][]{Marconi2007}{}{} {--} though more recent models have shown masses $\gtrsim0.7 M_{\odot}$ are required to facilitate observed first overtone-only pulsation \citep{Soszynski2019}.
Our analysis also shows the photometry of all 3 CHeB stars to be \emph{highly likely blended} or \emph{likely blended}, potentially contributing to their measured PHAT photometry being $\sim2$ magnitudes brighter than their isochrone-derived values and, in turn, producing their larger ellipse scaling factors (one $11\times$ and two $14\times$).
We therefore tentatively designate these 3 CHeB stars as BL Herculis variables based on the available PHAT photometry, with additional observations required to verify their photometry and examine their potential period-luminosity relation for Type II Cepheid characteristics.

We classify the single, \emph{Bronze} shHB star with ${\rm F475W}-{\rm F814W}\sim1.6$, ${\rm F814W}\sim-5.8$ and initial mass $9.8 M_{\odot}$ as a yellow supergiant. 
Its close proximity to the theoretical Cepheid instability strip calculated by \cite{Fiorentino2002}, as shown in Figures \ref{fig:MIST_CMD_GoldSilver} \& \ref{fig:MIST_CMD_all}, and the transition to helium-burning processes inherent to this variable star class make it consistent with a classical Cepheid variable.
We note that there is no one-to-one correspondence between the inferred masses of our variable stars and the masses of the models used to construct the instability strip boundaries in \citet{Fiorentino2002}. We primarily use these models for an approximate demarcation of the yellow belt on the CMD where unstable pulsation modes are excited.  

The remaining cluster-variable stars in our sample all exhibit bluer isochrone-derived colors (${\rm F475W}-{\rm F814W}\lesssim1.4$). 
We designate the single \emph{Bronze}-class, \emph{likely blended} MS variable star in this region of the CMD (cf.\ Figure ~\ref{fig:MIST_CMD_all}) with initial mass $\sim33~M_{\odot}$ as a $\beta$~Cephei variable. These variable stars undergo p-mode pulsations driven by the $\kappa$-mechanism at the iron opacity bump \citep[e.g. ][]{Moskalik1992,Miglio2007,Saio2013}.
The F475W and F814W light curves for this MS variable star are shown in Figure \ref{fig:lc_examples}.
The 4 shHB stars with ${\rm F475W}-{\rm F814W}<1$ and initial stellar masses $11~M_{\odot}< M_{init} <34~M_{\odot}$ comprise 3 \emph{Bronze} class results and one \emph{Gold} class result. 
We classify these 4 shHb stars as slowly pulsating B-type ``super giants'' (SPBsg) -- g-mode-pulsating variable stars distinct from $\alpha$~Cygni variables \citep{Saio2006,spbsg2013}. These authors examined stellar evolutionary models for $7-20~M_{\odot}$ stars, finding that g-mode excitation is possible in post-MS stars at a broad range of effective temperatures for masses $\gtrsim9~M_{\odot}$. It is possible that p-modes are excited simultaneously in some of these shHB stars as there is some overlap in the instability regions of the SPBsg and $\beta$~Cephei classes \citep[see e.g. ][]{Saio2006} -- those stars will therefore be hybrids of the two classes. 
An example light curve for one of the shHB stars in our sample is shown in Figure \ref{fig:lc_examples}.

The 6 blue CHeB stars comprise 1 \textit{Gold} and 5 \textit{Bronze} results -- the \emph{Gold}-class star having the same color (${\rm F475W}-{\rm F814W}=0$) and magnitude (${\rm F814W}=-7$) values as a \textit{Bronze} CHeB star.
These 6 CHeB stars are divided into 2 fainter stars, with ${\rm F814W} > -6.5$ and colors $1<{\rm F475W}-{\rm F814W}<1.5$, and 4 more luminous stars, with ${\rm F814W}<-6.5$ and broader $0<{\rm F475W}-{\rm F814W}<1$ color range.
The $\sim8\mbox{--}16~M_{\odot}$ initial mass range of these 6 CHeB stars indicates they may also be SPBsg stars, though in core helium burning phase rather than hydrogen shell burning (see \citealt{Ostrowski2015}). 
We note, however, that the median F814W amplitude of $0.21$ mag for the combined set of blue shHB and CHeB variable stars in our sample is larger than that for the single SPBsg known so far (HD~163899; see \citealt{Saio2006}). 
Processes such as mass loss, rotational mixing, core contraction etc.,\ can also lead to a blue-ward evolution of massive stars after the RSG phase before core-helium exhaustion \citep{Saio2013, Meynet2015}. 
In particular, \citet{Saio2013} found that many pulsation modes can be excited in these post-RSGs in the blue region, with such stars constituting the $\alpha$~Cygni variable class. 
The 2 fainter (${\rm F814W}>-6.5$ mag), \emph{Bronze}-class CHeB stars have initial masses in the range $\sim8 - 10 M_{\odot}$. 
Modelling has shown that stars in this mass range can undergo extensive blue loops; see, for example, Figure 4 of \citet{Ekstrom2012}. 
Examining the MIST-generated stellar evolutionary tracks of these fainter CHeB stars shows the most probable isochrone points for both stars reside on their respective blue loops.
We therefore classify the 4 more luminous, blue CHeB stars as SPBsg, while the 2 fainter, blue CHeB are likely $\alpha$~Cygni variables -- though it is unclear if their luminosity-to-mass ratio is large enough to excite pulsations (see \citealt{Saio2013}). Detailed frequency analysis via follow-up observations of these fainter stars would help to robustly identify their pulsation modes and verify their variable class type.

Finally, the 2 \emph{Bronze} WR stars in our sample have the largest initial stellar masses in our sample at $\sim52~M_{\odot}$ and $\sim66~M_{\odot}$. 
WR stars have high mass loss rates due to their strong stellar winds \citep{Poe1989}, with variability observed over ranges of hours to days \citep[e.g. ][]{Toala2022_WRVariability}{}. 
The variability of WR stars in M31 has previously been studied by \citet{Soraisam2020}, with the variability of WR stars attributed to a combination of line-driven instabilities producing inhomogeneities in their stellar wind, rotation-driven perturbations in their winds and pulsation instabilities \citep[][]{Poe1989,StLouis2009,Toala2022_WRVariability}{}.

\subsection{Caveats}
\label{subsec:caveats}

There are a number of important caveats to highlight when considering the results produced from this work. We summarise these key points below:

\begin{itemize}
    \item It is important to emphasise that the photometry of all cluster stars in the PHAT survey will still be affected by blending to some extent, including all the cluster-variable stars in our sample, regardless of blending probability.
    Ideally, the use of sophisticated de-blending software may allow for the flux from a single source to be obtained, however the variation in crowding across cluster environments would necessitate careful treatment of each detected source to accurately de-blend the measured photometry. This is beyond the scope of the present study. Nevertheless, it is notable that we were still able to detect 
    variable sources within these crowded cluster environments through our difference image analysis, despite the prevalence of blending in the measured PHAT photometry. 
    
    \item As noted previously -- and demonstrated by Figures \ref{fig:good_lcs}, \ref{fig:ambiguous_lcs} 
    \& \ref{fig:lc_examples} -- the sparseness of the PHAT survey photometry means only lower limits on stellar light curve amplitudes can be produced from PHAT survey data, impacting this study in two key ways. Firstly, some cluster-variable stars detectable by the PHAT survey are likely not to have been identified, either due to being at the minimum phase in their light curve when observed, or due to an insufficiently long baseline of observations being taken, precluding their detection. 
    Secondly, the lower limits on light curve amplitudes mean that the equilibrium point in color-magnitude space for a given cluster-variable star will be subject to greater uncertainty. 
    This will also affect identification of the most probable isochrone datum for the star, and therefore its initial mass and evolutionary phase (Section \ref{subsec:PosteriorCalc}). 
    Obtaining accurate light curves for the stars in the PHAT survey, for example through a long-baseline extension of the PHAT survey, would allow for these uncertainties to be reduced as well as provide huge potential for further variable star identification and subsequent classification. 
    
    \item The results presented in this work are also highly dependent upon the accuracy of literature cluster properties and the published isochrone models used. Inaccuracies in either of these dependencies will naturally impact the results obtained. The most notable effects would originate from the use of incorrect cluster properties based on solar- vs non-solar metallicity estimates and the inability of isochrone models to accurately account for the various mechanisms driving stellar variability. An attempt at the latter could be explored by building stellar variability into models of host cluster isochrones based on empirical measurements, such as from this work, and these modified isochrones then being used to examine expected variable behaviour over a given timescale. 
    Even with the dependencies described above, however, and despite the limitations inherent to the sparse sampling of the PHAT survey data, the variable stars identified in this study show remarkable alignment with the initial masses and evolutionary phases of known variable stars. 

    \item Similarly, and as noted in our investigation into the effects of uncertainty on host cluster properties, there is a general lack of formally-derived uncertainty estimates for host cluster age, metallicity and extinction in the literature. 
    Our analysis using the consistency between independent studies to derive uncertainty estimates in Section \ref{subsec:PosteriorCalc} provides an indication of the effects of varying these parameters for a representative host cluster -- however, the limited number of host clusters common to both studies precludes more thorough analysis across all clusters in our sample.  
    Our conclusions on the overall effects of varying cluster properties on the isochrone-derived variable star parameters are therefore limited without formal uncertainties for each host cluster.

    \item We also assume all cluster variable stars in our sample are single-star systems when generating isochrones and evolutionary tracks. 
    The evolution of binary or multi-star systems is expected to follow the same path as their single-star counterparts - provided their stellar orbits are sufficiently separated and therefore non-interacting. 
    When interactions occur -- for example, mass transfer or tidal interactions -- the evolutionary path of both stars involved can be significantly altered compared to single-star systems depending on their initial stellar and orbital properties, with such interactions contributing to the more exotic known star types \citep[see, for example, ][and references therein]{Hurley2002}. 
    The global binary fraction ranges from $10-50\%$ in globular clusters \citep{Sollima2007} and $35-70\%$ in open clusters \citep{Sollima2010}.  The binary fraction for massive stars is $1.0\pm0.2$ with $80-90\%$ likely to interact with a companion, while for lower mass, e.g., solar-type stars, the binary fraction is $\sim0.08$ with $15\pm3\%$ likely to experience binary interactions \citep{Moe2017}. Hence, our massive-star results are likely to be sensitive to the models used, but we do not expect our low-mass star results to be dramatically affected by using single-star models.  
    That said, the crowded nature of the cluster environments makes even verifying the multiplicity of star systems identified within our sample a challenge.

    \item In using a single set of derived literature age and metallicity values for each cluster in this study, we are also making the assumption that the cluster and its constituent stars are co-evolved. Observations of clusters show increasing evidence of multiple populations of distinct age and metallicity being produced within cluster environments (see, for example, \citealt{Milone2022}). Investigating this possibility and its potential effects on our results is beyond the scope of this work, however an extension of this study could be conducted should multiple populations be identified in the host clusters analysed in this work. 
    
\end{itemize}

\hfill \\
\section{Future Direction of Work}\label{sec:future}

We have specifically focused on variable stars in M31 stellar clusters in this work, generating isochrones using estimates of host cluster age, extinction and metallicity from which we have inferred the evolutionary phases and initial masses for the variable stars in our sample. 
Our next study will identify field variable stars in the M31 PHAT footprint, using the knowledge gained from applying the DI Pipeline in this work. 
Before proceeding with this next study, we first identify potential areas for improvement in our methodology by comparing our DI pipeline difference images to those produced by a cross-filtering convolution method \citep{GalYam2008} using the \texttt{photutils} `effective PSF' (ePSF) generation routine \citep{Anderson2000ePSF}. 
These methods were chosen due to ePSFs having supplanted the \texttt{TinyTim} \citep{Krist1993} Hubble PSF-generating software for ACS observations\footnote{\url{https://www.stsci.edu/hst/instrumentation/focus-and-pointing/focus/tiny-tim-hst-psf-modeling}}, 
while cross-filtering convolution allows for comparisons with DI pipeline results without extensively altering the OIS method to use ePSFs.

We select six science-template image pairs and their pipeline difference images for investigation. Each set of images is associated with a different cluster and selected to create a representative sample of potential DI pipeline behaviour, with 2 `good' (minimal artifacts), 2 `poor' (extended, brighter residuals) and 2 misaligned subtractions. 
Reference sources required for ePSF generation are obtained from background-subtracted science and template frames, obtained by reprojecting and aligning each pair of science and template frames in the same way described for the DI Pipeline in Section \ref{subsubsec:alignmentBackgroundmatching}, before estimating the background of each frame.
PSF reference sources are then detected in each background-subtracted frame using \texttt{find\_peaks} from \texttt{photutils} with a threshold value of 2. 
We visually assess cutouts of each potential reference source for crowding, contamination and stellar-like behaviour, excluding those sources contaminated within a 5 pixel radius of the peak. Isolated, stellar-like sources are used in \texttt{photutils}  \texttt{ePSFBuilder} to generate accurate ePSF models for the frame. 
We vary the \texttt{ePSFBuilder} sigma-clipping, maximum iteration and input cutout size (but no oversampling) to identify the best parameter values for accurate ePSF model generation, obtaining best results using 9 pixel diameter cutouts, $3\sigma$-clipping and 50 max. iterations.  
We then convolve the science frame with the template ePSF and vice versa, background matching the convolved frames by subtracting the convolved template frame from the convolved science frame and using this intermediate difference image to produce a \texttt{Background2D} map which is added to the convolved template frame. The difference image process is then repeated and the result compared to the DI pipeline difference image through visual inspection and image statistics measured within a $75\times75~{\rm pixel}$ region centred on the host cluster in each difference image. 

\begin{figure*}[tb]
    \centering
    \includegraphics[scale = 0.52]{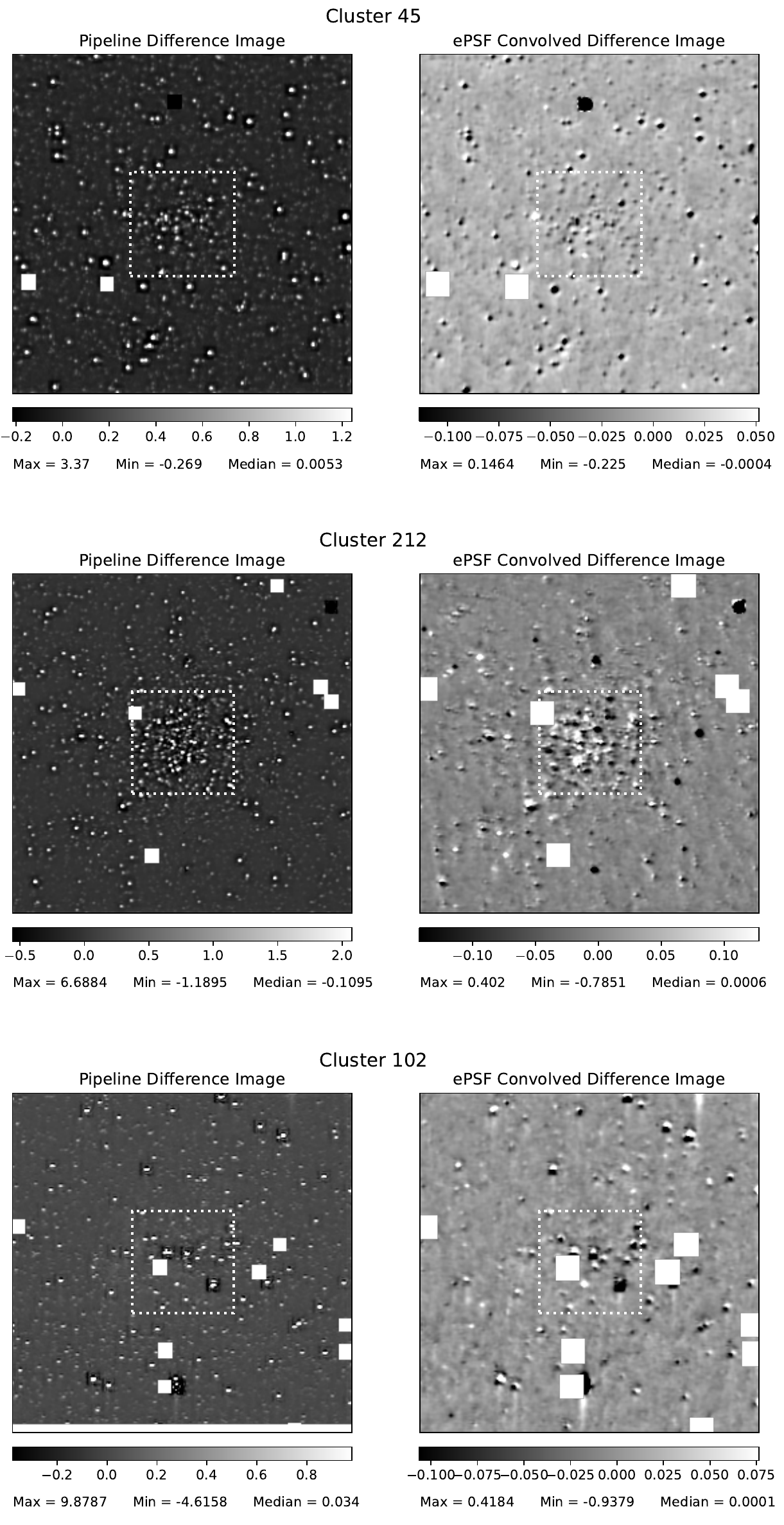}
    \caption{DI pipeline (left) and ePSF-based (right) difference images produced for 3 of the 6 test cases used to identify potential improvements using ePSF models. Areas surrounded by the white-dashed line denote the region of interest encapsulating each cluster within which image statistics are calculated.}
    \label{fig:ePSFComparison}
\end{figure*}

Figure \ref{fig:ePSFComparison} shows example DI pipeline vs convolved ePSF difference images for science-template frame pairs. 
We find the ePSF difference images show a general improvement in image statistics, with the strength of over- and under-subtraction substantially improved. That said, the ePSF results are strongly dependent on input PSF behaviour -- particularly for very crowded clusters, e.g. Cluster 212 in Figure \ref{fig:ePSFComparison}. 
Small deviations from stellar behaviour in input PSFs can produce inaccuracies in the ePSF model, leaving residual patterns around sources in the difference images. 
Even with this dependence on input PSF behaviour, the improvements shown by the ePSF results look promising. 
We will consider adopting and further refining the ePSF-based convolution method in upcoming studies searching for M31 variable field stars; this will require automation of input PSF selection due to the larger dataset to be analysed. 
We will also investigate the `proper image subtraction' (`\texttt{ZOGY}') method developed by \cite{Zackay2016}, including its in-built PSF extraction, as an alternative to the ePSF convolution method. 

Aside from applying our results to future studies, our results highlight the enormous potential of well sampled, long baseline observations. 
Though sparsely sampled, our analysis of M31 stellar clusters in PHAT survey has nevertheless produced valuable results and potential targets for follow-up observations. 
A purpose-built time-domain survey of a similar scope to the PHAT survey with a space mission would allow for many of the variable stars in this work to be verified, provide further constraints on the properties of longer-period variables and likely result in additional candidate and confirmed variable stars.
The upcoming \emph{Roman Observatory} holds enormous potential for such a study, as highlighted by the proposed \emph{RomAndromeda} survey \citep{RomAndromeda2023}.
Though a similarly sparse sampling to the PHAT survey, the proposed \emph{RomAndromeda} survey compliments PHAT and allows for shorter-period variable stars to be identified at magnitudes fainter than the PHAT survey (${\rm F146} < 25.3$, ${\rm F062} < 24.8$ AB mag; \citealt{RomAndromeda2023}) within the first year of \emph{Roman} operations. This could feasibly be expanded to identifying longer-period variables in M31 observations should the \emph{Roman} mission be extended and additional observing epochs obtained.

\section{Conclusions}\label{sec:con}

We exploit the photometric depth and spatial resolution of \emph{HST} to systematically identify variable stars within M31 stellar clusters using the PHAT survey data. 
Out of 294,981 stars in 2,753 M31 clusters, we identify 239 luminous (${\rm F814W}<19$~mag) candidate variable stars via statistical analysis of their sparse PHAT light curves. 
The highly crowded environments of M31 stellar clusters present a challenge for visually vetting the candidates. We therefore utilise difference imaging to verify PHAT light curve variability and confirm 86 cluster-variable stars, after removing 3 likely foreground stars through crossmatching with Gaia DR3. 
We use the MIST web interpolator to generate an isochrone for each host cluster based on properties available in the literature.
We match each confirmed variable star to the most probable point on its host cluster isochrone, finding initial stellar masses in the range $0.8\mbox{--}67~M_{\odot}$ for the 86 stars in our sample. 
In terms of evolutionary phase, we find the majority of the variable stars are in the post-AGB phase, while the remaining stars are distributed across the MS phase and more evolved hydrogen- and helium-burning phases.

For the first time, we use the initial mass and evolutionary phase inferred from host cluster isochrones to assign a variable class to each star in our sample.
We attribute the stars in more evolved hydrogen- and helium-burning phases to a mixture of $\beta$ Cephei variables, slowly pulsating B-type supergiants and candidate $\alpha$ Cygni variables.
We find 8 RSG variable stars in our sample, 5 of which have been previously identified by \cite{Johnson2012} which we confirm through their variability, alongside a candidate super-asymptotic giant branch star based on its red supergiant-like luminosity, red color and $\sim8 M_{\odot}$ initial mass.
We also identify numerous RV Tauri variables, along with smaller numbers of candidate Mira variables, a likely classical Cepheid, BL Herculis variables and Wolf Rayet stars.
We calculate the probability of measured PHAT photometry being blended for each of our cluster-variable stars, which we consider along with the quality of difference image source detections to inform our level of confidence in their attributed properties.
From this analysis, we characterise the properties of 12 cluster-variable stars in our sample -- 9 RV Tauri stars, 2 slowly pulsating B-type supergiant stars, and one red supergiant -- with higher confidence.

We caution that this work is dependent upon the accuracy of published isochrone models and cluster properties used, in addition to PHAT survey photometry of all cluster stars remaining subject to blending effects, and uncertainties in derived stellar properties due to the sparse PHAT light curve data. 
That said, our results show a remarkable consistency with known variable classes, indicating the potential for discoveries from a purpose-built, well-sampled PHAT-like time domain survey with an extended time baseline. 
The \emph{Roman Observatory} holds enormous potential for studying variable stars in M31, with the proposed \emph{RomAndromeda} survey \citep{RomAndromeda2023} allowing for short-period variables at fainter magnitudes than those in the PHAT survey to be identified within a year of \emph{Roman} operations commencing, with potential expansion to longer-period variables should the mission be extended.

\begin{longrotatetable}
\begin{deluxetable*}{cCCCCcccCc|CCCCCC} 
    \tabletypesize{\tiny}
    \tablecaption{\label{table:Confirmed} Properties of each of the 89 cluster-variable stars confirmed through DI Pipeline analysis, including host cluster properties from \citet{deMeulenaer2017}.  
    Star IDs (\emph{PHAT ID}) are those used in the PHAT survey and include the catalog-recorded RA and Dec of the star in decimal degrees. 
    Variable star F814W magnitudes and F475W-F814W colors are divided into observed (\emph{obs}) and isochrone-derived values (\emph{iso}; see Section \ref{subsec:PosteriorCalc}), the latter used in Figure \ref{fig:MIST_CMD_GoldSilver}.
    }
    \centering
    \tablehead{
        \multicolumn{10}{c}{Variable Star Properties} & \multicolumn{6}{|c}{Host Cluster Properties} \\ [-0.15cm] 
          \colhead{PHAT ID} &  \multicolumn{2}{c}{F814W} & \multicolumn{2}{c}{F475W - F814W} & \colhead{Stellar} & \colhead{Blending} & \colhead{Ranking} & \colhead{$M_{\text{init}}$} & \colhead{Phase} & \multicolumn{1}{|c}{Cluster ID} &  \colhead{RA} & \colhead{DEC} &  \colhead{Age} & \colhead{$A_v$} & \colhead{Z}\\ [-0.35cm]
        \colhead{} & \multicolumn{2}{c}{[mag]} & \multicolumn{2}{c}{} & \colhead{} & \colhead{} & \colhead{} & \colhead{[$M_\odot$]} & \colhead{} & \multicolumn{1}{|c}{} &  \colhead{[$^\circ$]} & \colhead{[$^\circ$]} &  \colhead{[Log(yr)]} & \colhead{[mag]} & \colhead{}  \\  [-0.3cm]    
          \colhead{} &  \colhead{obs} & \colhead{iso} & \colhead{obs} & \colhead{iso} & \colhead{} & \colhead{} & \colhead{} & \colhead{} & \colhead{} & \multicolumn{1}{|c}{} &  \colhead{} & \colhead{} &  \colhead{} & \colhead{} & \colhead{}
    }
    \startdata
PHAT11.1262019+41.843634 & -5.616 & -5.278 & 4.339 & 1.678 & Ambiguous & Least & Bronze & 1.57 & postAGB & 230 & 11.127033 & 41.843671 & 9.30 & 0.06 & -0.40 \\
PHAT10.9115373+41.380738 & -6.587 & -6.411 & 3.285 & 3.864 & Ambiguous & Least & Bronze & 7.77 & TPAGB & 390 & 10.911579 & 41.380726 & 7.65 & 0.12 & 0.02 \\
PHAT11.1047292+41.346899 & -7.422 & -7.879 & 0.967 & -0.214 & Good & Least & Gold & 33.23 & shHB & 403 & 11.104662 & 41.347108 & 6.75 & 0.45 & 0.02 \\
PHAT11.1050349+41.347152 & -6.106 & -6.558 & 0.555 & -0.378 & Ambiguous & Likely & Bronze & 33.20 & MS & 403 & 11.104662 & 41.347108 & 6.75 & 0.45 & 0.02 \\
PHAT10.9517333+41.448087 & -6.820 & -5.055 & 1.798 & 3.041 & Ambiguous & Highly & Bronze & 1.08 & TPAGB & 445 & 10.951771 & 41.448099 & 9.65 & 0.22 & -1.80 \\
PHAT10.9517775+41.448073 & -6.366 & -5.055 & 1.732 & 3.041 & Ambiguous & Highly & Bronze & 1.08 & TPAGB & 445 & 10.951771 & 41.448099 & 9.65 & 0.22 & -1.80 \\
PHAT10.9517795+41.448104 & -6.000 & -5.055 & 1.454 & 3.041 & Ambiguous & Likely & Bronze & 1.08 & TPAGB & 445 & 10.951771 & 41.448099 & 9.65 & 0.22 & -1.80 \\
PHAT11.1676358+41.250191 & -5.897 & -3.798 & 1.464 & 1.961 & Ambiguous & Highly & Bronze & 0.99 & CHeB & 483 & 11.167649 & 41.250178 & 9.80 & 0.18 & -2.20 \\
PHAT11.1676294+41.250149 & -6.507 & -3.798 & 1.531 & 1.961 & Ambiguous & Likely & Bronze & 0.99 & CHeB & 483 & 11.167649 & 41.250178 & 9.80 & 0.18 & -2.20 \\
PHAT11.1676700+41.250131 & -5.686 & -3.798 & 1.415 & 1.961 & Ambiguous & Highly & Bronze & 0.99 & CHeB & 483 & 11.167649 & 41.250178 & 9.80 & 0.18 & -2.20 \\
PHAT11.6125244+42.031279 & -5.865 & -4.958 & 1.608 & 1.694 & Ambiguous & Highly & Bronze & 1.14 & postAGB & 540 & 11.612541 & 42.031319 & 9.65 & 0.12 & -0.80 \\
PHAT11.6125277+42.031316 & -6.359 & -4.958 & 1.706 & 1.694 & Ambiguous & Highly & Bronze & 1.14 & postAGB & 540 & 11.612541 & 42.031319 & 9.65 & 0.12 & -0.80 \\
PHAT11.1383029+41.886371 & -6.211 & -6.368 & 1.026 & 1.290 & Ambiguous & Highly & Bronze & 9.15 & CHeB & 609 & 11.138301 & 41.886391 & 7.50 & 0.56 & 0.02 \\
PHAT11.1381453+41.886387 & -7.120 & -6.673 & 2.228 & 3.836 & Ambiguous & Least & Bronze & 9.19 & shHeB & 609 & 11.138301 & 41.886391 & 7.50 & 0.56 & 0.02 \\
PHAT10.9923715+41.410639 & -5.779 & -4.958 & 1.744 & 1.694 & Good & Likely & Silver & 1.14 & postAGB & 626 & 10.992397 & 41.410646 & 9.65 & 0.12 & -0.80 \\
PHAT10.7148817+41.391778 & -6.826 & -7.067 & 3.274 & 0.067 & Good & Least & Gold & 15.14 & CHeB & 662 & 10.715145 & 41.391583 & 7.10 & 0.20 & 0.02 \\
PHAT10.7152697+41.391582 & -6.859 & -7.813 & 82.188 & 4.031 & Ambiguous & Least & Bronze & 16.35 & shHeB & 662 & 10.715145 & 41.391583 & 7.10 & 0.20 & 0.02 \\
PHAT10.9139786+41.168986 & -7.127 & -7.498 & 0.450 & 1.038 & Ambiguous & Least & Bronze & 13.49 & CHeB & 693 & 10.914089 & 41.169140 & 7.20 & 0.33 & 0.02 \\
PHAT11.4280933+41.756429 & -6.588 & -7.270 & 0.761 & -0.054 & Ambiguous & Least & Bronze & 18.17 & shHB & 711 & 11.427535 & 41.756568 & 7.00 & 0.68 & 0.02 \\
PHAT10.9665814+41.263916 & -6.148 & -5.976 & 0.564 & 1.327 & Ambiguous & Least & Bronze & 8.19 & CHeB & 742 & 10.966655 & 41.264048 & 7.60 & 0.53 & 0.02 \\
PHAT11.1754618+41.731831 & -5.704 & -5.025 & 1.671 & 3.861 & Good & Highly & Bronze & 1.04 & TPAGB & 757 & 11.175425 & 41.731819 & 9.75 & 0.11 & -1.00 \\
PHAT11.1932671+41.488258 & -6.142 & -4.743 & 1.581 & 1.678 & Ambiguous & Highly & Bronze & 1.02 & postAGB & 859 & 11.193255 & 41.488278 & 9.90 & 0.06 & -0.40 \\
PHAT11.2445990+41.910174 & -6.492 & -6.901 & 1.645 & 3.217 & Ambiguous & Likely & Bronze & 11.85 & shHeB & 980 & 11.244653 & 41.910184 & 7.30 & 0.57 & 0.02 \\
PHAT11.2446190+41.910229 & -6.533 & -6.326 & 1.055 & 0.768 & Ambiguous & Likely & Bronze & 11.20 & shHB & 980 & 11.244653 & 41.910184 & 7.30 & 0.57 & 0.02 \\
PHAT11.2445324+41.910222 & -6.196 & -6.713 & 2.356 & 2.959 & Good & Likely & Silver & 11.84 & CHeB & 980 & 11.244653 & 41.910184 & 7.30 & 0.57 & 0.02 \\
PHAT11.1798774+41.868042 & -7.034 & -7.179 & 3.081 & 4.074 & Ambiguous & Least & Bronze & 11.86 & shHeB & 1026 & 11.179935 & 41.868007 & 7.30 & 0.25 & 0.02 \\
PHAT10.8783069+41.354624 & -5.735 & -5.263 & 1.762 & 1.569 & Ambiguous & Highly & Bronze & 1.01 & postAGB & 1052 & 10.878286 & 41.354559 & 9.75 & 0.15 & -1.80 \\
PHAT10.8783503+41.354581 & -5.924 & -5.263 & 1.907 & 1.569 & Ambiguous & Highly & Bronze & 1.01 & postAGB & 1052 & 10.878286 & 41.354559 & 9.75 & 0.15 & -1.80 \\
PHAT10.9942367+41.505036 & -5.937 & -5.070 & 1.697 & 1.626 & Ambiguous & Highly & Bronze & 1.16 & postAGB & 1137 & 10.994251 & 41.505051 & 9.60 & 0.14 & -1.00 \\
PHAT10.9942370+41.505079 & -5.837 & -5.070 & 1.810 & 1.626 & Ambiguous & Highly & Bronze & 1.16 & postAGB & 1137 & 10.994251 & 41.505051 & 9.60 & 0.14 & -1.00 \\
PHAT11.1970592+41.473290 & -5.925 & -6.358 & 0.528 & -0.500 & Ambiguous & Least & Bronze & 51.83 & WR & 1230 & 11.197085 & 41.473312 & 6.65 & 0.38 & 0.02 \\
PHAT10.7799717+41.196765 & -5.637 & -4.836 & 2.098 & 1.664 & Ambiguous & Likely & Bronze & 1.03 & postAGB & 1309 & 10.779986 & 41.196790 & 9.95 & 0.14 & -0.20 \\
PHAT10.8142021+41.190278 & -5.758 & -5.074 & 1.623 & 1.721 & Good & Highly & Bronze & 1.16 & postAGB & 1416 & 10.814255 & 41.190281 & 9.60 & 0.13 & -1.00 \\
PHAT10.8142534+41.190277 & -6.687 & -5.074 & 1.553 & 1.721 & Good & Least & Gold & 1.16 & postAGB & 1416 & 10.814255 & 41.190281 & 9.60 & 0.13 & -1.00 \\
PHAT11.4220892+41.739675 & -7.219 & -7.658 & 3.353 & 4.095 & Ambiguous & Least & Bronze & 15.00 & shHeB & 1423 & 11.422059 & 41.739605 & 7.15 & 0.48 & 0.02 \\
PHAT10.7151784+41.234343 & -5.637 & -4.668 & 1.734 & 1.673 & Good & Highly & Bronze & 0.90 & postAGB & 1642 & 10.715190 & 41.234370 & 10.00 & 0.08 & -0.80 \\
PHAT11.0146309+41.510745 & -6.213 & -4.612 & 1.469 & 1.334 & Good & Highly & Bronze & 0.86 & postAGB & 1680 & 11.014600 & 41.510742 & 10.10 & 0.08 & -0.60 \\
PHAT10.7899451+41.359062 & -5.983 & -4.635 & 2.394 & 1.652 & Good & Likely & Silver & 0.86 & postAGB & 1797 & 10.789821 & 41.358924 & 10.10 & 0.31 & -0.60 \\
PHAT10.7898296+41.358822 & -5.911 & -4.635 & 2.450 & 1.652 & Good & Highly & Bronze & 0.86 & postAGB & 1797 & 10.789821 & 41.358924 & 10.10 & 0.31 & -0.60 \\
PHAT10.7897138+41.358912 & -5.856 & -4.635 & 2.187 & 1.652 & Good & Highly & Bronze & 0.86 & postAGB & 1797 & 10.789821 & 41.358924 & 10.10 & 0.31 & -0.60 \\
PHAT10.7896724+41.358971 & -5.865 & -4.635 & 2.363 & 1.652 & Good & Likely & Silver & 0.86 & postAGB & 1797 & 10.789821 & 41.358924 & 10.10 & 0.31 & -0.60 \\
PHAT10.8795858+41.304103 & -5.832 & -4.763 & 1.645 & 1.534 & Ambiguous & Highly & Bronze & 0.99 & postAGB & 1802 & 10.879593 & 41.304081 & 9.85 & 0.09 & -0.80 \\
PHAT10.8796413+41.304089 & -6.007 & -4.763 & 1.495 & 1.534 & Ambiguous & Highly & Bronze & 0.99 & postAGB & 1802 & 10.879593 & 41.304081 & 9.85 & 0.09 & -0.80 \\
PHAT11.1813291+41.443916 & -6.845 & -7.067 & 0.051 & 0.067 & Ambiguous & Least & Bronze & 15.14 & CHeB & 1825 & 11.181320 & 41.443749 & 7.10 & 0.21 & 0.02 \\
PHAT11.4242884+41.759313 & -5.884 & -4.836 & 1.878 & 1.664 & Ambiguous & Likely & Bronze & 1.03 & postAGB & 1920 & 11.424344 & 41.759294 & 9.95 & 0.07 & -0.20 \\
PHAT11.4242968+41.759278 & -6.389 & -4.836 & 1.983 & 1.664 & Ambiguous & Highly & Bronze & 1.03 & postAGB & 1920 & 11.424344 & 41.759294 & 9.95 & 0.07 & -0.20 \\
PHAT11.4243211+41.759318 & -6.103 & -4.836 & 1.821 & 1.664 & Good & Likely & Silver & 1.03 & postAGB & 1920 & 11.424344 & 41.759294 & 9.95 & 0.07 & -0.20 \\
PHAT11.4243543+41.759330 & -6.063 & -4.836 & 1.884 & 1.664 & Good & Highly & Bronze & 1.03 & postAGB & 1920 & 11.424344 & 41.759294 & 9.95 & 0.07 & -0.20 \\
PHAT10.8380804+41.235576 & -5.641 & -4.683 & 2.767 & 1.660 & Ambiguous & Likely & Bronze & 0.89 & postAGB & 2012 & 10.838010 & 41.235617 & 10.10 & 0.08 & -0.40 \\
PHAT10.8582937+41.358786 & -5.541 & -4.710 & 1.740 & 1.710 & Ambiguous & Highly & Bronze & 0.95 & postAGB & 2125 & 10.858304 & 41.358815 & 10.00 & 0.06 & -0.40 \\
PHAT11.0284722+41.674479 & -5.555 & -5.661 & -0.136 & -0.378 & Ambiguous & Least & Bronze & 22.40 & shHB & 2143 & 11.028640 & 41.674386 & 6.90 & 0.07 & 0.02 \\
PHAT11.2413408+41.489948 & -5.619 & -5.148 & 1.641 & 1.175 & Good & Highly & Bronze & 1.30 & postAGB & 2348 & 11.241335 & 41.489975 & 9.55 & 0.07 & -0.40 \\
PHAT10.8938331+41.161436 & -5.923 & -6.264 & 0.221 & -0.505 & Ambiguous & Least & Bronze & 66.15 & WR & 2672 & 10.893817 & 41.161411 & 6.60 & 0.36 & 0.02 \\
PHAT10.7833409+41.305080 & -5.753 & -4.890 & 1.949 & 1.731 & Good & Highly & Bronze & 1.33 & postAGB & 2683 & 10.783367 & 41.305101 & 9.70 & 0.09 & 0.20 \\
PHAT11.1534065+41.876700 & -6.605 & -6.909 & 3.882 & 3.238 & Ambiguous & Likely & Bronze & 11.85 & shHeB & 2805 & 11.153462 & 41.876728 & 7.30 & 0.30 & 0.02 \\
PHAT10.9480593+41.452227 & -6.173 & -4.907 & 2.097 & 0.968 & Ambiguous & Highly & Bronze & 1.17 & postAGB & 3088 & 10.948064 & 41.452235 & 9.70 & 0.09 & -0.40 \\
PHAT10.7637717+41.356039 & -7.180 & -4.878 & 1.929 & 1.705 & Good & Highly & Bronze & 1.04 & postAGB & 3510 & 10.763763 & 41.356043 & 10.00 & 0.06 & 0.00 \\
PHAT10.7637208+41.356041 & -6.125 & -4.879 & 1.658 & 1.613 & Good & Highly & Bronze & 1.04 & postAGB & 3510 & 10.763763 & 41.356043 & 10.00 & 0.06 & 0.00 \\
PHAT10.6854970+41.244921 & -5.909 & -5.001 & 2.039 & 1.686 & Good & Highly & Bronze & 1.17 & postAGB & 3704 & 10.685397 & 41.244901 & 9.65 & 0.20 & -0.60 \\
PHAT10.6853726+41.244896 & -6.304 & -5.001 & 1.815 & 1.686 & Good & Highly & Bronze & 1.17 & postAGB & 3704 & 10.685397 & 41.244901 & 9.65 & 0.20 & -0.60 \\
PHAT10.6854787+41.244965 & -6.148 & -5.001 & 2.231 & 1.686 & Ambiguous & Likely & Bronze & 1.17 & postAGB & 3704 & 10.685397 & 41.244901 & 9.65 & 0.20 & -0.60 \\
PHAT10.6854764+41.244798 & -6.378 & -5.001 & 2.027 & 1.686 & Good & Likely & Silver & 1.17 & postAGB & 3704 & 10.685397 & 41.244901 & 9.65 & 0.20 & -0.60 \\
PHAT10.6854201+41.244921 & -6.500 & -5.001 & 1.952 & 1.686 & Ambiguous & Highly & Bronze & 1.17 & postAGB & 3704 & 10.685397 & 41.244901 & 9.65 & 0.20 & -0.60 \\
PHAT10.6854053+41.244830 & -6.364 & -4.998 & 1.741 & 1.600 & Ambiguous & Highly & Bronze & 1.17 & postAGB & 3704 & 10.685397 & 41.244901 & 9.65 & 0.20 & -0.60 \\
PHAT10.6854095+41.244870 & -6.450 & -5.001 & 1.656 & 1.686 & Good & Highly & Bronze & 1.17 & postAGB & 3704 & 10.685397 & 41.244901 & 9.65 & 0.20 & -0.60 \\
PHAT10.6853291+41.244909 & -5.719 & -4.998 & 1.716 & 1.600 & Good & Likely & Silver & 1.17 & postAGB & 3704 & 10.685397 & 41.244901 & 9.65 & 0.20 & -0.60 \\
PHAT10.6855112+41.244854 & -6.368 & -5.001 & 1.989 & 1.686 & Ambiguous & Likely & Bronze & 1.17 & postAGB & 3704 & 10.685397 & 41.244901 & 9.65 & 0.20 & -0.60 \\
PHAT10.6853552+41.244936 & -5.785 & -4.998 & 1.846 & 1.600 & Good & Highly & Bronze & 1.17 & postAGB & 3704 & 10.685397 & 41.244901 & 9.65 & 0.20 & -0.60 \\
PHAT10.6853832+41.244965 & -5.876 & -5.001 & 2.052 & 1.686 & Good & Highly & Bronze & 1.17 & postAGB & 3704 & 10.685397 & 41.244901 & 9.65 & 0.20 & -0.60 \\
PHAT10.6853209+41.244828 & -6.286 & -4.998 & 1.806 & 1.600 & Good & Highly & Bronze & 1.17 & postAGB & 3704 & 10.685397 & 41.244901 & 9.65 & 0.20 & -0.60 \\
PHAT10.6852844+41.244888 & -5.723 & -5.001 & 1.826 & 1.686 & Good & Highly & Bronze & 1.17 & postAGB & 3704 & 10.685397 & 41.244901 & 9.65 & 0.20 & -0.60 \\
PHAT10.6853654+41.244792 & -5.932 & -5.001 & 1.794 & 1.686 & Ambiguous & Highly & Bronze & 1.17 & postAGB & 3704 & 10.685397 & 41.244901 & 9.65 & 0.20 & -0.60 \\
PHAT10.6853373+41.244875 & -6.141 & -5.001 & 1.823 & 1.686 & Good & Highly & Bronze & 1.17 & postAGB & 3704 & 10.685397 & 41.244901 & 9.65 & 0.20 & -0.60 \\
PHAT10.6853590+41.244846 & -6.093 & -5.001 & 1.830 & 1.686 & Good & Likely & Silver & 1.17 & postAGB & 3704 & 10.685397 & 41.244901 & 9.65 & 0.20 & -0.60 \\
PHAT10.7494354+41.268271 & -7.016 & -4.999 & 1.914 & 1.648 & Good & Least & Gold & 1.19 & postAGB & 3801 & 10.749432 & 41.268276 & 9.80 & 0.07 & 0.00 \\
PHAT11.1832836+41.440709 & -5.724 & -5.802 & 1.503 & 1.573 & Ambiguous & Least & Bronze & 9.81 & shHB & 3909 & 11.182722 & 41.439833 & 7.40 & 0.07 & 0.02 \\
PHAT10.7485055+41.322014 & -6.496 & -4.896 & 1.918 & 1.660 & Good & Highly & Bronze & 1.07 & postAGB & 3983 & 10.748530 & 41.322039 & 9.95 & 0.08 & 0.00 \\
PHAT10.7485231+41.322048 & -6.628 & -4.896 & 1.860 & 1.660 & Good & Highly & Bronze & 1.07 & postAGB & 3983 & 10.748530 & 41.322039 & 9.95 & 0.08 & 0.00 \\
PHAT11.1712113+41.875049 & -7.620 & -7.179 & 3.784 & 4.074 & Ambiguous & Least & Bronze & 11.86 & shHeB & 4434 & 11.171222 & 41.875064 & 7.30 & 0.13 & 0.02 \\
PHAT11.1838224+41.443995 & -6.805 & -6.948 & 0.149 & -0.001 & Ambiguous & Least & Bronze & 15.14 & CHeB & 4775 & 11.182554 & 41.443443 & 7.10 & 0.17 & 0.02 \\
PHAT10.6385785+41.295080 & -5.875 & -4.912 & 2.070 & 1.571 & Ambiguous & Highly & Bronze & 1.09 & postAGB & 5002 & 10.638539 & 41.295119 & 10.00 & 0.08 & 0.20 \\
PHAT10.6385439+41.295103 & -6.625 & -4.907 & 2.152 & 1.740 & Good & Highly & Bronze & 1.09 & postAGB & 5002 & 10.638539 & 41.295119 & 10.00 & 0.08 & 0.20 \\
PHAT10.7550717+41.269539 & -5.691 & -5.484 & 2.009 & 1.634 & Good & Highly & Bronze & 1.29 & postAGB & 5105 & 10.755098 & 41.269581 & 9.75 & 0.09 & 0.20 \\
PHAT10.7550909+41.269571 & -6.522 & -5.491 & 1.976 & 1.316 & Good & Highly & Bronze & 1.29 & postAGB & 5105 & 10.755098 & 41.269581 & 9.75 & 0.09 & 0.20 \\
PHAT10.6480674+41.242849 & -6.339 & -5.409 & 1.531 & 3.499 & Ambiguous & Highly & Bronze & 1.25 & TPAGB & 5117 & 10.647973 & 41.242830 & 9.45 & 0.25 & -1.80 \\
PHAT10.6920206+41.293412 & -5.895 & -4.896 & 1.977 & 1.660 & Good & Highly & Bronze & 1.07 & postAGB & 8052 & 10.692013 & 41.293419 & 9.95 & 0.07 & 0.00 \\
\enddata
\label{table:finalisochroneresults}
\end{deluxetable*}
\end{longrotatetable}

\section{Acknowledgments} \label{Acknowledgements}
We thank the anonymous reviewer for their valuable feedback.
Based on observations with the NASA/ESA Hubble Space Telescope
obtained from the Mikulski Archive for Space Telescopes (MAST) at the Space Telescope Science Institute (STScI), which is operated by the Association of Universities for Research in Astronomy, Incorporated, under NASA contract NAS5-26555. Support for Program number HST-AR-17064 was provided through a grant from the STScI under NASA contract NAS5-26555. PT, SZ, and JL conducted this research under the auspices of the Science Internship Program (SIP) at UCSC.

\emph{Software:} 
Astro Data Lab \citep{Fitzpatrick2014DataLab},
ARTES \citep{stolker17}, PICASO \citep{batalha19}, pandas \citep{mckinney2010data}, NumPy \citep{walt2011numpy}, IPython \citep{perez2007ipython}, Jupyter \citep{kluyver2016jupyter}, matplotlib \citep{Hunter:2007}, astroalign \citep{Astroalign}, astropy \citep{astropy:2018}, optimal image subtraction (OIS) \citep{martinberoiz2020}, scikit learn \citep{scikit-learn}

\bibliography{main}{}
\bibliographystyle{aasjournal}

\appendix  \label{Appendix}

\section{Astro Data Lab Access}

The NSF NOIRLab \emph{Astro Data Lab} \citep{Fitzpatrick2014DataLab} is an online astronomical data storage and analysis service, accessible at \url{https://datalab.noirlab.edu/}. 
A free Astro Data Lab user account is required to access the data used in this work, which can be generated at \url{https://datalab.noirlab.edu/account/register.html}. The account will then need to be verified by the Astro Data Lab Team before the service can be accessed. 

Further information on Astro Data Lab can be found at \url{https://datalab.noirlab.edu/docs/manual/index.html} and via the User Forum at \url{https://datalab.noirlab.edu/help/}. 

\section{Accessing data from this work}

The data used in this work are stored in the public directory on Astro Data Lab's JupyterLab service. 
A \texttt{JupyterLab} session is opened using the ``Launch a Jupyter Notebook'' button or under the Quick Start menu on the Astro Data Lab homepage, as shown:
\begin{center}
\includegraphics[width = \textwidth]{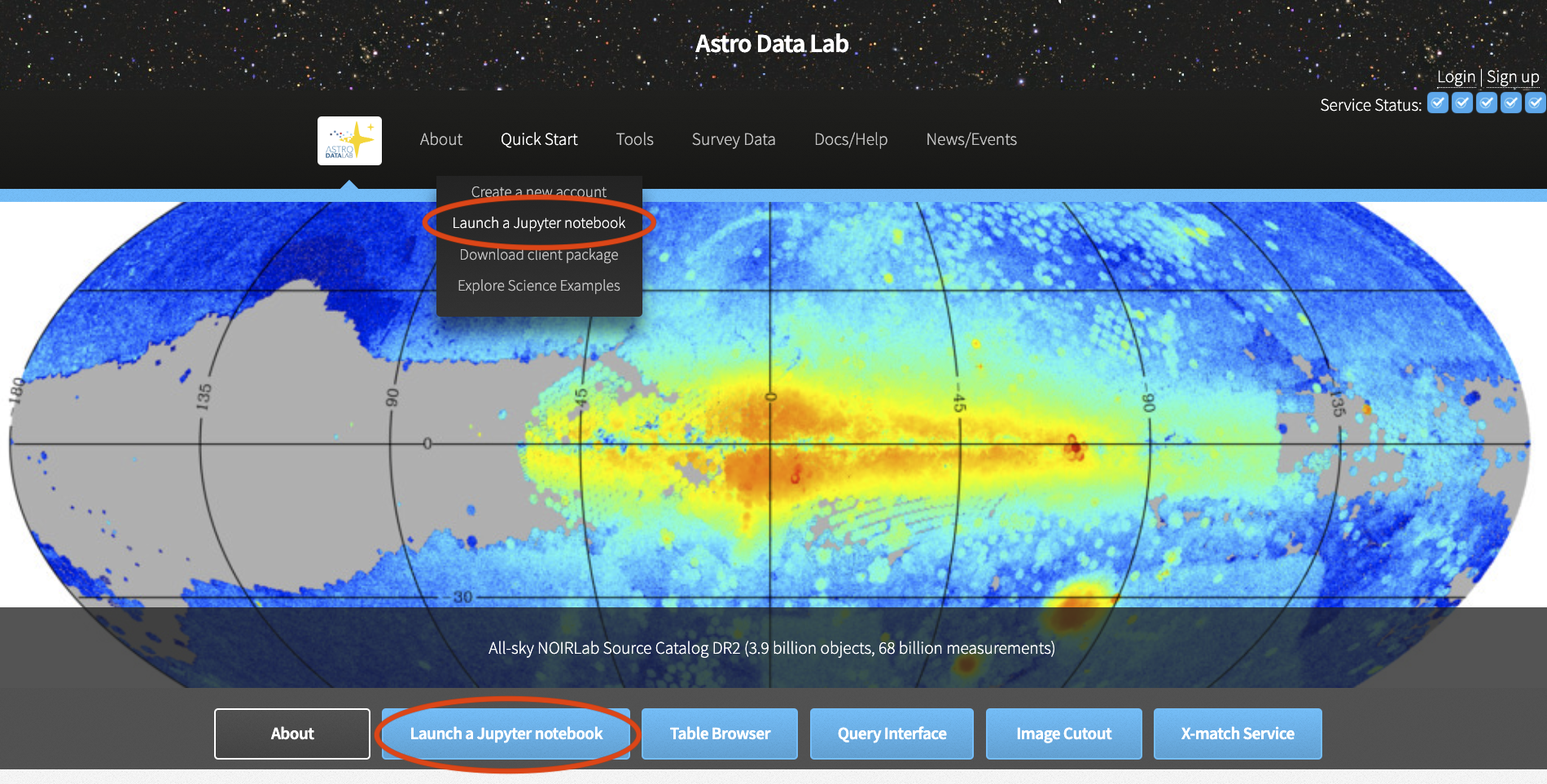}
    \label{fig:OpenJupyter}
\end{center}

In JupyterLab, a Python 3 environment and kernel can be initiated from the Launcher window to access the data directory:
\begin{center}
\includegraphics[width = \textwidth]{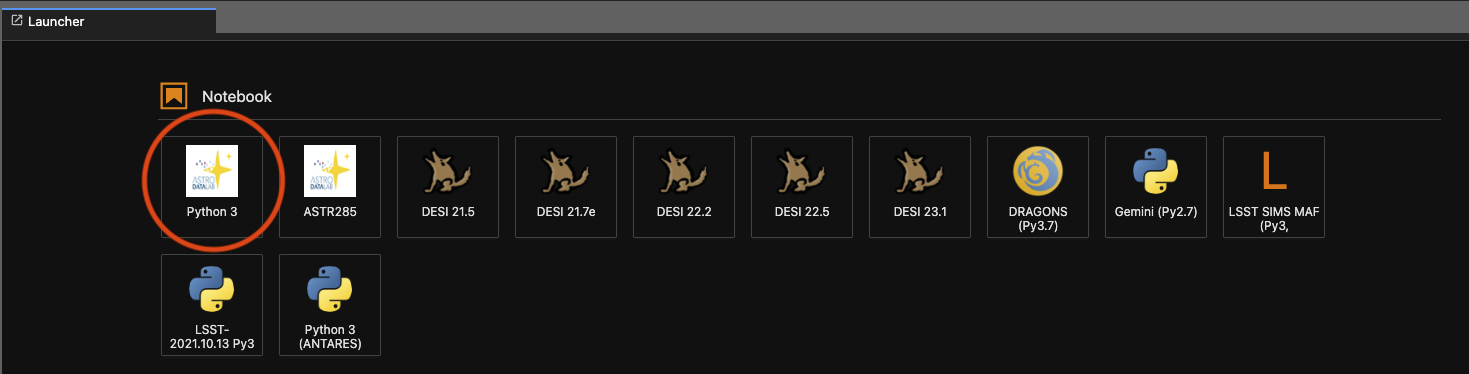}
    \label{fig:OpenPython}
\end{center}

The data used in this work are stored in the public directory \texttt{aspatel$\colon$//public/data\_products}, accessible through the Astro Data Lab client using the following Python 3 cell commands:

\begin{python} 

from dl import authClient as ac, storeClient as sc
auth_token = ac.login('YOUR__USERNAME','YOUR__PASSWORD')
sc.ls("aspatel://public/data_products")
\end{python}

\noindent using your Astro Data Lab username and password in place of \texttt{YOUR\_\_USERNAME} and \texttt{YOUR\_\_PASSWORD}, respectively. 
(Further information on using and accessing data from the Astro Data Lab StoreClient, including commands, is provided at \url{https://datalab.noirlab.edu/docs/manual/UsingAstroDataLab/ClientInterfaces/StoreClient/StoreClient.html}.)

Running the above cell will return a list of directories and files: \texttt{README, lcs.dat, tab2.dat, tab3.dat}.
\begin{itemize}
    \item \texttt{tab2.dat} contains the catalog of 239 candidate cluster-variable stars identified through light curve analysis described in Section \ref{subsec:lightcurveanalysis}. 
    \item \texttt{tab3.dat} contains the final sample of 86 confirmed cluster-variable stars along with their inferred evolutionary phases, initial masses and host cluster properties. 
    \item \texttt{lcs.dat} contains the light curve data for all 239 candidate cluster-variable stars identified through light curve analysis described in Section \ref{subsec:lightcurveanalysis}.  
\end{itemize}
All of the details regarding these tables are provided in the \texttt{README} file. 
For further context, the snippet from \texttt{ReadMe} in Figure \ref{fig:readme_ex} demonstrates how the tables and directories are organized:

\begin{figure}[!h]
    \centering
    \includegraphics[width=\textwidth]{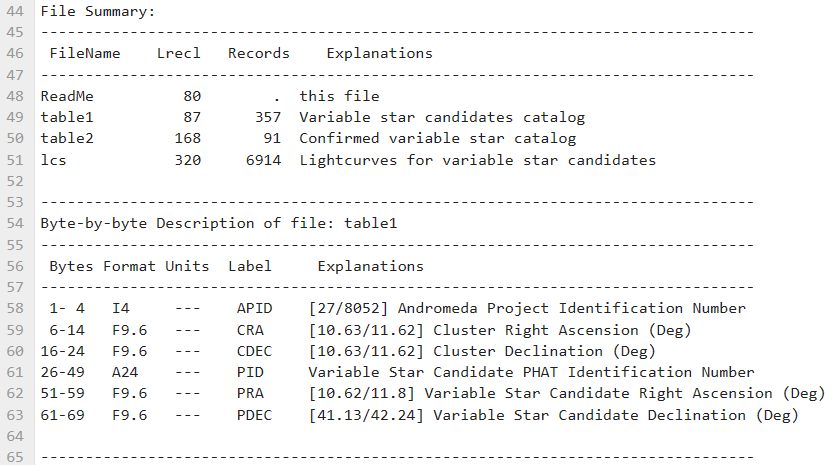}
    \caption{Example portion of the README text.}
    \label{fig:readme_ex}
\end{figure}

The data, tables and \texttt{README.txt} files can be copied from \texttt{aspatel$\colon$//public/data\_products} into another directory under a different user with the command:
\begin{python}
        sc.get("aspatel://public/data_products/" , "../DESTINATION__DIRECTORY")
\end{python}
using the name of your directory to store all data products in place of \texttt{../DESTINATION\_\_DIRECTORY}.  

\section{Accessing additional PHAT survey data}

Additional data from the \texttt{phot\_meas} and \texttt{phot\_mod} tables, which hold the \emph{per-exposure} and \emph{combined} (across exposures) photometric measurements from the PHAT survey, respectively, stored on Astro Data Lab can be accessed using the \texttt{QueryClient} in JupyterLab. 
The example code block below demonstrates how to retrieve all data from \texttt{phot\_meas} (\texttt{query\_all}) vs.\ specific column values (\texttt{query\_cols}) for a given PHAT target, returning the results of each query as \emph{pandas} dataframes. 
Our \texttt{phot\_meas} queries include \texttt{magvega < 99.0}, to filter exposures with non-detections, and the sharpness threshold \texttt{sharp\string^2 < 0.2} to exclude measurements affected by cosmic ray strikes, bad pixels, etc. as described in Section \ref{subsec:lightcurvecalibration}. 
The columns available for query are described at \url{https://datalab.noirlab.edu/query.php?name=phat_v2.phot_mod} for \texttt{phot\_mod} and \url{https://datalab.noirlab.edu/query.php?name=phat\_v2.phot_meas} for \texttt{phot\_meas}. 
The \texttt{phot\_mod} table can be queried by changing the \texttt{phot\_meas} entry in the code block, along with any relevant table column headers specified in the query. 
\begin{python}
phat_id = "PHAT_10.9517775+41.448073"

query_all = """SELECT *
                    FROM phat_v2.phot_meas
                    WHERE objid = '
                """ 
try:
    result1 = qc.query(auth_token, sql=query_all, timeout=400) 
except Exception as e:
    print(e)
df_all = helpers.utils.convert(result1,'pandas')

query_cols = """SELECT mjd, magvega, magerr,sharp
                    FROM phat_v2.phot_meas
                    WHERE objid = '
                """ 
try:
    result2 = qc.query(auth_token, sql=query_cols, timeout=400) 
except Exception as e:
    print(e)
df_cols = helpers.utils.convert(result2,'pandas')

\end{python}

\newpage
\section{Foreground Star Identification} \label{App:ForegroundStars}

We cross-match our initial sample of 376 luminous (F814W $< 19$~mag) PHAT cluster stars (see Table~\ref{tab1:number_breakdown}) with the Gaia DR3 catalog available through \textit{Astro Data Lab}\footnote{\url{https://datalab.noirlab.edu/gaia.php}} using a search radius of $0\farcs11$, comparable to the FWHM of the PSF for both ACS \citep{Santiago2010} as well as Gaia \citep{Fabricius-2016}. We obtain 204 PHAT stars with Gaia counterparts.

We designate potential foreground stars in the resulting sample using the same criteria described by \cite{Barmby2023}, with foreground stars having either: $\mathrm{i)}$ significant non-zero parallax \textit{and} proper motion inconsistent with that of M31, or $\mathrm{ii)}$ significant non-zero proper motion also inconsistent with that of M31. We adopt a significance level of  $n=10$ below. 
We determine the significant non-zero parallax condition by adopting the zeropoint-corrected parallax signal-to-noise ratio described in \cite{Barmby2023}:
\begin{equation}
    |\frac{\bar{\omega} + 0.017}{\sigma_{\bar{\omega}}}| > n
\end{equation}
\noindent where 
$\bar{\omega}$ and $\sigma_{\bar{\omega}}$ are the parallax and parallax error in mas, respectively; $0.017$~mas is the median parallax zeropoint of Gaia DR3 determined by \cite{Lindegren2021}.
The significant non-zero proper motion condition is determined following:
\begin{equation}
    \left[\mu_{ra},\mu_{dec}\right] \mathrm{Cov^{-1}} \left[\frac{\mathrm{\mu_{ra}}}{\mathrm{\mu_{dec}}}\right] > n^{2}
\end{equation}
\noindent which can be queried directly from the Gaia DR3 databases, as detailed in \cite{Barmby2023}, where 
$\mu_{ra}$ and $\mu_{dec}$ are the right ascension and declination proper motions of the star in mas/yr, respectively.
We use the proper motion of M31 measured by \cite{Salomon2021} ($48.9 \pm 10.5$~mas and $-36.9 \pm 8.1$~mas in right ascension and declination, respectively) to check for inconsistency between stellar proper motions with that of M31.

Of the 204 luminous cluster stars in the PHAT catalog with crossmatched counterparts in Gaia DR3, 142 stars have parallax and proper motion measurements in Gaia DR3, of which 7 stars meet the foreground star criteria defined above. 
Our analysis of these stars described in Section \ref{SubSubSec:DISourceDetection} reveals 3 of the forground stars -- $\mathrm{PHAT\_11.1765404+41.919860}$, 
$\mathrm{PHAT\_11.3058193+41.626605}$ and
$\mathrm{PHAT\_11.0161713+41.276141}$ -- have difference image source detections. We thus exclude these 3 stars from our 89 confirmed cluster-variable stars identified in Section \ref{SubSubSec:DISourceDetection}. 
Of the remaining 86 cluster-variable stars, we note that 47 of them are included in the above sample of 204 stars with Gaia DR3 counterparts, one of which -- $\mathrm{PHAT\_10.9665814+41.263916}$ -- has also been observed to be variable with Gaia.

\end{document}